\documentclass{aa}
\usepackage{graphicx,subfig}
\usepackage{txfonts}
\begin{document}
\title{A search for water maser emission toward obscured post-AGB star and planetary nebula candidates}

\author{J.F. G\'omez \inst{1}
	\and J.R. Rizzo \inst{2}
	\and O. Su\'arez \inst{3}
	\and A. Palau \inst{4}
	\and L.F. Miranda \inst{1}
	\and M.A. Guerrero \inst{1}
	\and G. Ramos-Larios \inst{5}
	\and J.M. Torrelles \inst{6,7}
                }

\offprints{J.F. G\'omez}

\institute{Instituto de Astrof\'{\i}sica de Andaluc\'{\i}a, CSIC, Glorieta de la Astronom\'{\i}a s/n, E-18008 Granada, Spain, jfg@iaa.es
\and
Centro de Astrobiolog\'{\i}a, INTA-CSIC, Carretera M-108, km. 4, E-28850 Torrej\'on de Ardoz, Madrid, Spain
\and
UMR 6525 H.Fizeau, Universit\'e de Nice Sophia Antipolis, CNRS, OCA. Parc
  Valrose, F-06108 Nice Cedex 2, France
\and
Centro de Radioastronom\'{\i}a y Astrof\'{\i}sica, UNAM, Campus
Morelia, Apdo. Postal 3-72, Morelia, Michoac\'an 58089, M\'exico
\and Instituto de Astronom\'{\i}a y Meteorolog\'{\i}a, CUCEI, Universidad de Guadalajara, Av. Vallarta No. 2602, Col. Arcos Vallarta, 44130 Guadalajara,
Jalisco, M\'exico
\and Institut de Ci\`encies de l'Espai (CSIC)/IEEC, Barcelona, Spain
\and
Institut de Ci\`encies del Cosmos, Universitat de Barcelona (associated unit of CSIC through the ICE), Mart\'{\i} i Franqu\`es 1, E-08028 Barcelona, Spain}
\date{Received <date>/Accepted <date>}

\abstract
%   % context heading (optional)
  {Water maser emission at 22 GHz is a useful probe for studying the transition between the nearly spherical mass loss in the asymptotic giant branch (AGB) to a collimated one in the post-AGB phase. In their turn, collimated jets in the post-AGB phase could determine the shape of planetary nebulae once photoionization starts.} %leave it empty if necessary  
   % aims heading (mandatory)
{We intend to find new cases of post-AGB stars and planetary nebulae (PNe) with water maser emission, including some especially interesting and rare types: water fountains (evolved objects with high velocity
 collimated jets traced by water masers) or water-maser-emitting PNe.  Since previous studies have shown a higher detection rate of water maser emission in evolved objects that are optically obscured, we selected a sample that contains a significant fraction of post-AGB and young PN candidate sources showing signs of strong obscuration.}
%   % methods heading (mandatory)
{We searched for water maser emission in 133 evolved objects using the radio telescopes in Robledo de Chavela, Parkes, and Green Bank.} 
%   % results heading (mandatory)
    {We detected water maser emission in 15 sources of our sample, of which seven are reported here for the first time (IRAS 13483$-$5905, IRAS 14249$-$5310, IRAS 15408$-$5413, IRAS 17021$-$3109, IRAS 17348$-$2906, IRAS 17393$-$2727, and IRAS 18361$-$1203). We identified three water fountain candidates:  IRAS 17291$-$2147, with a total velocity spread of $\simeq 96$ km s$^{-1}$ in its water maser components and  two sources (IRAS 17021$-$3109 and IRAS 17348$-$2906) that show water maser emission whose velocity lies outside the velocity range covered by OH masers. We have also identified IRAS 17393$-$2727 as a possible new water-maser-emitting PN.}
%   % conclusions heading (optional), leave it empty if necessary
    {The detection rate is higher in obscured objects (14\%) than in those with optical counterparts (7\%), which is consistent with previous results. Water maser emission seems to be common  in objects that are bipolar in the near-IR (43\% detection rate in such sources). The water maser spectra of water fountain candidates like IRAS 17291$-$2147 show significantly fewer maser components than others (e.g., IRAS 18113$-$2503). We speculate that most post-AGBs may show water maser emission with wide enough velocity spread ($\geq 100$ km  s$^{-1}$) when observed with enough sensitivity and/or for long enough periods of time. Therefore, it may be necessary to single out a special group of ``water fountains'', probably defined by their high maser luminosities. We also suggest that the presence of both water and OH masers in a PN is a better tracer of its youth, than is the presence of just one of these species.}

\authorrunning{G\'omez et al.}
\titlerunning{water maser emission in obscured evolved stars}
\maketitle
%
%________________________________________________________________
%
\section{Introduction}

\label{intro}
Maser emission is present in different astrophysical environments (e.g., star-forming regions, evolved stars, active galactic nuclei), and it is a signpost of energetic phenomena, since strong shocks or intense radiation fields are required to produce the population inversion of energy levels necessary for maser amplification. 
In the particular case of evolved stars, maser emission is common in the circumstellar structures of O-rich sources during the asymptotic giant branch (AGB) phase \citep{lew89}, characterized by strong mass loss.

 Especially conspicuous is the presence of double-peaked OH maser profiles \citep[see, e.g.,][]{sev97}. It is widely accepted that maser emission in these sources traces the expanding circumstellar envelope \citep{rei77}. The velocity separation between the OH peaks is $\sim 30$ km s$^{-1}$, and it provides a good estimate of the expansion velocity of the envelope (the expansion velocity should be half of the OH peak separation) at the location of the OH maser emission \citep[$\ga 10^3$ AU,][]{rei81}, whereas the mean OH velocity accurately traces the stellar velocity along the line of sight. SiO and H$_2$O masers are also widespread in AGB stars, and they seem to arise from gas volumes closer to the star than the OH emission. These inner zones may have higher expansion velocities, but they also show larger velocity gradients. Since maser emission requires a sufficient column density of molecules with coherent velocities along the line of sight, SiO and H$_2$O masers tend to arise from gas moving in directions closer to the plane of the sky than for emitting OH masers, and the observed velocity along the line of sight is lower. The result is that the velocity components in the spectra of SiO and H$_2$O masers are contained within the velocity range of OH.

Maser emission is also present in subsequent phases \citep{eng02}, but with much lower prevalence. When the AGB mass loss stops and the star enters the post-AGB phase, the maser emission from the expanding envelope is expected to disappear on timescales of around 10, 100, and 1000 yr for SiO, H$_2$O, and OH masers, respectively \citep{lew89,gom90}. However, although the mass-loss rates during the post-AGB phase are much lower \citep[$\la 10^{-7}$ M$_\odot$ yr$^{-1}$,][]{vas94,blo95b} than during the AGB \citep[up to $\simeq 10^{-4}$ M$_\odot$ yr$^{-1}$,][]{vas93,blo95a}, it may take the form of high-velocity, collimated jets, which also carry enough energy to pump maser emission. 
Therefore, this emission is an excellent tool for studying the transition from nearly spherical to collimated mass loss. This is a key issue, since the effect of jets on the circumstellar envelopes during the post-AGB phase may determine the shape of PNe once ionization starts.

The paradigmatic examples of collimated jets traced by maser emission are the small group of sources called ``water fountains'' \citep{ima07a}. These are objects (most of them post-AGB stars) whose H$_2$O maser emission traces high-velocity ($\ga 100$ km s$^{-1}$), collimated jets, with very short dynamical ages ($\la 100$ yr). These objects are usually identified by the large velocity spread in their water maser components \citep[up to $\simeq 500$ km s$^{-1}$][]{gom11}, amply surpassing the velocity range of OH emission. 
A firm confirmation of their nature as ``water fountains'' requires interferometric observations, to ascertain that there is no confusion of maser components from different sources within a single-dish beam  and to determine the bipolar nature of the maser emission, as expected from a jet. 
The short dynamical ages of ``water fountains'' jets suggest that they may represent one of the first manifestations of collimated mass loss in evolved stars. Thirteen ``water fountains'' have been confirmed so far \citep{gom15b}.

When the central star is hot enough to photoionize the surrounding envelope, the source enters the planetary nebula (PN) phase. The presence of maser emission is uncommon in PNe, and it may be restricted to the very first stages of this phase. No PN has been found to harbor SiO masers, while only six have been confirmed so far as OH-maser emitters \citep{usc12}, usually known as OHPNe \citep{zij89}, and five have H$_2$O masers \citep[H$_2$O-PNe,][]{mir01,ideg04,gom08,usc14,gom15a}. Of these, two harbor both OH and H$_2$O masers. Maser emission in PNe seems to be exclusively associated to bipolar objects \citep{usc14}, and it has been suggested that maser-emitting PNe may have ``water fountains'' as their precursors \citep{gom08}. However, as opposed to ``water fountains'', OH and H$_2$O maser emission in PNe shows very low velocities \citep[with the notable exception of IRAS 15103$-$5754,][]{gom15a}, and may primarily trace circumstellar toroidal structures. In any case, masers in PNe may not be the remnant of those in the AGB phase, but seem to be related to mass-loss processes started in subsequent phases. This is because, on the one hand, the short timescale of survival of H$_2$O maser emission after the AGB (100 yr) would prevent its detection in PNe. On the other hand, while OH masers pumped in the AGB may still survive in extremely young PNe, some relatively evolved PNe are also OH emitters \citep{usc12}, indicating 
that new OH masers can appear after the end of the AGB.

In summary, maser emission is a good tracer of energetic mass-loss processes in objects on their way to becoming PNe. 
With this in mind, we undertook an extensive and sensitive search for water maser emission in post-AGB stars and young PNe \citep{sua07,sua09}. From these past observations, it was evident that water maser emission appears to be more common in optically obscured sources. In this paper we present a new sensitive survey for water maser emission. A significant portion of the sample consists of  post-AGB and young PN candidate sources showing signs of strong obscuration.

\section{Source sample}

The observed sources are listed in Table \ref{observadas}. They comprise 
most of the sources in \citet{rl1}. They are post-AGB stars and PN candidates with the IRAS color criteria of \citet{sua06} and with signs of strong optical obscuration. We have also included some optically visible post-AGB stars from \citet{sua06} that were not included in our previous water maser observations of \citet{sua07,sua09} or for which those observations had poor sensitivity. There are previous water maser observations of some sources in our sample, which are also cited in Table \ref{observadas} \citep[including those by][who also selected 76 sources from Ramos-Larios et al. 2009 as part of their sample of 164 post-AGB candidates]{yoo14}, although our data have  significantly higher sensitivity (over one order of magnitude more in most cases). A total of 133 sources were observed.

\section{Observations}

We observed the $6_{16}-5_{23}$ transition of H$_2$O (rest frequency = 22235.08 MHz) using three different telescopes: the DSS-63 antenna (70 m diameter) at the Madrid Deep Space Communications Complex (MDSCC) near Robledo de Chavela (Spain), the 64 m antenna at the Parkes Observatory of the Australia Telescope National Facility (ATNF), and  the 100 m Robert C. Byrd Green Bank Telescope (GBT) of the National Radio Astronomy Observatory. The observed positions, rms noise per spectral channel, and observing dates are listed in Table \ref{observadas}.
All spectra were centered at $V_{\rm LSR} = 0.0$ km s$^{-1}$. The only exceptions are the spectra toward IRAS 18113$-$2503, which were formed from two independent integrations centered on different velocities, to cover the whole velocity range of detected components better.
In the case of the Robledo data, for which Doppler tracking was not available, the central velocity could vary within $V_{\rm LSR} = 0.0\pm 0.5$ km s$^{-1}$. All spectra were corrected for the elevation-dependent gain of each telescope and for atmospheric opacity.

\onltab{
\renewcommand{\thefootnote}{\alph{footnote}}
\onecolumn
\begin{longtab}
\begin{longtable}{ccccllllc}
\caption{Observed sources \label{observadas}}\\
\hline \hline 
IRAS name & RA (J2000) &  Dec (J2000)& Observation date & Telescope\footnotemark[1] & rms\footnotemark[2] (Jy) & Previous\footnotemark[3] (Jy) & References\footnotemark[4]  & Image\footnotemark[5] \\
\hline
\endfirsthead
\caption{Observed sources (continued)}\\
\hline\hline
IRAS name & RA (J2000) &  Dec (J2000)& Observation date & Telescope\footnotemark[1] & rms\footnotemark[2] (Jy) & Previous\footnotemark[3] (Jy) & References\footnotemark[4]  & Image\footnotemark[5] \\
\hline
\endhead
\hline
\endfoot
07582$-$4059 &  07 59 57.7 & $-41$ 07 23 & 14-SEP-2008  & PKS & 0.16 & & & V\\
08143$-$4406 &  08 16 03.0 & $-44$ 16 04 & 14-SEP-2008  & PKS & 0.18 & & & V\\
08213$-$3857 &  08 23 12.1 & $-39$ 07 08 & 14-SEP-2008  & PKS & 0.12 & & & V \\
09055$-$4629 &  09 07 19.5 & $-46$ 41 23 & 12-SEP-2008  & PKS & 0.09 & & & N\\
09102$-$5101 &  09 11 57.3 & $-51$ 14 24 & 12-SEP-2008  & PKS & 0.06 & & & V\\
             &             &             & 14-SEP-2008  & PKS & 0.08 \\
09119$-$5150 &  09 13 33.0 & $-52$ 02 41 & 14-SEP-2008  & PKS & 0.09 & & & V\\
09362$-$5413 &  09 37 51.8 & $-54$ 27 09 & 14-SEP-2008  & PKS & 0.11 & & & V\\
09378$-$5117 &  09 39 37.0 & $-51$ 31 29 & 14-SEP-2008  & PKS & 0.08 & & & V\\
10194$-$5625 &  10 21 15.2 & $-56$ 40 32 & 14-SEP-2008  & PKS & 0.09 & & & N,O\\
11444$-$6150 &  11 46 54.0 & $-62$ 07 09 & 13-SEP-2008  & PKS & 0.07 & & & N,O\\
11488$-$6432 &  11 51 17.3 & $-64$ 49 12 & 12-SEP-2008  & PKS & 0.11 & & & V\\
11544$-$6408 &  11 56 57.8 & $-64$ 25 17 & 12-SEP-2008  & PKS & 0.08 & & & V\\
11549$-$6225 &  11 57 30.8 & $-62$ 42 12 & 12-SEP-2008  & PKS & 0.08 & & & N,O\\
12067$-$4508 &  12 09 23.8 & $-45$ 25 35 & 15-SEP-2008  & PKS & 0.13 & $<1.2$ & 1 & V\\
12302$-$6317 &  12 33 07.0 & $-63$ 33 43 & 15-SEP-2008  & PKS & 0.12 & $<0.7$ & 2 & V\\
12360$-$5740 &  12 38 53.1 & $-57$ 56 31 & 13-SEP-2008  & PKS & 0.10 & & & V\\
13010$-$6012 &  13 04 05.5 & $-60$ 28 46 & 14-SEP-2008  & PKS & 0.16 & & & V\\
13245$-$5036 &  13 27 36.1 & $-50$ 52 06 & 14-SEP-2008  & PKS & 0.15 & & & V\\
13266$-$5551 &  13 29 51.0 & $-56$ 06 54 & 14-SEP-2008  & PKS & 0.15 & & & V\\
13313$-$5838 &  13 34 37.4 & $-58$ 53 32 & 14-SEP-2008  & PKS & 0.15 & & & V\\
13398$-$5951 &  13 43 12.5 & $-60$ 07 03 & 12-SEP-2008  & PKS & 0.12 & & & N\\
13404$-$6059 &  13 43 50.3 & $-61$ 14 30 & 13-SEP-2008  & PKS & 0.10 & & & N,O\\
13421$-$6125 &  13 45 34.0 & $-61$ 40 03 & 13-SEP-2008  & PKS & 0.09 & & & V\\
13428$-$6232 &  13 46 20.5 & $-62$ 48 00 & 15-SEP-2008  & PKS & 0.13 & & & V\\
13483$-$5905 &  13 51 43.7 & $-59$ 20 15 & 13-SEP-2008  & PKS & 0.09 & & & V\\ %ch%
14104$-$5819 &  14 14 00.5 & $-58$ 33 58 & 13-SEP-2008  & PKS & 0.09 & & & N,O\\
14249$-$5310 &  14 28 24.7 & $-53$ 24 04 & 13-SEP-2008  & PKS & 0.07 & & & N\\ %ch%
14325$-$6428 &  14 36 34.4 & $-64$ 41 31 & 15-SEP-2008  & PKS & 0.23 & & & V\\
14331$-$6435 &  14 37 10.1 & $-64$ 48 05 & 15-SEP-2008  & PKS & 0.15 & & & V\\
14346$-$5952 &  14 38 24.6 & $-60$ 04 53 & 15-SEP-2008  & PKS & 0.14 & & & V\\
14429$-$4539 &  14 46 13.8 & $-45$ 52 05 & 15-SEP-2008  & PKS & 0.14 & & & V\\
14521$-$5300 &  14 55 45.7 & $-53$ 12 30 & 15-SEP-2008  & PKS & 0.07 & & & V\\
15038$-$5533 &  15 07 34.7 & $-55$ 44 50 & 15-SEP-2008  & PKS & 0.09 & & & N\\
15229$-$5433 &  15 26 40.5 & $-54$ 44 17 & 12-SEP-2008  & PKS & 0.08 & & & V\\
15284$-$6026 &  15 32 37.1 & $-60$ 37 04 & 15-SEP-2008  & PKS & 0.08 & & & N,O\\ 
15408$-$5413 &  15 44 39.8 & $-54$ 23 05 & 15-SEP-2008  & PKS & 0.09 & & & N,O\\
15408$-$5657 &  15 44 48.3 & $-57$ 07 08 & 15-SEP-2008  & PKS & 0.09 & $<1.5$ & 1 & N\\ %ch%
15452$-$5459 &  15 49 11.3 & $-55$ 08 51 & 12-SEP-2008  & PKS & 0.08 & $6.5\pm 0.3$ & 3,4 & N\\ %ch%
15531$-$5704 &  15 57 10.9 & $-57$ 13 21 & 12-SEP-2008  & PKS & 0.09 & & & N,O\\
16206$-$5956 &  16 25 02.6 & $-60$ 03 32 & 14-SEP-2008  & PKS & 0.17 & & & V\\
16209$-$4714 &  16 24 33.9 & $-47$ 21 30 & 12-SEP-2008  & PKS & 0.08 & $<0.4$ & 3 & V\\
16245$-$3859 &  16 27 53.7 & $-39$ 05 46 & 12-SEP-2008  & PKS & 0.11 & & & N,O\\
16279$-$4757 &  16 31 38.7 & $-48$ 04 06 & 14-SEP-2008  & PKS & 0.12 & & & V\\
16296$-$4507 &  16 33 12.5 & $-45$ 13 43 & 12-SEP-2008  & PKS & 0.08 & & & V\\
16279$-$8158 &  16 37 51.6 & $-82$ 04 49 & 12-SEP-2008  & PKS & 0.09 & & & V\\
16507$-$4810 &  16 54 30.9 & $-48$ 15 24 & 12-SEP-2008  & PKS & 0.09 & $<0.4$ & 3 & V\\
16517$-$3626 &  16 55 06.2 & $-36$ 31 32 & 12-SEP-2008  & PKS & 0.09 & & & V\\
16558$-$3417 &  16 59 10.5 & $-34$ 22 05 & 12-SEP-2008  & PKS & 0.08 & & & M\\
16567$-$3838 &  17 00 09.0 & $-38$ 43 09 & 15-SEP-2008  & PKS & 0.10 & & & N,O\\
17021$-$3109 &  17 05 23.3 & $-31$ 13 18 & 05-JUN-2008  & ROB & 0.14 & & & N,O \\
			 &			   &			 & 21-MAR-2010  & GBT & 0.022 \\
17021$-$3054 &  17 05 24.1 & $-30$ 58 14 & 06-JUN-2008  & ROB & 0.10 & & & N \\
			 &			   &			 & 21-MAR-2010  & GBT & 0.022 \\
17052$-$3245 &  17 08 33.2 & $-32$ 49 44 & 13-SEP-2008  & PKS & 0.10 & & & N,O \\
17067$-$3759 & 	17 10 08.3 & $-38$ 03 22 & 13-SEP-2008  & PKS & 0.09 & & & N \\
17097$-$3624 & 	17 13 05.1 & $-36$ 27 53 & 13-SEP-2008  & PKS & 0.14 & $<0.4$ & 3 & N,O \\
17149$-$3053 &  17 18 10.9 & $-30$ 56 39 & 12-SEP-2008  & PKS & 0.08 & $<0.9$ & 5 & N \\
			 & 			   &			 & 03-MAR-2010  & GBT & 0.023 \\
17150$-$3224 &  17 18 19.8 & $-32$ 27 21 & 12-SEP-2008  & PKS & 0.08 & $<0.4$ & 3 & V\\
17158$-$4049 & 	17 19 19.6 & $-40$ 52 37 & 13-SEP-2008  & PKS & 0.09 & & & N \\
17175$-$2819 & 17 20 42.5  & $-28$ 22 37 & 03-MAR-2010  & GBT & 0.020 & $<1.1$ & 5 & V\\
17233$-$2602 & 17 26 28.7  & $-26$ 04 58 & 06-JUN-2008  & ROB & 0.08 & $<0.7$ & 5 & V\\
			 & 			   &			 & 21-MAR-2008  & GBT & 0.019 \\
17291$-$2147 & 17 32 10.1  & $-21$ 49 59 & 06-JUN-2008  & ROB & 0.07 & $1.3\pm 0.7$ & 5 & V\\
			 & 			   & 			 & 21-MAR-2010  & GBT & 0.011  \\
17301$-$2538 & 17 33 14.1  & $-25$ 40 24 & 03-MAR-2010  & GBT & 0.019 & $<1.0$ & 5 & V\\
17348$-$2906 & 17 38 04.2  & $-29$ 08 23 & 03-MAR-2010  & GBT & 0.022 & & & V\\
17359$-$2902 & 17 39 08.0  & $-29$ 04 06 & 03-MAR-2010  & GBT & 0.020 & $<0.4$ & 3,5,6 & N\\
17360$-$2142 & 17 39 05.9  & $-21$ 43 52 & 01-MAR-2010  & GBT & 0.011 & $<0.6$ & 5 & V\\
17361$-$4159 & 17 39 44.3  & $-42$ 00 39 & 13-SEP-2008  & PKS & 0.13 & & & N \\
17376$-$3448 &  17 40 56.3 & $-34$ 50 00 & 15-SEP-2008  & PKS & 0.07 & & & N \\
17382$-$2531 & 17 41 20.1  & $-25$ 32 53 & 03-MAR-2010  & GBT & 0.019 & $<0.7$ & 5 & N \\
17385$-$2413 & 17 41 38.3  & $-24$ 14 41 & 01-MAR-2010  & GBT & 0.011 & $<0.6$ & 5 & M\\
17385$-$3332 &  17 41 52.2 & $-33$ 33 40 & 15-SEP-2008  & PKS & 0.07  & $<0.4$ & 3 & V\\
17393$-$2727 & 17 42 32.2  & $-27$ 28 28 & 03-MAR-2010  & GBT & 0.020 & $<0.4$ & 3,5,6 & N,O\\
17404$-$2713 & 17 43 37.2  & $-27$ 14 46 & 03-MAR-2010  & GBT & 0.019 & $<0.4$ & 3,5 & N\\
			 & 			   &			 & 21-MAR-2010  & GBT & 0.021\\
17479$-$3032 & 17 51 12.5  & $-30$ 33 44 & 03-MAR-2010  & GBT & 0.021 & $<0.4$ & 3 & N\\
17482$-$2501 & 17 51 22.5  & $-25$ 01 51 & 03-MAR-2010  & GBT & 0.018 & $<0.4$ & 3,5 & N\\
17487$-$1922 & 17 51 44.7  & $-19$ 23 42 & 06-JUN-2008  & ROB & 0.06 & $<0.7$ & 5 & V\\
			 & 			   &			 & 21-MAR-2010  & GBT & 0.017\\
17499$-$3520 &  17 53 20.4 & $-35$ 21 10 & 15-SEP-2008  & PKS & 0.07 & & & V\\ 
17506$-$2955 & 17 53 49.3  & $-29$ 55 35 & 03-MAR-2010  & GBT & 0.021 & $<0.4$ & 3 & N\\
17516$-$2525 & 17 54 43.4  & $-25$ 26 30 & 01-MAR-2010  & GBT & 0.011 & $<0.3$ & 5,7 & V\\
17540$-$2753 & 17 57 14.1  & $-27$ 54 16 & 01-MAR-2010  & GBT & 0.012 & & & N\\
17543$-$3102 &  17 57 33.6 & $-31$ 03 03 & 15-SEP-2008  & PKS & 0.09 & $<0.4$ & 3 & N\\
17548$-$2753 & 17 57 57.8  & $-27$ 53 21 & 03-MAR-2010  & GBT & 0.019 & $<0.4$ & 3 & N\\
17550$-$2120 & 17 58 04.2  & $-21$ 21 09 & 03-MAR-2010  & GBT & 0.021 & $<0.4$ & 3,5 & N,O\\
17550$-$2800 & 17 58 10.6  & $-28$ 00 26 & 03-MAR-2010  & GBT & 0.017 & $<0.8$ & 5 & V\\
17552$-$2030 & 17 58 16.3  & $-20$ 30 22 & 01-MAR-2010  & GBT & 0.011 & & & N\\
17560$-$2027 & 17 59 04.5  & $-20$ 27 24 & 01-MAR-2010  & GBT & 0.010 & $<0.4$ & 3,5 & V\\
17596$-$3952 &  18 03 06.7 & $-39$ 51 53 & 13-SEP-2008  & PKS & 0.11 & & & V\\
18011$-$1847 & 18 04 02.7  & $-18$ 47 10 & 06-JUN-2008  & ROB & 0.07 & $<0.8$ & 5 & N,O \\
			 & 			   & 			 & 21-MAR-2010  & GBT & 0.018\\
18015$-$1352 & 18 04 22.2  & $-13$ 51 49 & 06-JUN-2008  & ROB & 0.06  & $<0.5$ & 5 & N,O \\
			 & 			   & 			 & 21-MAR-2010  & GBT & 0.024\\
18016$-$2743 & 18 04 45.8  & $-27$ 43 11 & 01-MAR-2010  & GBT & 0.012 & & & N,O\\
18039$-$1903 & 18 06 53.3  & $-19$ 03 09 & 01-MAR-2010  & GBT & 0.014 & $11.3\pm 1.2$ & 5 & N,O\\
18049$-$2118 & 18 07 54.8  & $-21$ 18 09 & 01-MAR-2010  & GBT & 0.010 & $<4$ & 5 & N,O\\
18051$-$2415 & 18 08 12.8  & $-24$ 14 36 & 03-MAR-2010  & GBT & 0.018 & $<0.4$ & 3 & M\\
18071$-$1727 & 18 10 05.9  & $-17$ 26 35 & 01-MAR-2010  & GBT & 0.013 & $<6$ & 6 & N,O\\
18083$-$2155 & 18 11 18.9  & $-21$ 55 05 & 01-MAR-2010  & GBT & 0.015 & $<1.4$ & 5 & M\\
18087$-$1440 & 18 11 34.6  & $-14$ 39 56 & 03-MAR-2010  & GBT & 0.014 & $<0.21$ & 3,5,6 & N\\
18105$-$1935 & 18 13 32.2  & $-19$ 35 03 & 01-MAR-2010  & GBT & 0.014 & $<0.9$ & 5 & N\\
18113$-$2503 & 18 14 26.3  & $-25$ 02 56 & 03-MAR-2010  & GBT & 0.013 & $105.814\pm 0.003$ &5,8 & M\\
			 & 			   & 			 & 21-MAR-2010  & GBT & 0.018\\
18135$-$1456 & 18 16 26.1  & $-14$ 55 13 & 01-MAR-2010  & GBT & 0.014 & $5.3\pm 0.4$ & 3,5,9,10 & N,O\\
18183$-$2538 & 18 21 24.7  & $-25$ 36 35 & 01-MAR-2010  & GBT & 0.012 & $<1.2$ & 5 & V\\
18199$-$1442 & 18 22 50.8  & $-14$ 40 49 & 01-MAR-2010  & GBT & 0.013 & $<0.8$ & 5 & N,O\\
18229$-$1127 & 18 25 45.0  & $-11$ 25 56 & 06-JUN-2008  & ROB & 0.06  & $<2.4$ & 5 & V\\
			 & 			   & 			 & 21-MAR-2010  & GBT & 0.015\\
18236$-$0447 & 18 26 20.3  & $-04$ 45 42 & 01-MAR-2010  & GBT & 0.013  & $<0.12$ & 5,10 & N,O\\
18355$-$0712 & 18 38 15.4  & $-07$ 09 52 & 01-MAR-2010  & GBT & 0.012 & $<0.4$ & 3,5,6 & N\\
18361$-$1203 & 18 38 58.8  & $-12$ 00 44 & 01-MAR-2010  & GBT & 0.012 & $<1.3$ & 5 & V\\
			 & 			   & 		     & 21-MAR-2010  & GBT & 0.016\\
18385$+$1350 & 18 40 52.1  & $+13$ 52 54 & 07-MAR-2010  & GBT & 0.015 & $<0.9$ & 5 & N\\
18434$-$0042 & 18 46 04.4  & $-00$ 38 55 & 01-MAR-2010  & GBT & 0.011 & $<0.9$\footnotemark[6] & 5 & N\\
18454$+$0001 & 18 48 01.5  & $+00$ 04 48 & 01-MAR-2010  & GBT & 0.012 & $<1.2$ & 5 & V\\
18470$+$0015 & 18 49 39.1  & $+00$ 18 52 & 01-MAR-2010  & GBT & 0.013 & $<1.5$ & 5 & N\\
18485$+$0642 & 18 50 58.9  & $+06$ 45 55 & 06-JUN-2008  & ROB & 0.05  & $<0.4$ & 3,5 & V\\
18514$+$0019 & 18 53 58.0  & $+00$ 23 25 & 01-MAR-2010  & GBT & 0.012 & $<1.3$ & 5 & N\\
18529$+$0210 & 18 55 26.3  & $+02$ 14 49 & 03-MAR-2010  & GBT & 0.018 & & & M,O\\
			 & 			   & 			 & 07-MAR-2010  & GBT & 0.014\\
18580$+$0818 & 19 00 25.2  & $+08$ 22 47 & 07-MAR-2010  & GBT & 0.013 & $<0.8$ & 5 & N,O\\
18596$+$0315 & 19 02 06.4  & $+03$ 20 15 & 07-MAR-2010  & GBT & 0.014 & $18\pm 6$& 3,9,11,12,13 & N,O\\
19006$+$1022 & 19 02 59.9  & $+10$ 26 35 & 20-JUN-2008  & ROB & 0.06 & $<0.7$ & 5 & N\\
			 & 			   & 			 & 21-MAR-2010  & GBT & 0.013\\
19011$+$1049 & 19 03 30.7  & $+10$ 53 53 & 07-MAR-2010  & GBT & 0.013 & & & M\\
19013$+$0629 & 19 03 44.4  & $+06$ 34 12 & 20-JUN-2008  & ROB & 0.06  & $<0.8$ & 5 & N,O\\
19015$+$1256 & 19 03 52.6  & $+13$ 01 21 & 07-MAR-2010  & GBT & 0.013 & $<1.0$ & 5 & N\\
19071$+$0857 & 19 09 29.7  & $+09$ 02 23  & 09-OCT-2010\footnotemark[7]	 & VLA & 0.0007 & $2.1\pm 0.5$ & 5 & N\\
19075$+$0432 & 19 10 00.0  & $+04$ 37 06 & 07-MAR-2010  & GBT & 0.013 & $<0.8$ & 5,14 & V\\
19079$-$0315 & 19 10 32.5  & $-03$ 10 16 & 20-JUN-2008  & ROB & 0.10  & $<0.8$ & 5 & V\\
			 & 			   & 			 & 21-MAR-2010  & GBT & 0.013\\
19094$+$1627 & 19 11 44.6  & $+16$ 32 54 & 07-MAR-2010  & GBT & 0.013 & $<1.4$ & 5 & V\\
19134$+$2131 & 19 15 35.4  & $+21$ 36 33 & 07-MAR-2010  & GBT & 0.013 & $6.0\pm 2.4$ & 15,16,17,18,19 & N\\
19176$+$1251 & 19 19 55.9  & $+12$ 57 35 & 20-JUN-2008  & ROB & 0.08  & $<0.6$ & 5 & N\\
19178$+$1206 & 19 20 14.1  & $+12$ 12 22 & 07-MAR-2010  & GBT & 0.012 & $<0.4$ & 20 & N,O\\
19190$+$1102 & 19 21 25.3  & $+11$ 08 40 & 07-MAR-2010  & GBT & 0.013 & $96.\pm 0.4$ & 11,21,22 & N\\
19319$+$2214 & 19 34 03.4  & $+22$ 21 14 & 28-JUN-2008  & ROB & 0.06 & $3.7\pm 0.5$& 20 & N\\
			 & 			   & 			 & 07-MAR-2010  & GBT & 0.013\\
19374$+$2359 & 19 39 35.4  & $+24$ 06 25 & 07-MAR-2010  & GBT & 0.013 & $38\pm 15$ & 5,23\footnotemark[8] & V\\
20035$+$3242 & 20 05 29.6  & $+32$ 51 35 & 24-SEP-2008  & ROB & 0.08 & $<1.9$ & 5 & N\\
			 & 			   & 			 & 07-MAR-2010  & GBT & 0.010\\
20042$+$3259 & 20 06 10.6  & $+33$ 07 51 & 28-JUN-2008  & ROB & 0.08 & $<2.5$ & 5 & N\\
			 & 			   & 			 & 07-MAR-2010  & GBT & 0.012\\
20214$+$3749 & 20 23 19.2  & $+37$ 58 52 & 29-JUL-2008  & ROB & 0.06 & $<0.6$ & 5 & N\\
			 & 			   & 			 & 07-MAR-2010  & GBT & 0.011\\
20244$+$3509 & 20 26 25.4  & $+35$ 19 14 & 24-SEP-2008  & ROB & 0.12 & $<0.7$ & 5 & V\\
			 & 			   & 			 & 21-MAR-2010  & GBT & 0.013\\
20461$+$3853 & 20 48 04.6  & $+39$ 05 01 & 29-JUL-2008  & ROB & 0.06  & $<1.7$ & 5 & V\\
			 & 			   & 			 & 21-MAR-2010  & GBT & 0.013\\
21525$+$5643 & 21 54 15.0  & $+56$ 57 23 & 29-JUL-2008  & ROB & 0.05  & $<1.6$ & 5 & N\\
			 & 			   & 			 & 21-MAR-2010  & GBT & 0.014\\
21554$+$6204 & 21 56 58.3  & $+62$ 18 43 & 29-JUL-2008  & ROB & 0.05  & $<0.11$ & 5,10,24 & N,O\\
			 & 			   & 			 & 21-MAR-2010  & GBT & 0.014\\
\end{longtable}
\tablebib{
(1)~\cite{deg89}; (2) \cite{sua09}; (3) \cite{dea07}; (4) \cite{cer13}; (5) \cite{yoo14}; (6) \cite{gom90}; (7) \cite{sua07}; (8) \cite{gom11}; (9) \cite{eng86}; (10) \cite{yun13}; (11) \cite{bra94}; (12) \cite{gom94}; (13) \cite{eng02}; (14) \cite{lik89}; (15) \cite{eng84}; (16) \cite{lik92}; (17) \cite{com90}; (18) \cite{ima04}; (19) \cite{ima07b}; (20) \cite{eng96}; (21) \cite{lik89}; (22) \cite{day10}; (23) \cite{han98}; (24) \cite{zuc87} }
\tablefoot{ 
\tablefoottext{a}{GBT: Green Bank. PKS: Parkes. ROB: Robledo. VLA: Very Large Array.}
\tablefoottext{b}{One-sigma noise level per spectral channel.}
\tablefoottext{c}{Previous water maser observations of the sources reported in the literature. For detections, we give the highest reported flux density and its $2\sigma$ uncertainty. For non-detections, the lowest reported $3\sigma$ upper limit is given.}
\tablefoottext{d}{References for previous water maser observations.}
\tablefoottext{e}{Visibility of sources in optical and infrared images. V: Sources with optical counterpart in the Digital Sky Survey \citep[as mentioned in][]{rl1,rl2}, or with optical spectrum in \cite{sua06}. N: Sources detected only at near-infrared wavelenghts or longer \citep{rl1,rl2}. M: Sources detected only at mid-infared wavelenghts or longer \citep{rl1,rl2}. O: Sources with strong obscuration, based on their infrared colours \citep[table 9 in][]{rl2}.}
\tablefoottext{f}{\cite{mig99} reported a water maser of 56 Jy, which was incorrectly labelled as IRAS 18434$-$0042. The maser is $\simeq 2^\circ$ away from this infrared source and it is associated with the star-forming region W43S instead.}
\tablefoottext{g}{We observed IRAS 19071+0857 with Robledo and GBT, but it was contaminated by the strong maser emission from W49A, which spilled into the telescope sidelobes. The reported detection by \cite{yoo14} is also most likely contaminated. The upper limit listed in this table corresponds to observations taken with the Very Large Array (Su\'arez et al. in preparation), which confirmed that the maser emission was not associated with IRAS 19071+0857.}
\tablefoottext{h}{Labeled as 19375+2359 in \cite{han98}}
}
\end{longtab}
}

\subsection{Robledo de Chavela}

The DSS-63 antenna at MDSCC has a diameter of 70 m, providing a half-power beam width of $\simeq 42''$ at this frequency. We used a 384-channel spectrometer covering a bandwidth of 16 MHz ($\simeq 216$ km s$^{-1}$ with $\simeq 0.6$ km s$^{-1}$ spectral resolution).  Spectra were taken in position-switching mode, and only left-handed circular polarization was available. The total integration time was typically  30 minutes (on + off). Initial calibration, along with gain and opacity corrections, was carried out with procedures written in Interactive Data Language (IDL). Subsequent data reduction was performed using the CLASS package, which is part of the GILDAS software.

\subsection{Parkes}

The 64m antenna at the Parkes Observatory has a beamwidth of $\simeq 1.3'$ at this frequency.  Both right and left circular polarization
were observed simultaneously. As a backend, we used
the Multibeam Correlator, covering a bandwidth of 64 MHz
(i.e., velocity coverage $\simeq 863$ km s$^{-1}$) with 2048 spectral channels
for each polarization, thus yielding a spectral resolution of
31.25 kHz (0.42 km s$^{-1}$).  The observations were taken
in position-switching mode with a total integration time (on+off)
of 30 min. 
Data reduction was carried out with the libraries of the ATNF Spectral line Analysis Package (ASAP) included within the Common Astronomy Software Applications (CASA) package. 

\subsection{Green Bank}

The 1.3 cm receiver of the GBT comprised four beams, arranged in two pairs that could be tuned independently. We used one such pair, with a separation of $178.8''$ between the beams, to simultaneously observe an on- and off-source position in dual polarization. Antenna nodding between the two beams was used to subtract atmospheric and instrumental contributions. We selected two spectral windows, one centered on the water line and the other one comprising three methanol lines \citep[whose results were presented in][]{gom14}. The bandwidth of each spectral window was 50 MHz ($\simeq 674$  km s$^{-1}$ velocity coverage) sampled over 8192 channels ($\sim 0.082$ km s$^{-1}$ velocity resolution). The
half-power beam width of this 100m telescope was $34''$ at the frequency of the water line.
The integration time per source was approximately four minutes, and all of it was effectively on-source time, because of the use of the antenna-nodding with dual beams. The data was reduced with the GBTidl package. Final spectra were smoothed up to a resolution of 0.33 km s$^{-1}$.

\section{Results}

%\subsection{Water maser detections}

We have detected water maser emission in 15 sources of our sample, of which seven are reported here for the first time (IRAS 13483$-$5905, IRAS 14249$-$5310, IRAS 15408$-$5413, IRAS 17021$-$3109, IRAS 17348$-$2906, IRAS 17393$-$2727, and IRAS 18361$-$1203). We also detected water maser emission in the spectrum taken toward IRAS 19071$+$0857 with both Robledo and GBT telescopes.  However, follow-up observations with the Very Large Array (VLA) have shown that the maser emission did not arise from this source, but it was emission from the star-forming region W49A (Su\'arez et al. in preparation), located $\sim 11'$ away, which spilled into the sidelobes of the single-dish telescope. We believe that the detection reported by \citet{yoo14} toward  IRAS 19071$+$0857 is likewise contaminated by W49A.

In what follows we discuss the individual sources where H$_2$O maser emission is detected. The spectra observed toward these sources are shown in Figure \ref{fig_detections},  while the main parameters of the H$_2$O masers are listed in Table \ref{detections}.

\renewcommand{\thefootnote}{\alph{footnote}}
\begin{table*}
\caption{Water maser detections \label{detections}}
\centering
\begin{tabular}{ccccccccc}
\hline \hline 
IRAS  & $V_{\rm peak}$\footnotemark[1] & $V_{\rm min}$\footnotemark[2] & $V_{\rm max}$\footnotemark[3] & $S_{\rm peak}$\footnotemark[4] & $\int S_\nu dv$\footnotemark[5] & Date \\
      & (km s$^{-1}$)  & (km s$^{-1}$) & (km s$^{-1}$) & (Jy)           & (Jy km s$^{-1}$)                 \\
\hline
13483$-$5905 & $+24.0\pm 0.4$  & $-27.0$ & $+27.0$ & $2.78\pm 0.17$  & $21.3\pm 0.9$  & 13-SEP-2008\\
14249$-$5310 & $-10.1\pm 0.4$  & $-19.8$ & $-9.7$  & $0.42\pm 0.14$  & $0.42\pm 0.20$ & 13-SEP-2008\\
15408$-$5413 & $-82.6\pm 0.4$  & $-85.1$ & $-75.9$ & $52.44\pm 0.17$ & $59.0\pm 0.5$  & 15-SEP-2008\\
15452$-$5459 & $-76.3\pm 0.4$  & $-79.2$ & $-35.4$ & $1.62\pm 0.15$  & $10.9\pm 0.8$  & 12-SEP-2008\\
17021$-$3109 & $+13.5\pm 0.6$  & $+13.2$ & $+13.5$ & $0.26\pm 0.04$  & $0.20\pm 0.06$ & 21-MAR-2010\\
17291$-$2147 & $-15.0\pm 0.6$  & $-18.8$ & $-13.9$ & $0.59\pm 0.13$  & $2.62\pm 0.17$ & 06-JUN-2008\\
			 & $-14.5\pm 0.3$  & $-28.0$ & $+69.5$ & $3.381\pm 0.022$& $5.49\pm 0.07$ & 21-MAR-2010\\
17348$-$2906 & $+1.0\pm 0.3$   & $+0.7$  & $+9.2$  & $0.23\pm 0.04$  & $0.52\pm 0.10$ & 03-MAR-2010\\
17393$-$2727 & $-107.6\pm 0.3$ & $-108.0$& $-106.7$& $1.07\pm 0.04$  & $0.90\pm 0.04$ & 03-MAR-2010\\
18039$-$1903 & $+153.1\pm 0.3$ & $+150.8$& $+165.9$& $3.98\pm 0.03$  & $5.38\pm 0.07$ & 01-MAR-2010\\
18113$-$2503 & $-147.8\pm 0.3$ & $-154.1$& $+334.5$\footnotemark[6]& $119.67\pm 0.03$& $845.9\pm 0.3$ & 03-MAR-2010\\
			 & $-147.6\pm 0.3$ & $-154.2$& $+345.5$& $111.92\pm 0.04$& $812.8\pm 0.4$ & 21-MAR-2010\\
18361$-$1203 & $+12.5\pm 0.3$  & $+11.8$ & $+13.5$ & $0.101\pm 0.023$& $0.19\pm 0.04$ & 01-MAR-2010\\
			 & $+12.8\pm 0.3$  & $+11.8$ & $+13.5$ & $0.14\pm 0.03$  & $0.15\pm 0.04$ & 21-MAR-2010\\
18596$+$0315 & $+61.6\pm 0.3$  & $+59.6$ & $+62.2$ & $0.72\pm 0.03$  & $0.69\pm 0.04$ & 07-MAR-2010 \\
19134$+$2131 & $-9.2\pm 0.3$   & $-124.8$& $-7.9$  & $3.60\pm 0.03$  & $5.51\pm 0.17$ & 07-MAR-2010 \\
19190$+$1102 & $+59.3\pm 0.3$  & $-55.3$ & $+90.5$ &$25.64\pm 0.03$ & $164.94\pm 0.19$& 07-MAR-2010 \\
19319$+$2214 & $+15.1\pm 0.6$  & $+15.1$ & $+15.1$ &$0.35\pm 0.12$  & $0.35\pm 0.14$ & 28-JUN-2008\\
			 & $+15.5\pm 0.3$  & $+8.8$  & $+33.6$ &$0.547\pm 0.024$& $1.95\pm 0.08$ & 07-MAR-2010\\
%ROB deltaV=0.5632
%GBT deltaV=0.3292
\hline
\end{tabular}
\tablefoot{
\tablefoottext{a}{LSR velocity of the peak emission in each spectrum. The uncertainty is the channel width.}
\tablefoottext{b}{Minimum LSR velocity of the detected emission.}
\tablefoottext{c}{Maximum LSR velocity of the detected emission.}
\tablefoottext{d}{Flux density of the peak emission, with $2\sigma$ uncertainty.}
\tablefoottext{e}{Velocity-integrated flux density of the spectrum, with $2\sigma$ uncertainty.}
\tablefoottext{f}{There is detected emission very close to the redshifted edge of the bandpass, so we probably missed some components beyond that edge.}
}
\end{table*}

\onlfig{
\begin{figure*}
\centering
\includegraphics[width=0.3\textwidth,angle=-90]{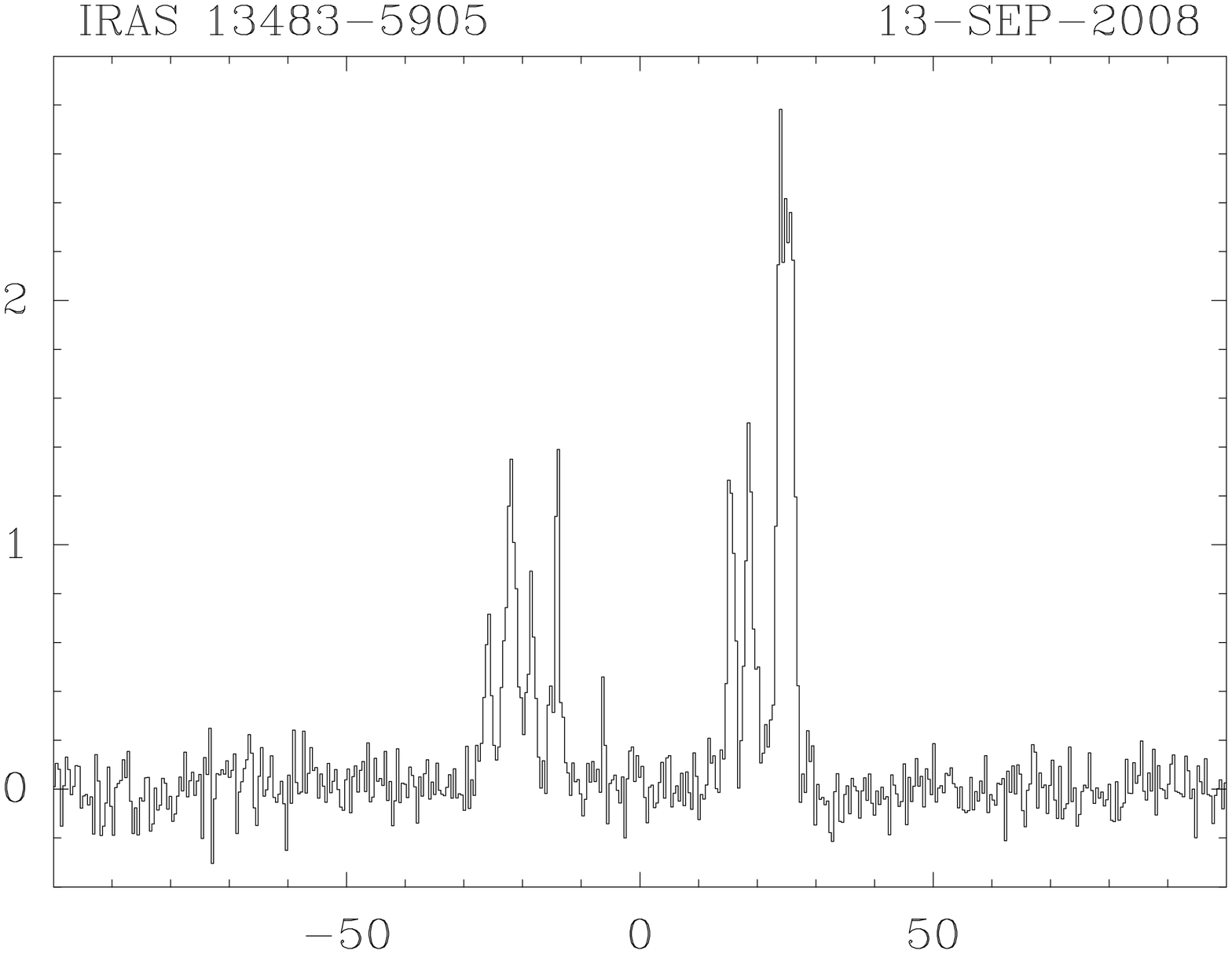}
\includegraphics[width=0.3\textwidth,angle=-90]{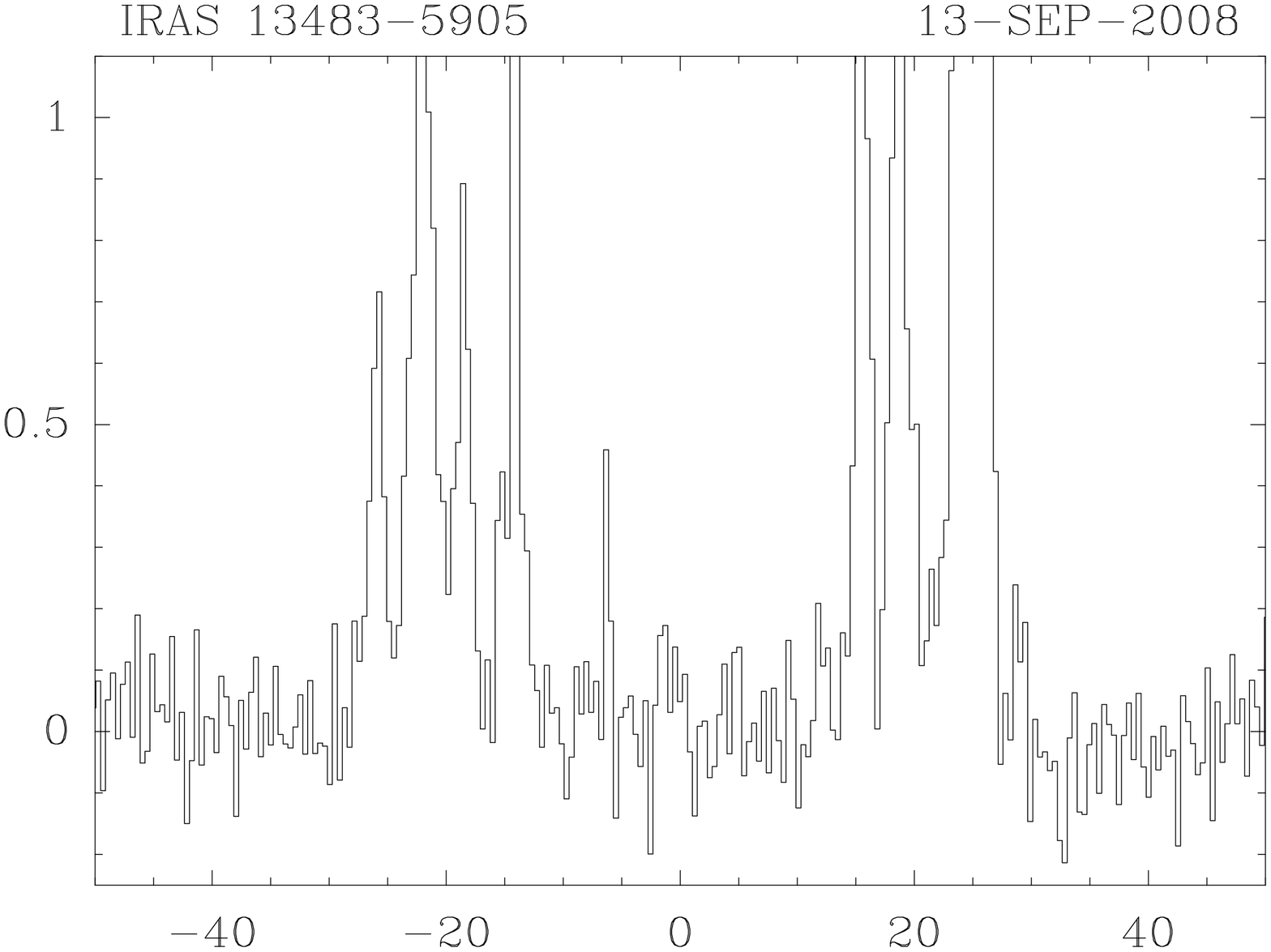}
\includegraphics[width=0.3\textwidth,angle=-90]{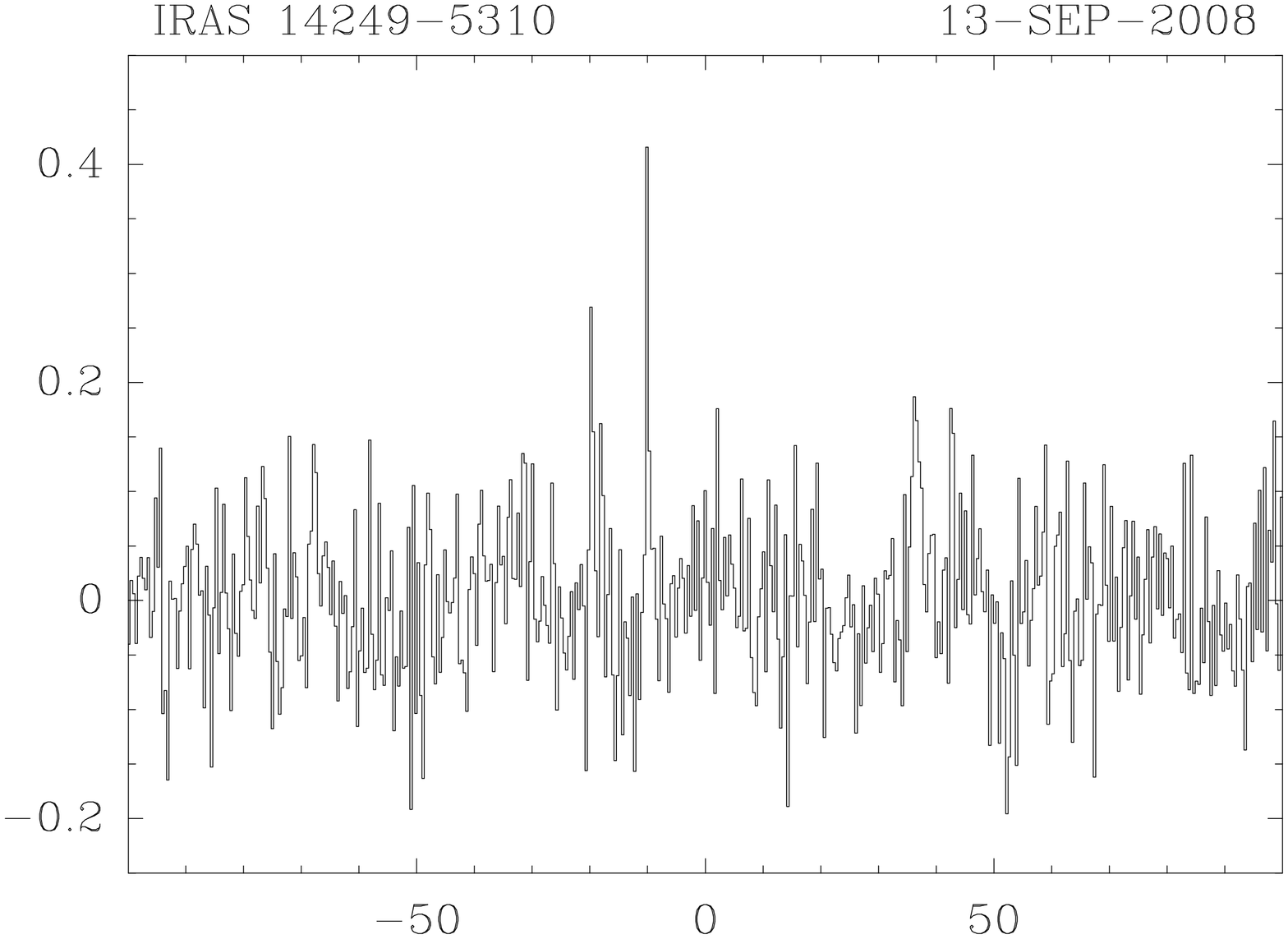}
\includegraphics[width=0.3\textwidth,angle=-90]{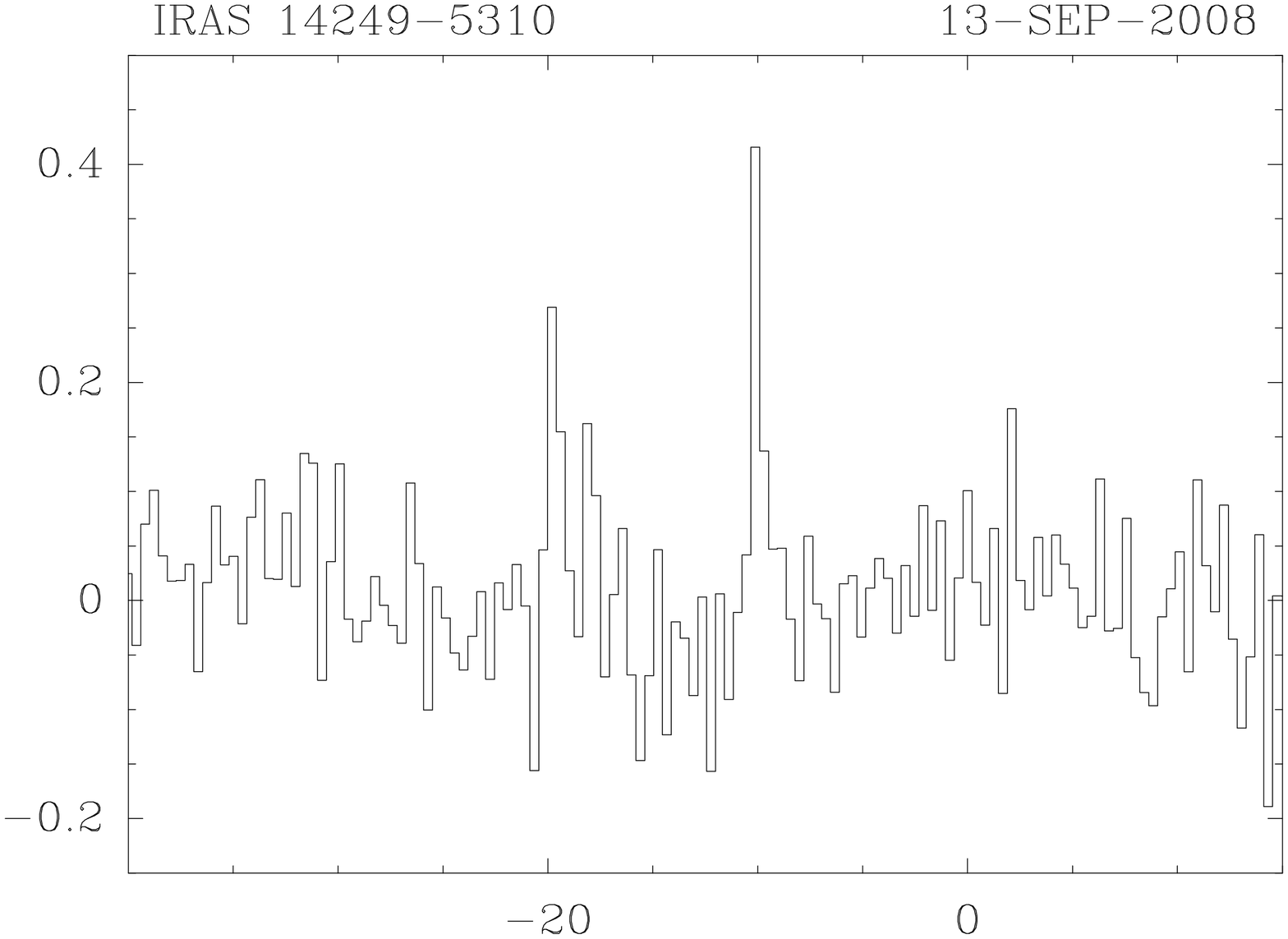}
\includegraphics[width=0.3\textwidth,angle=-90]{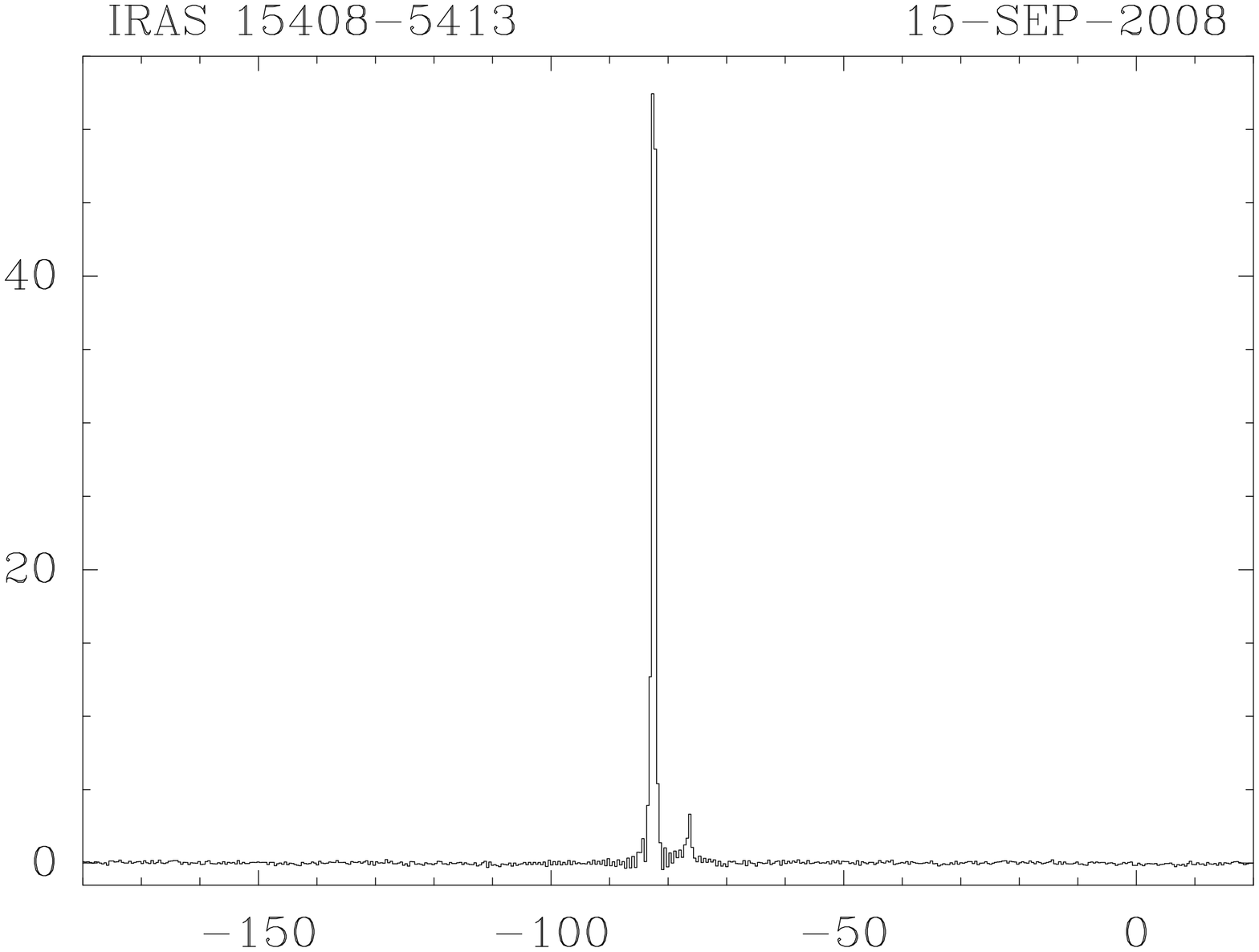}
\includegraphics[width=0.3\textwidth,angle=-90]{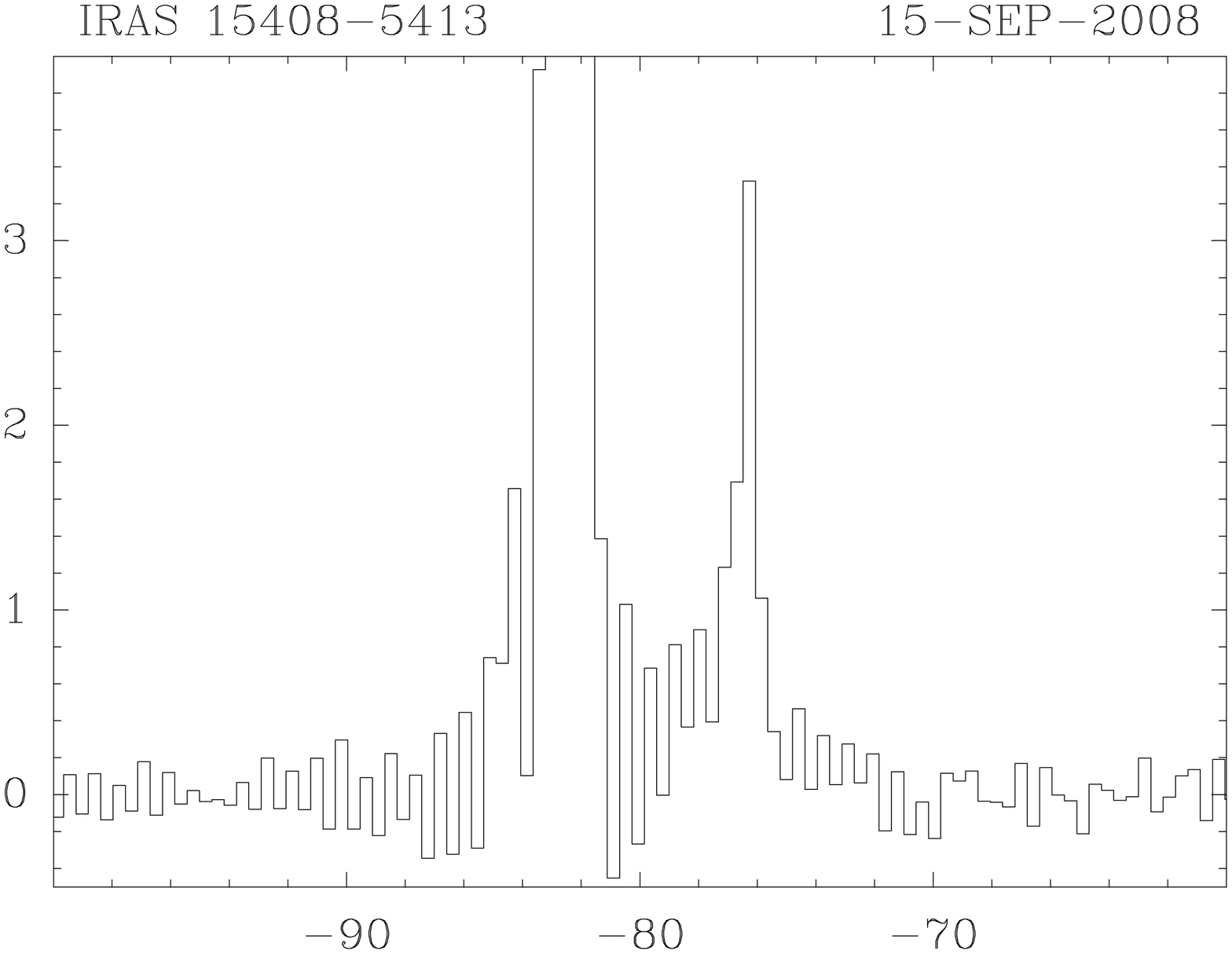}
\includegraphics[width=0.3\textwidth,angle=-90]{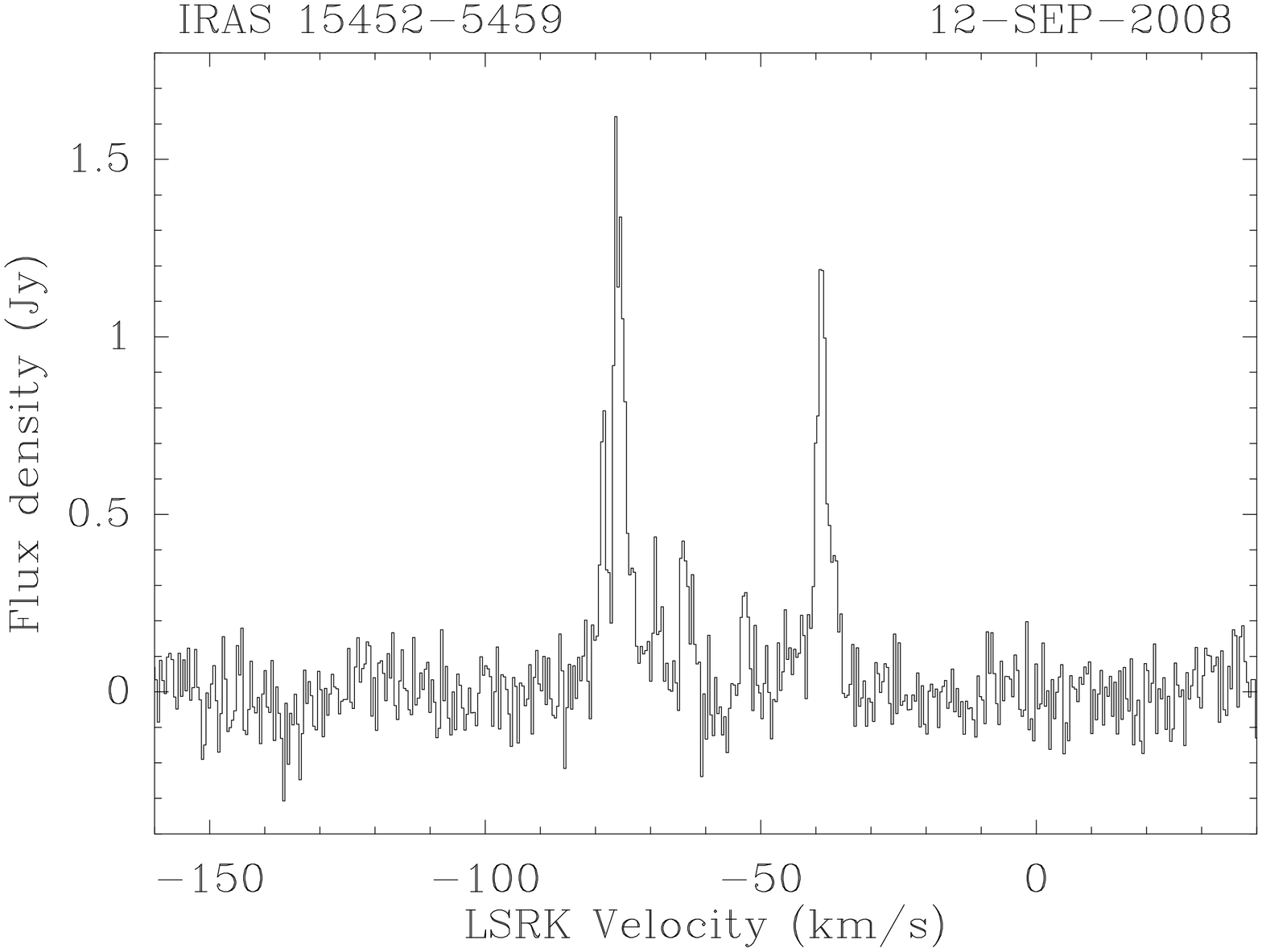}
\includegraphics[width=0.3\textwidth,angle=-90]{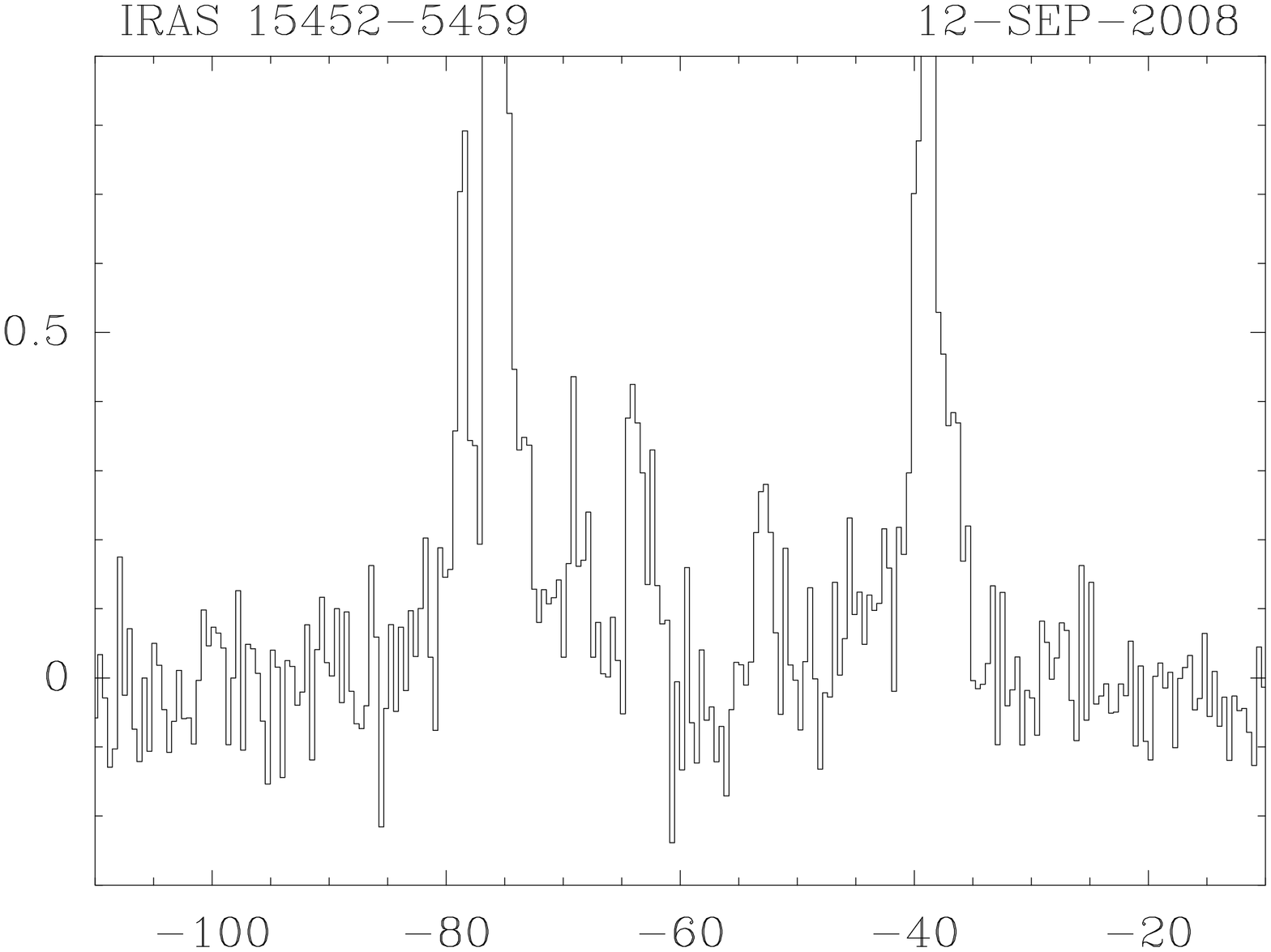}
\caption{Water maser spectra of the detected sources. Note that some objects were observed more than once. The x-axis is the velocity with respect to the local standard of rest (kinematical definition) in km s$^{-1}$, and the y-axis is flux density in Jansky. The spectra in the left column show the full scale in flux density, and a range of 200 km s$^{-1}$ in velocity (except in the case of IRAS 18113$-$2503, which is shown with a velocity range of 600 km s$^{-1}$). The spectra in the right column are the same as the corresponding ones in the left column, but with a restricted scale in flux density and/or velocity, to better show weak or narrow spectral features.}
\label{fig_detections}
\end{figure*}

\begin{figure*}[!t]
\centering
\ContinuedFloat
\includegraphics[width=0.3\textwidth,angle=-90]{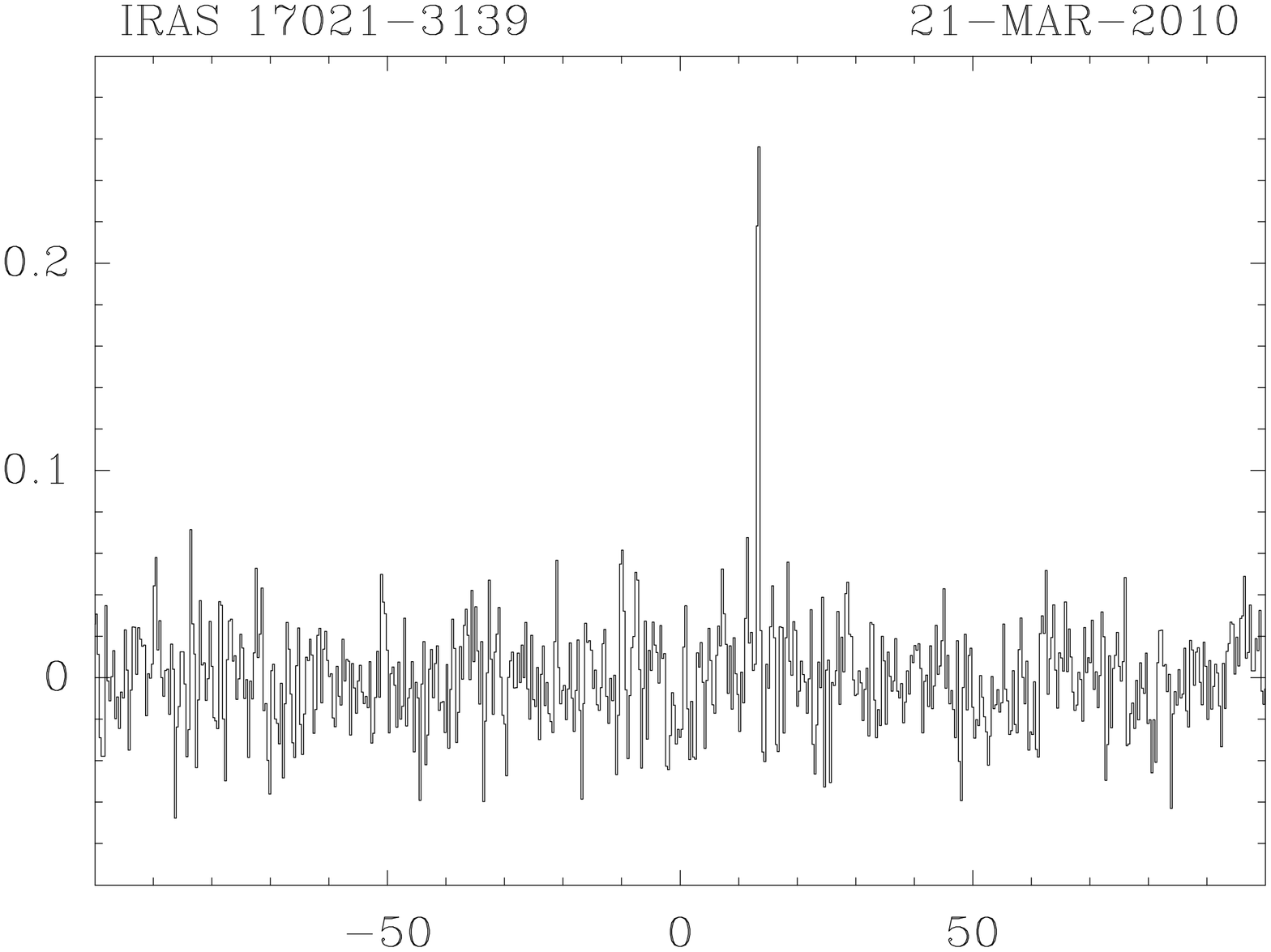}
\includegraphics[width=0.3\textwidth,angle=-90]{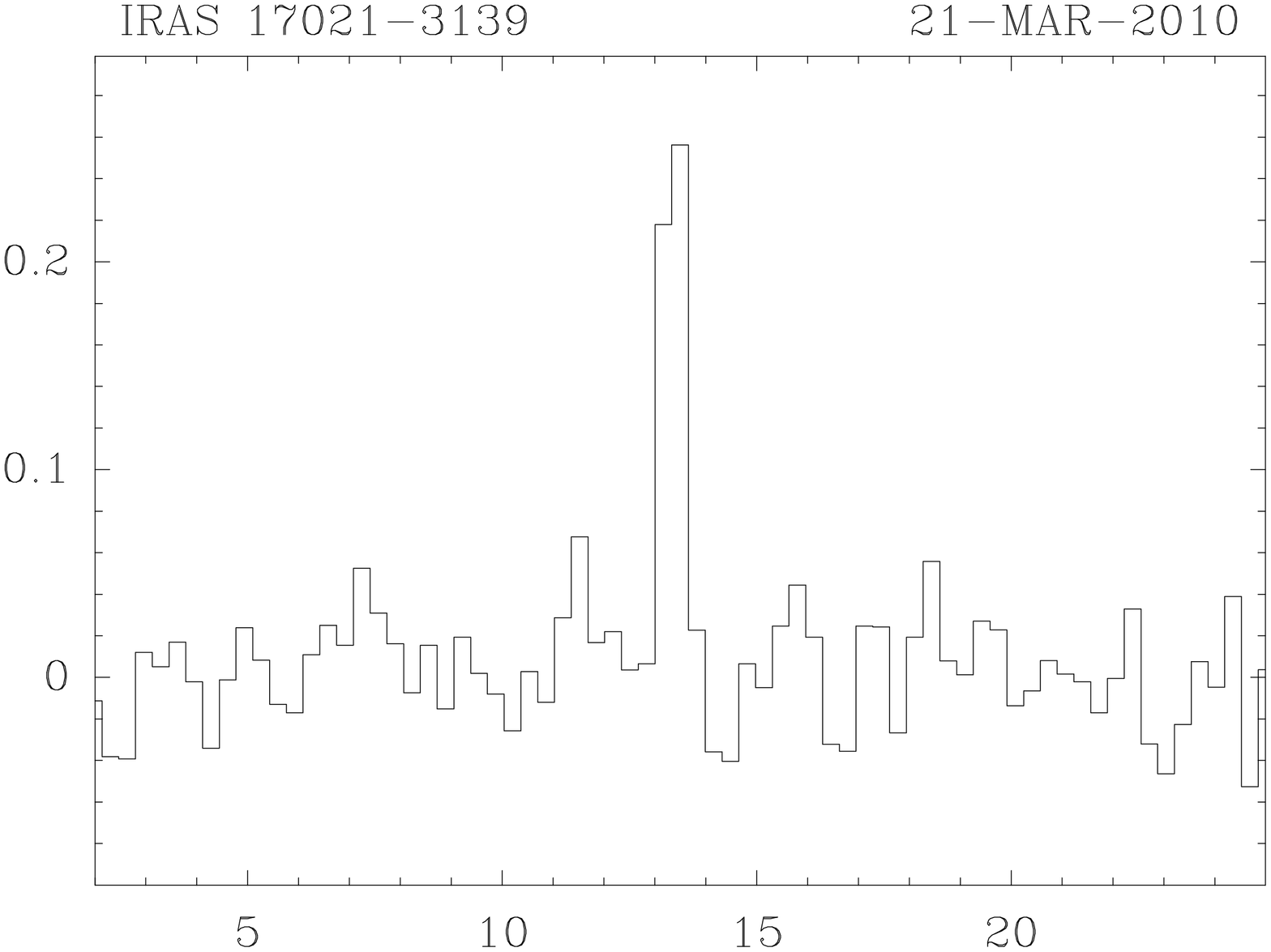}
\includegraphics[width=0.3\textwidth,angle=-90]{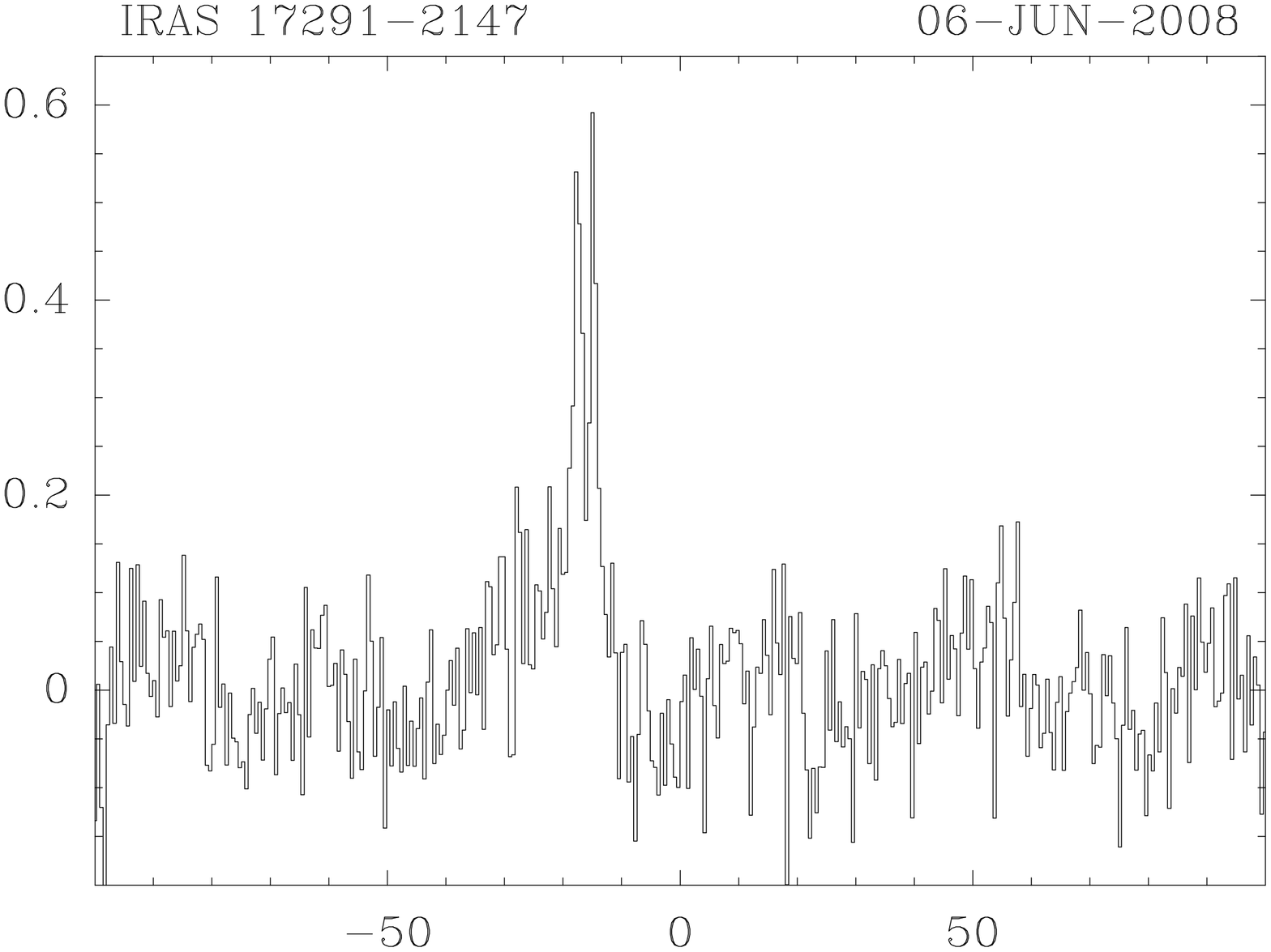}
\includegraphics[width=0.3\textwidth,angle=-90]{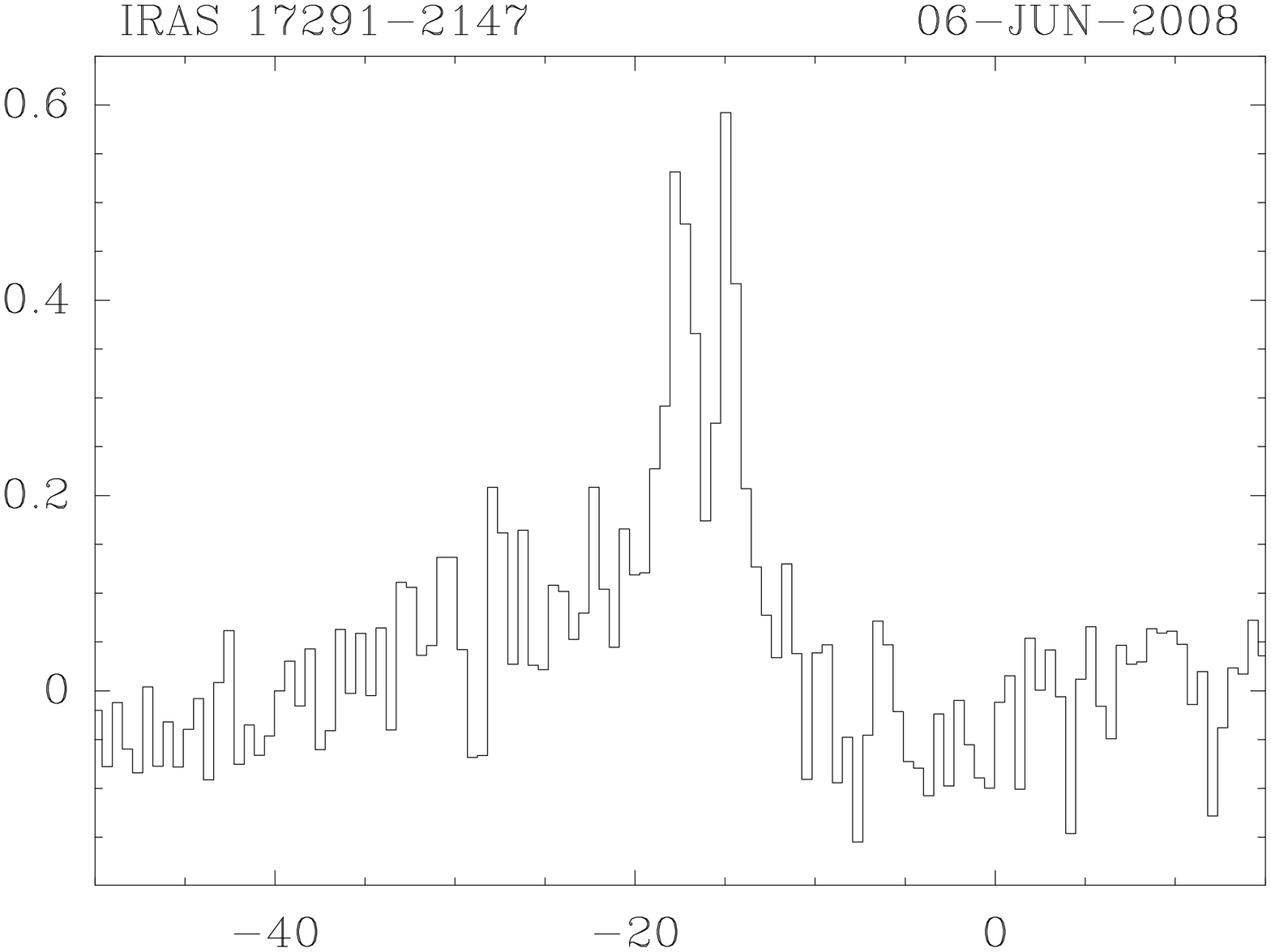}
\includegraphics[width=0.3\textwidth,angle=-90]{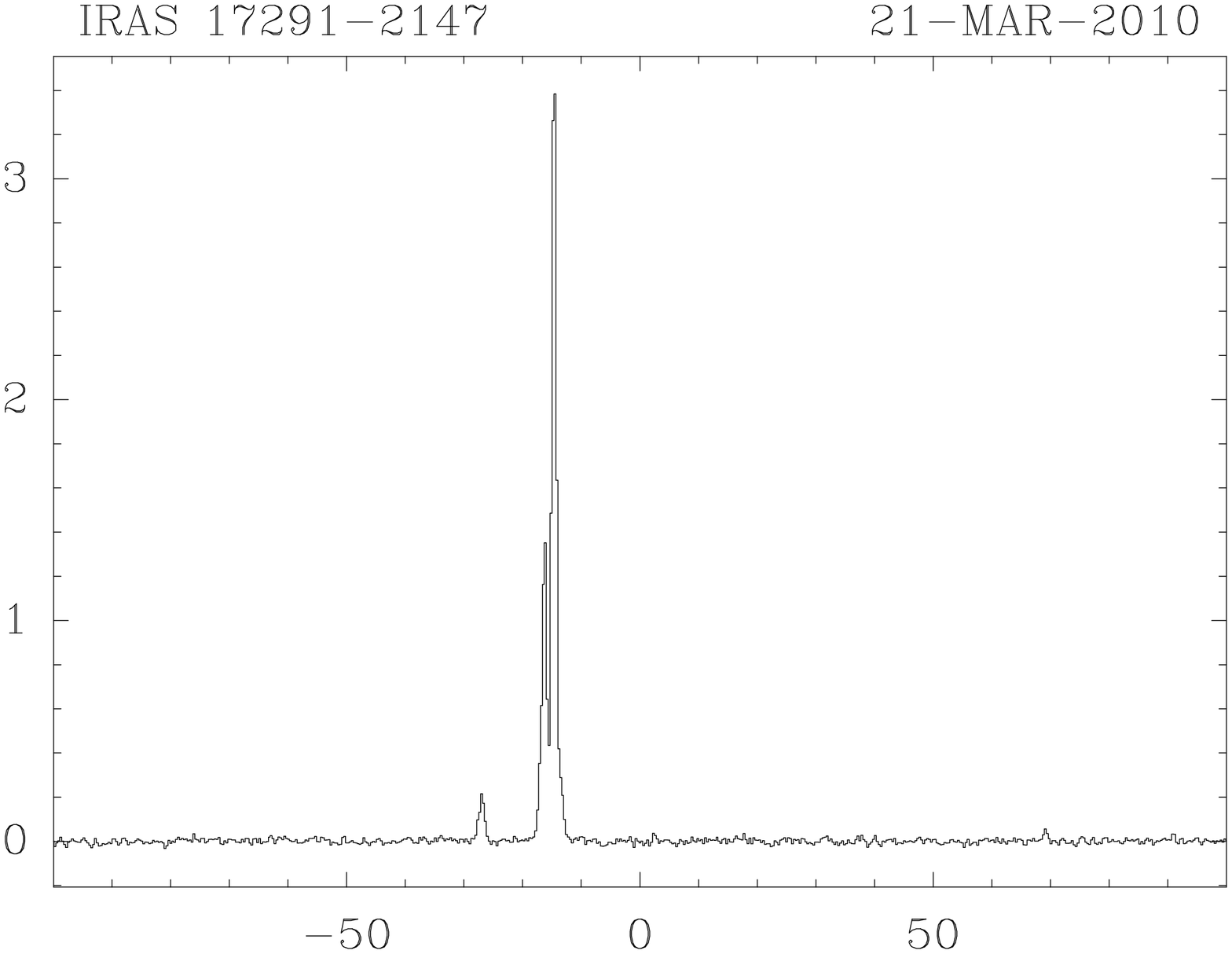}
\includegraphics[width=0.3\textwidth,angle=-90]{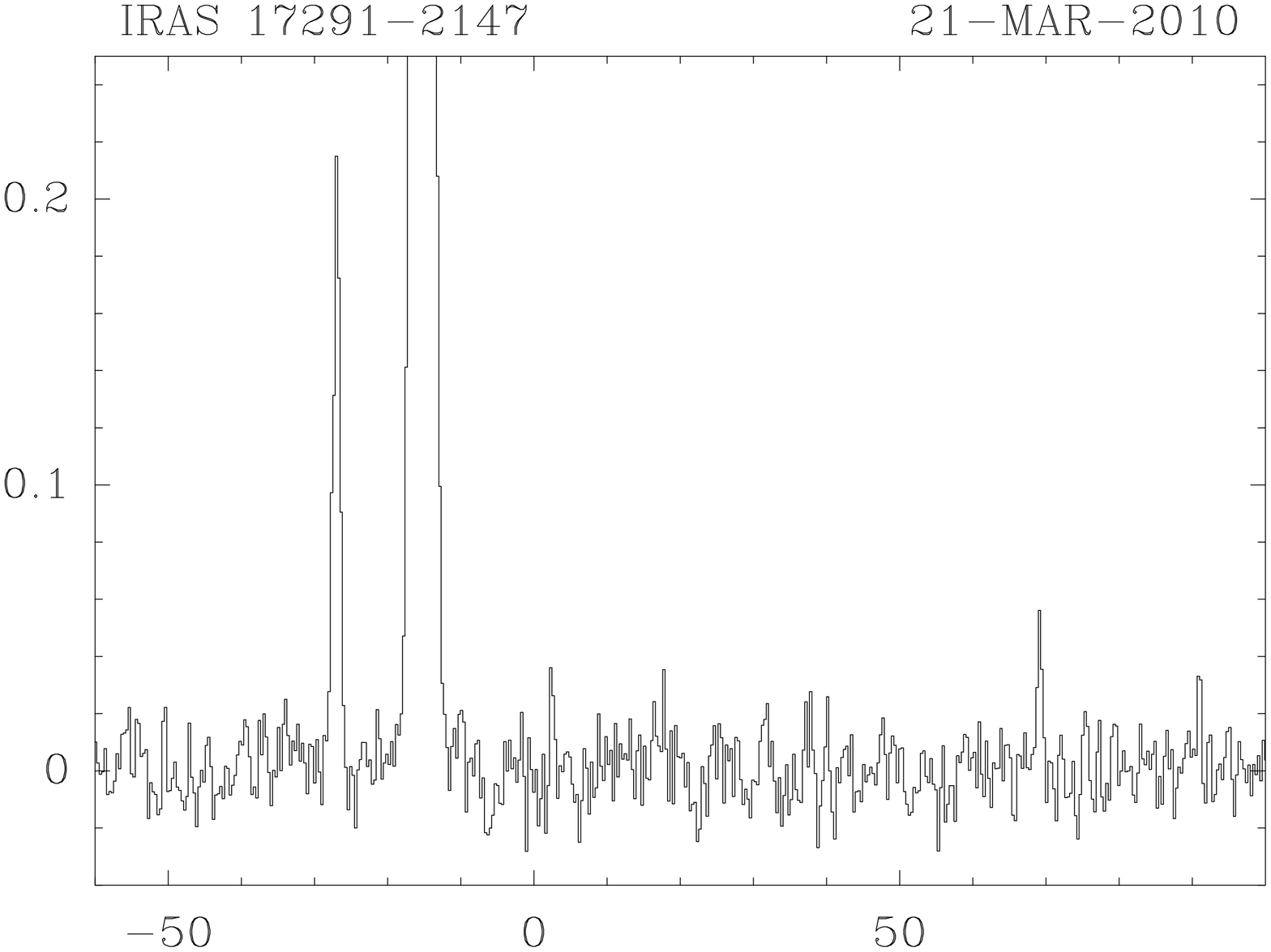}
\includegraphics[width=0.3\textwidth,angle=-90]{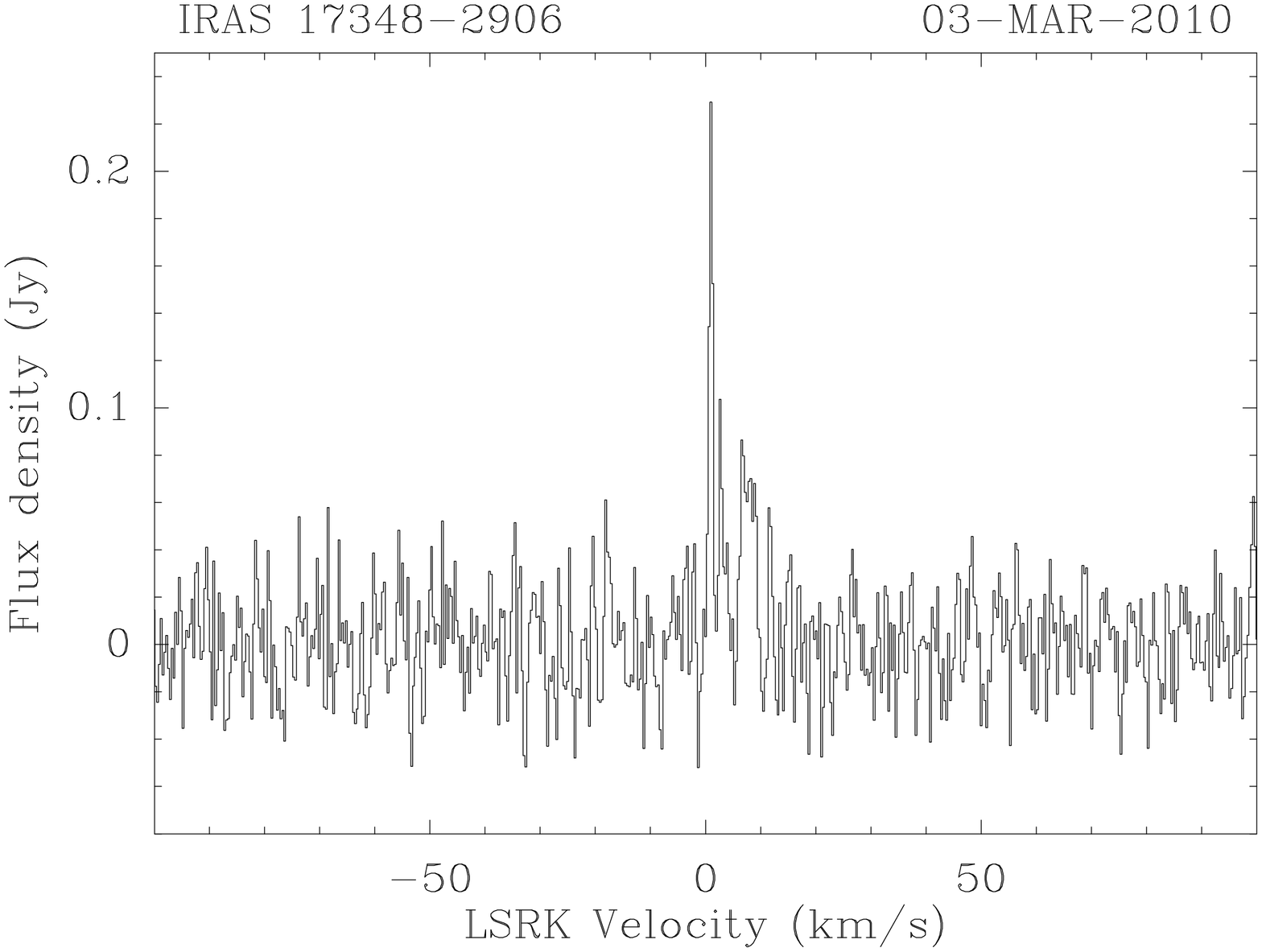}
\includegraphics[width=0.3\textwidth,angle=-90]{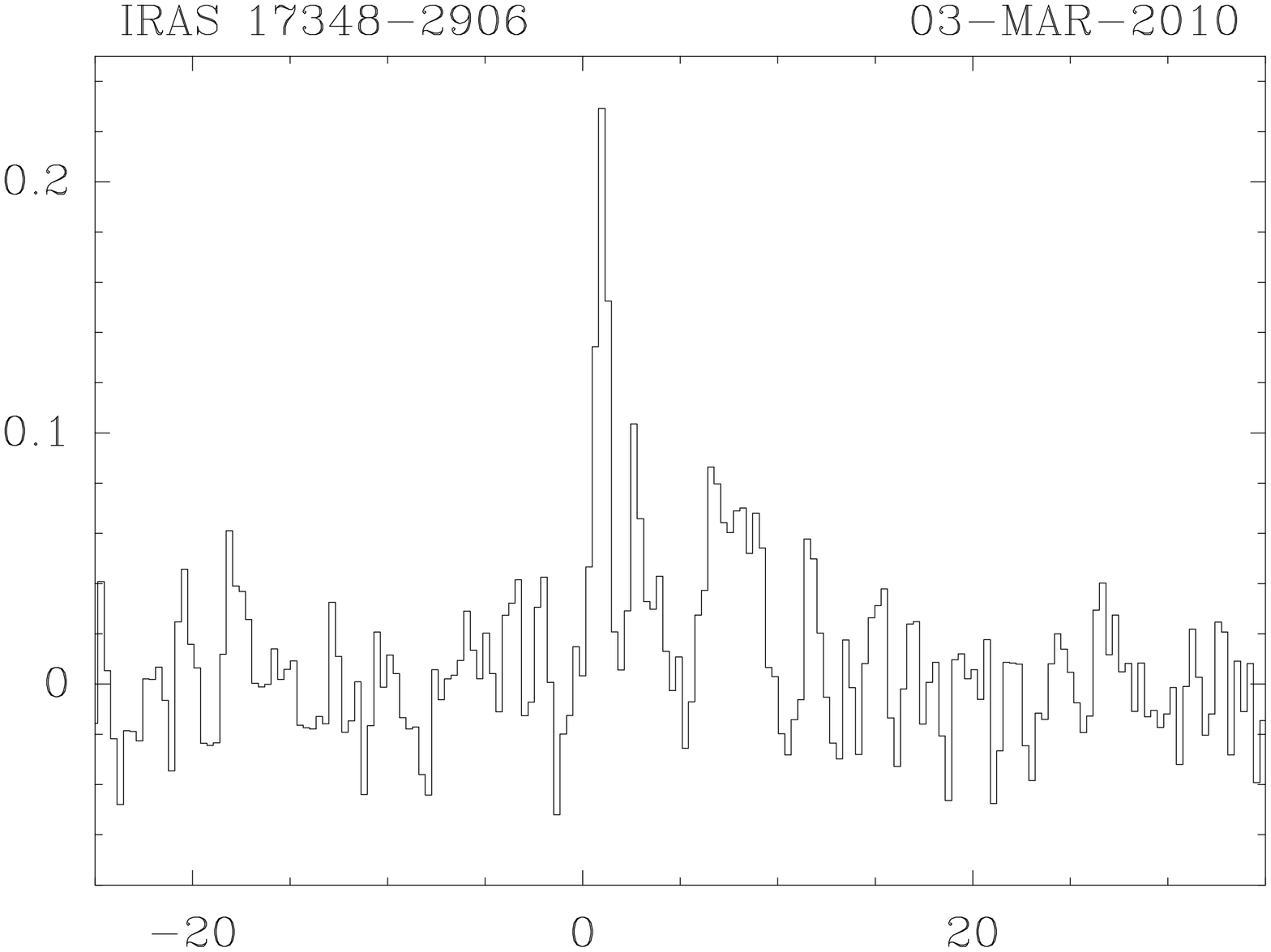}
\caption{
continued.}
\end{figure*}

\begin{figure*}[!t]
\centering
\ContinuedFloat
\includegraphics[width=0.3\textwidth,angle=-90]{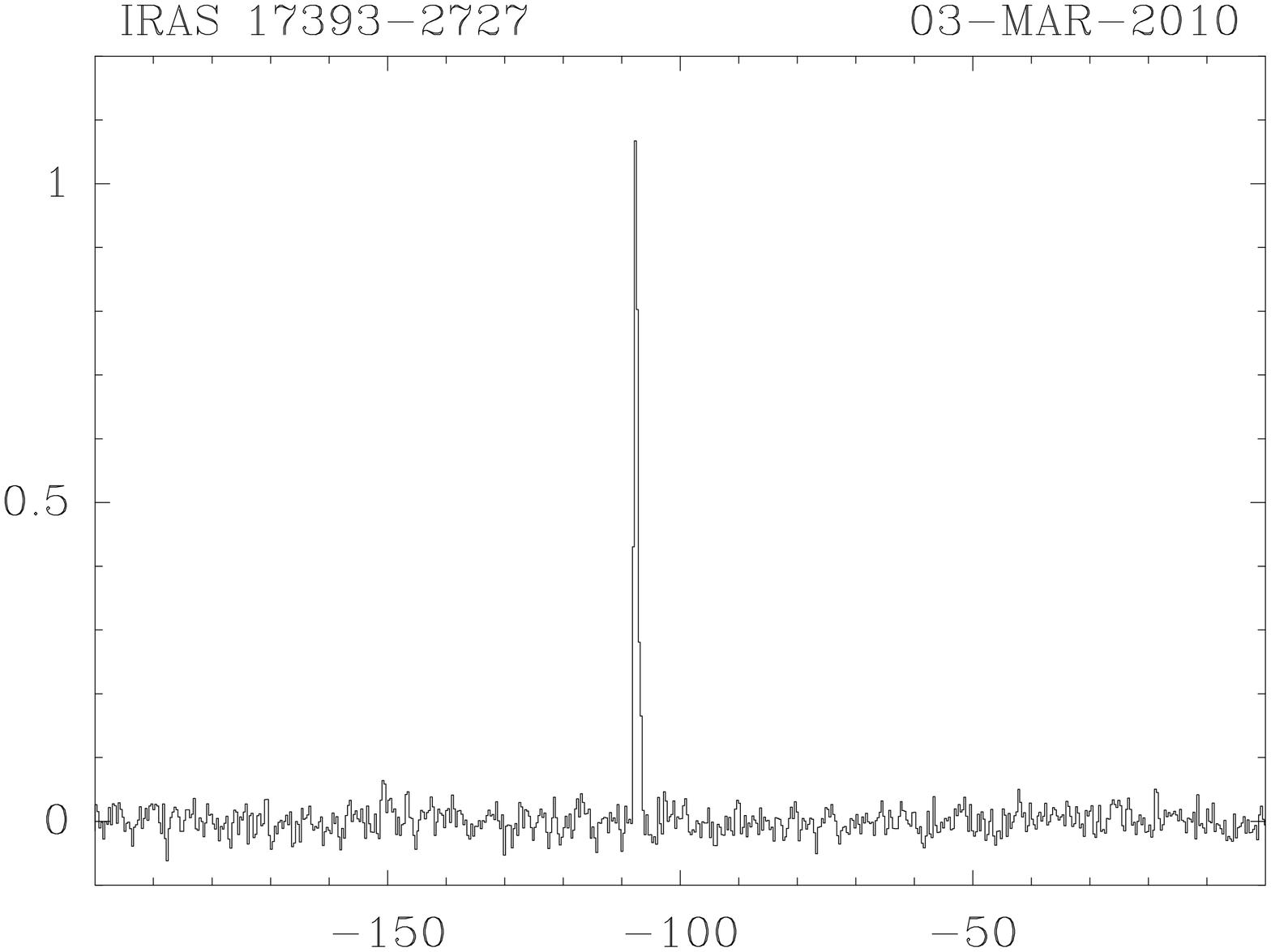}
\includegraphics[width=0.3\textwidth,angle=-90]{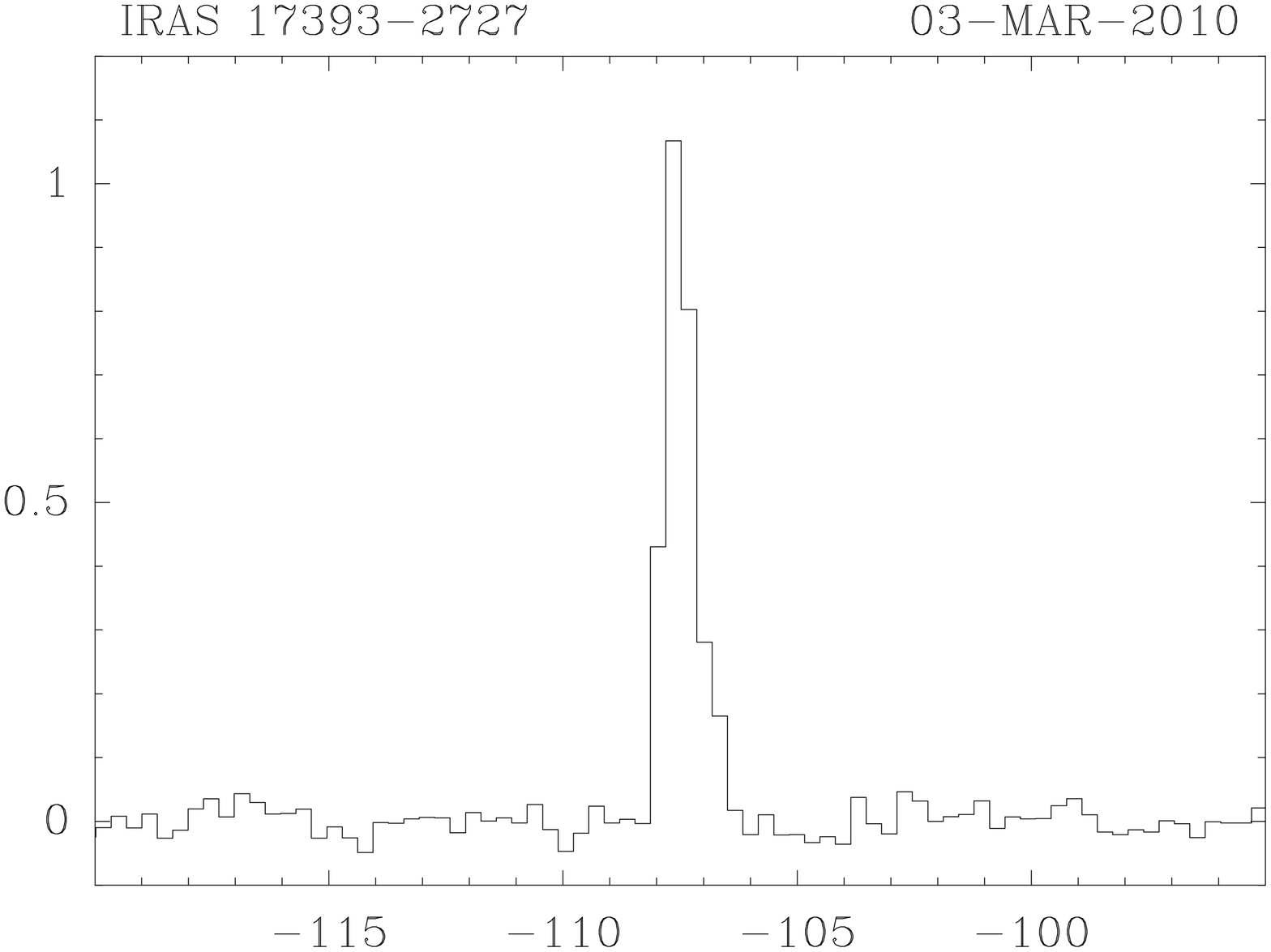}
\includegraphics[width=0.3\textwidth,angle=-90]{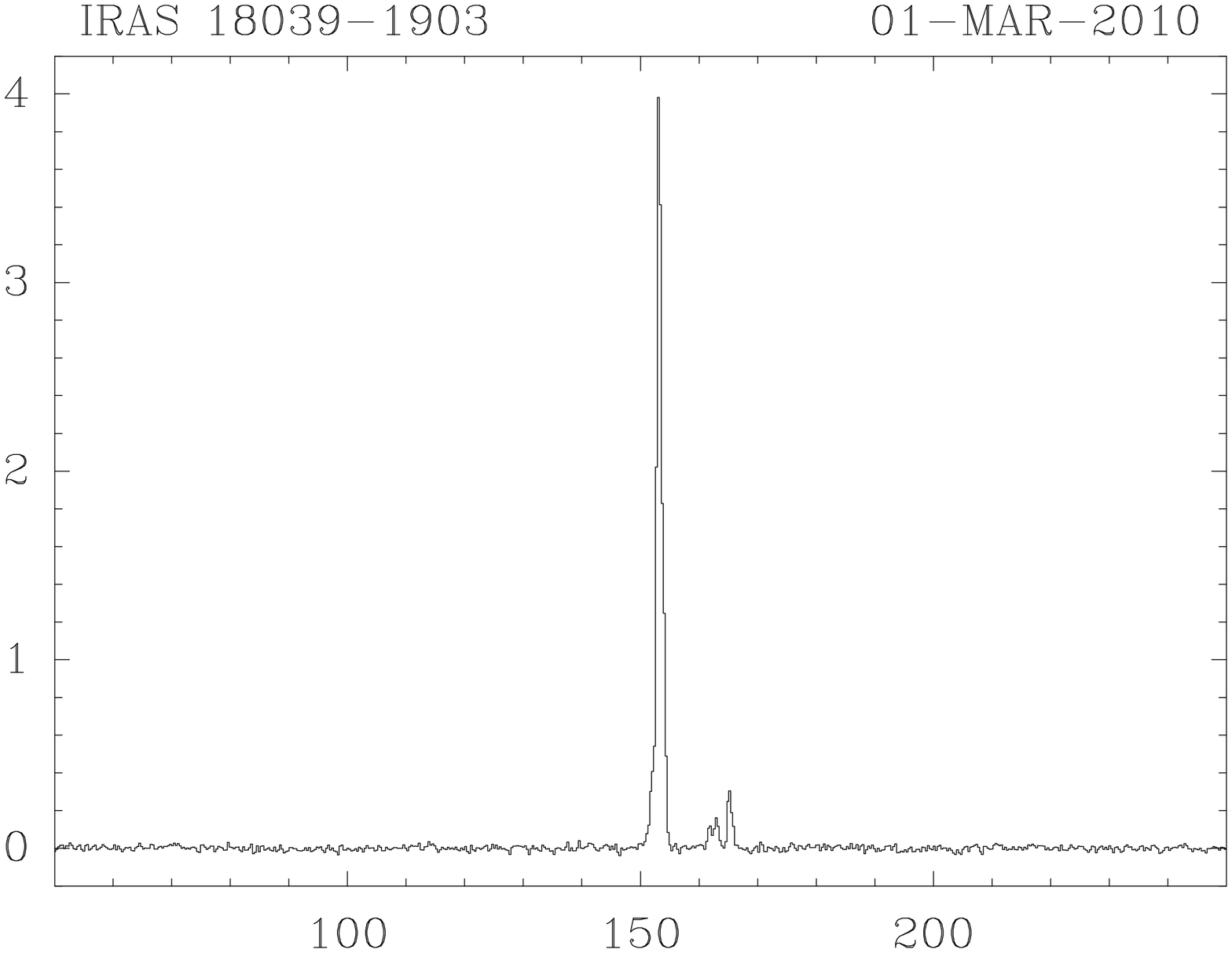}
\includegraphics[width=0.3\textwidth,angle=-90]{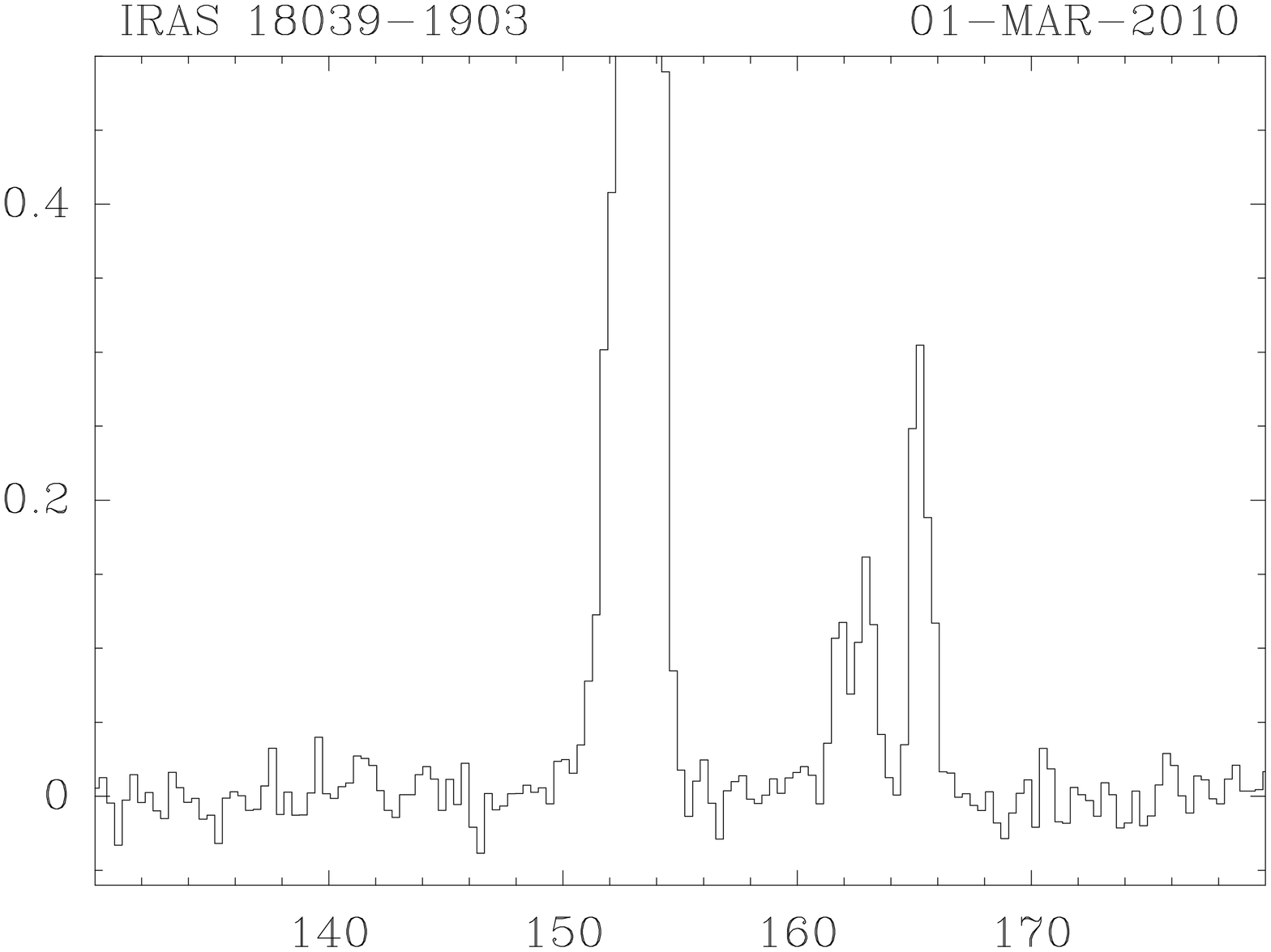}
\includegraphics[width=0.3\textwidth,angle=-90]{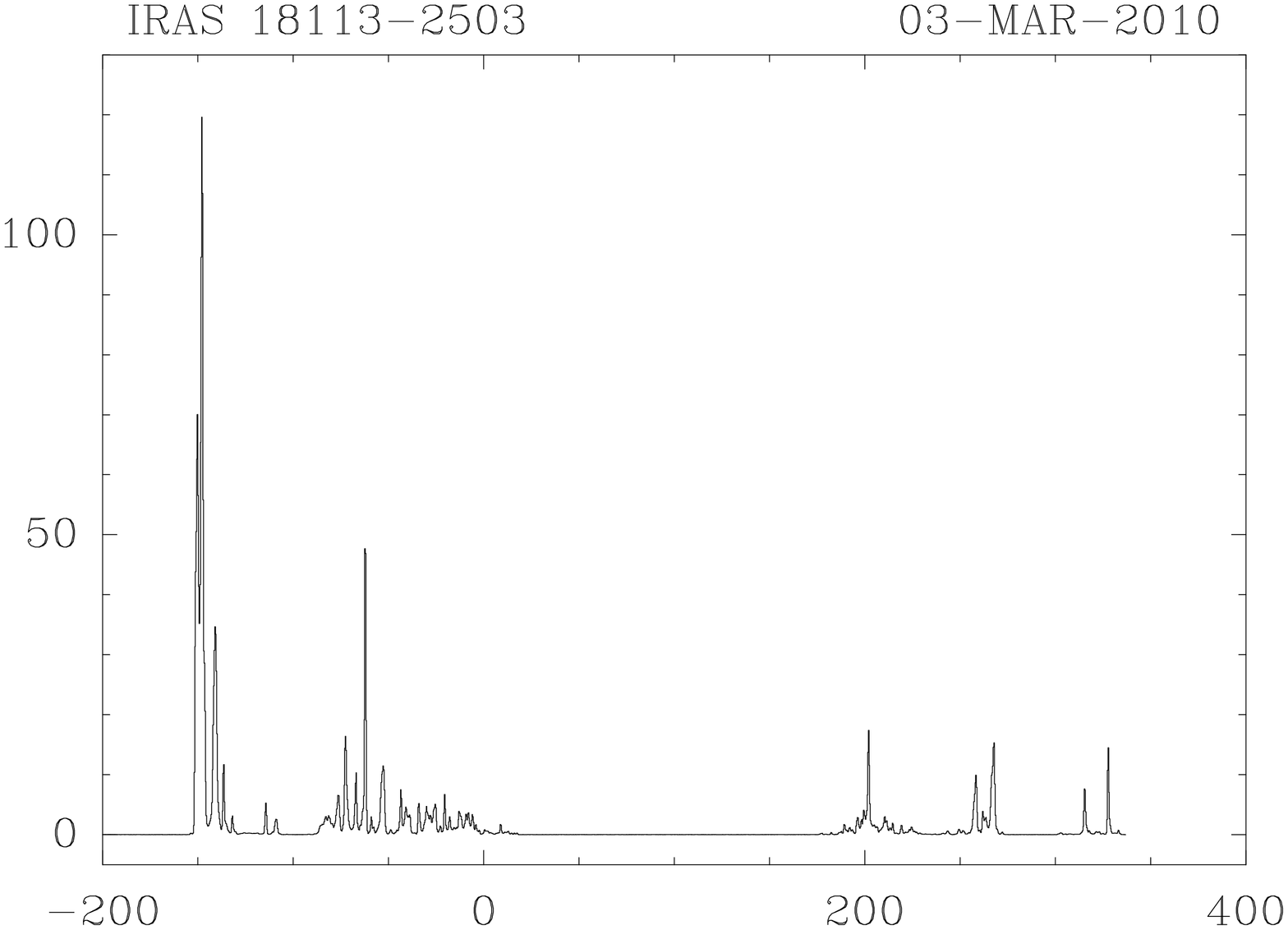}
\includegraphics[width=0.3\textwidth,angle=-90]{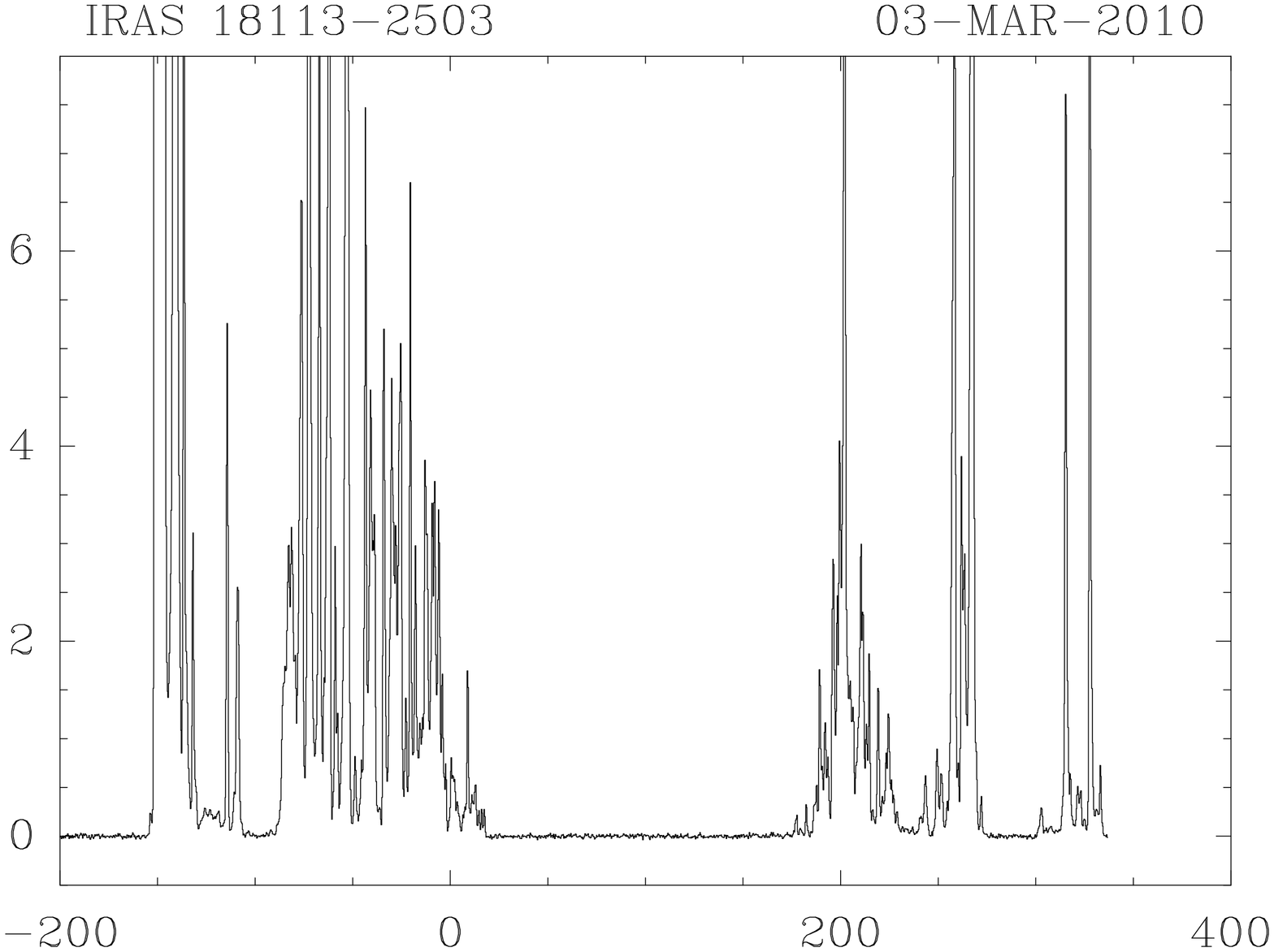}
\includegraphics[width=0.3\textwidth,angle=-90]{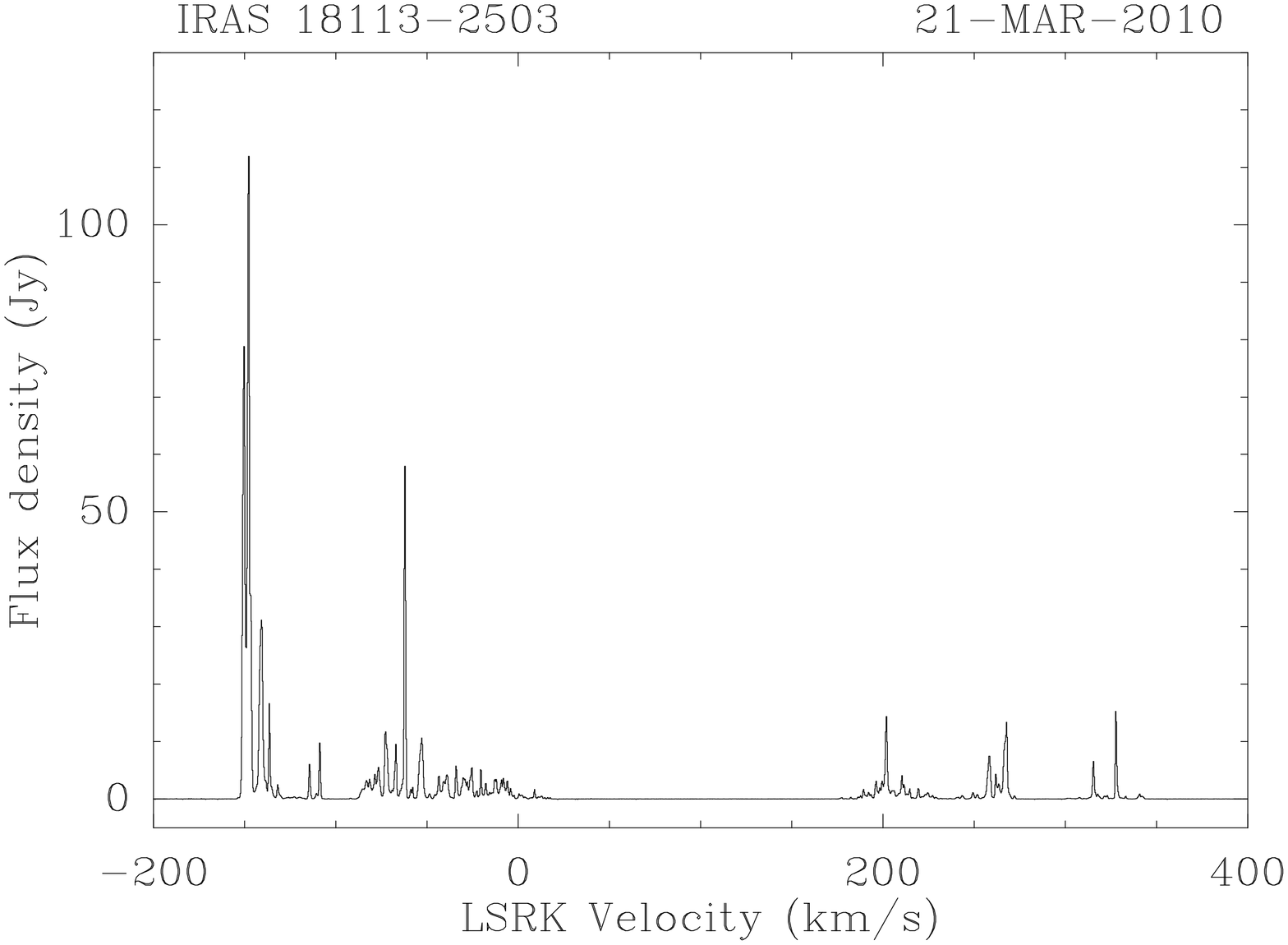}
\includegraphics[width=0.3\textwidth,angle=-90]{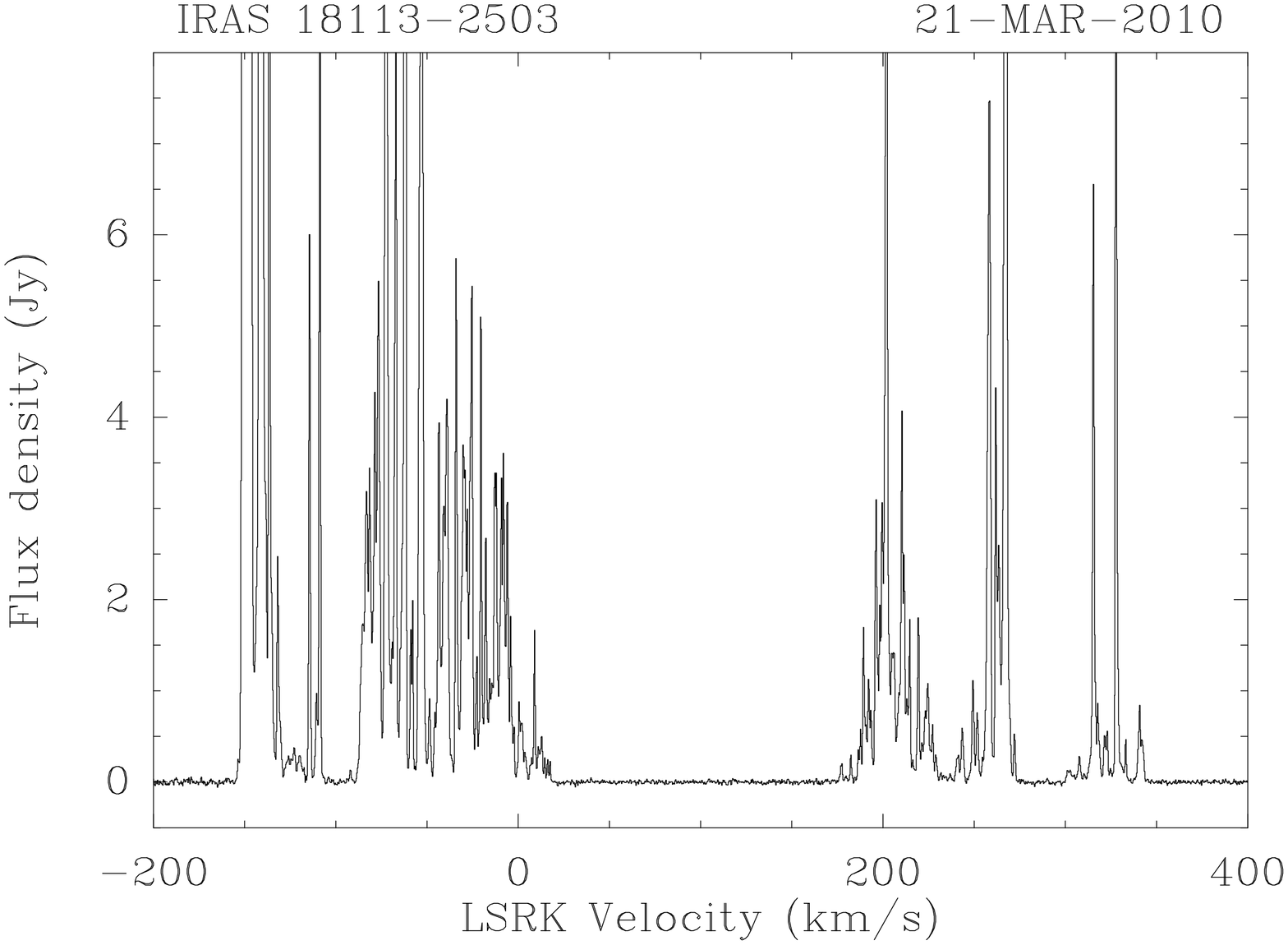}
\caption{
continued.}
\end{figure*}

\begin{figure*}[!t]
\centering
\ContinuedFloat
\includegraphics[width=0.3\textwidth,angle=-90]{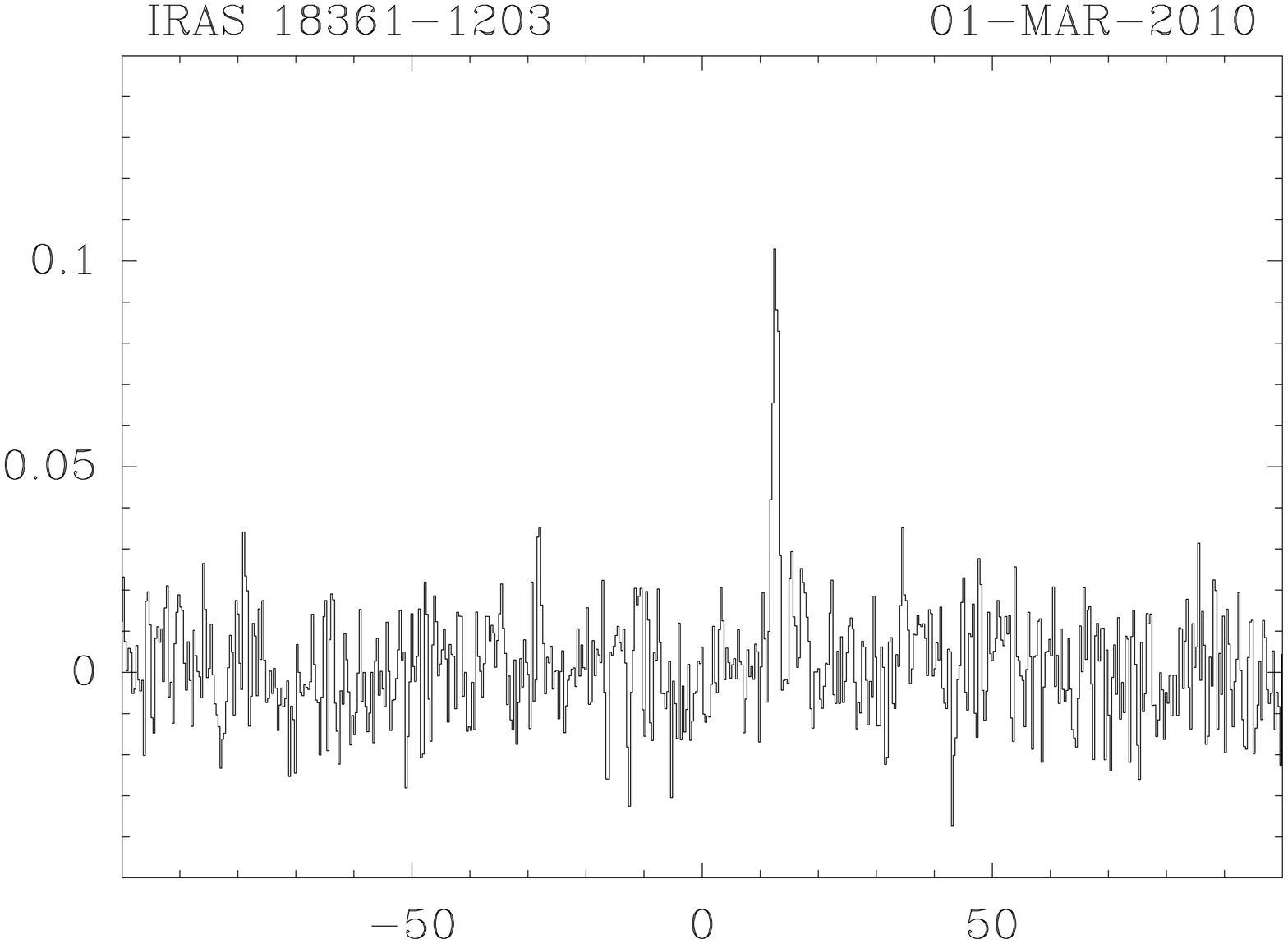}
\includegraphics[width=0.3\textwidth,angle=-90]{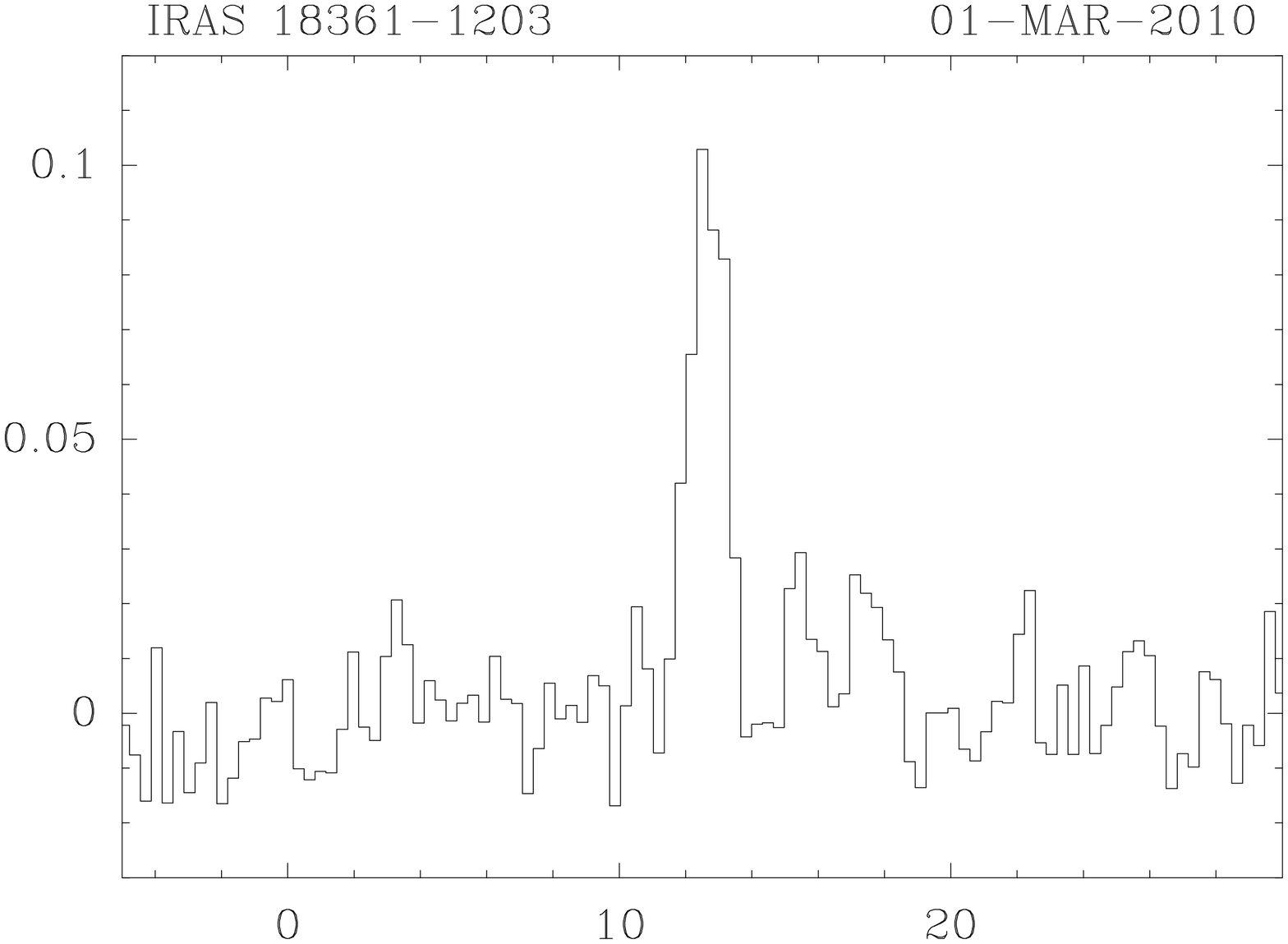}
\includegraphics[width=0.3\textwidth,angle=-90]{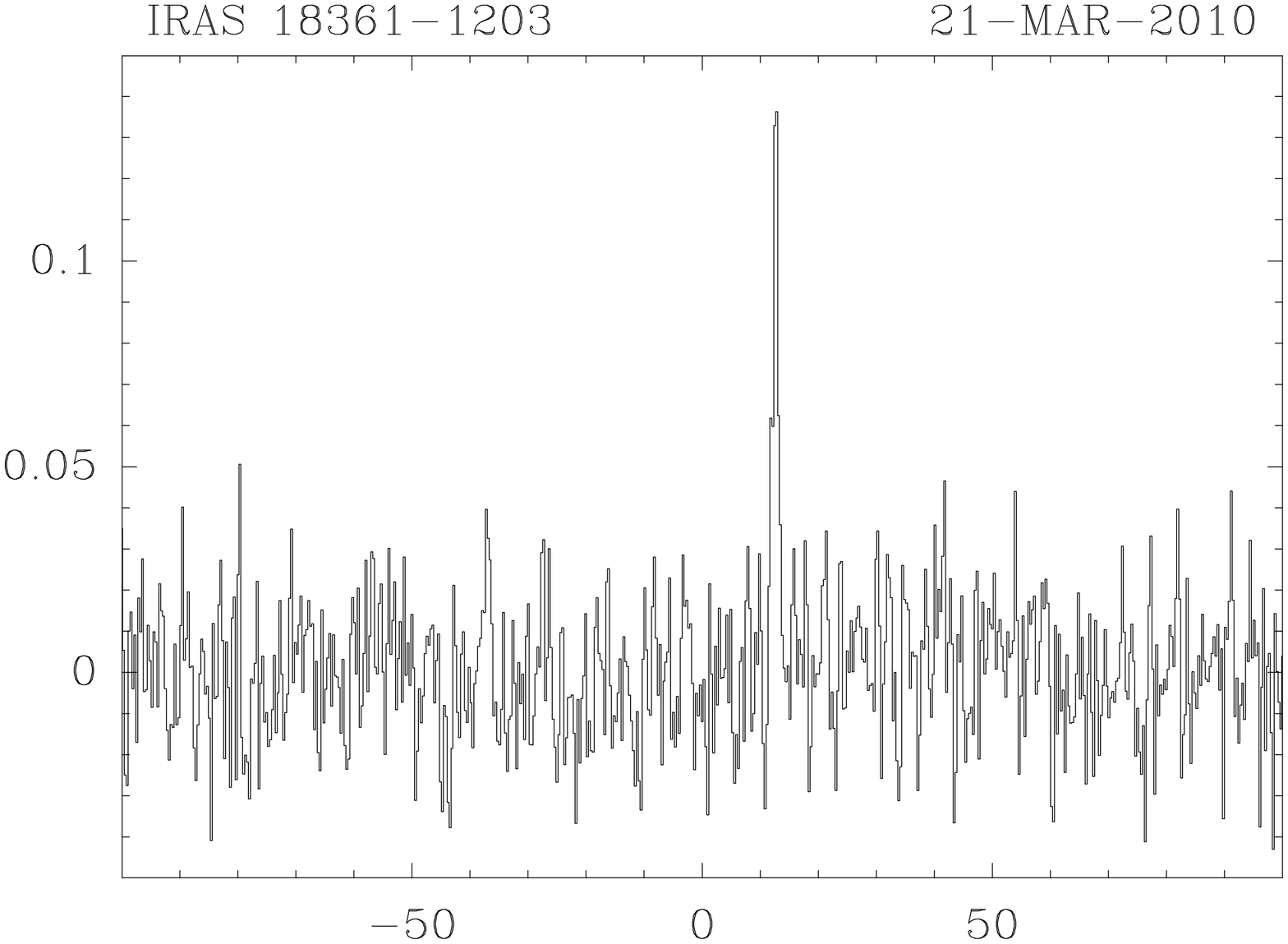}
\includegraphics[width=0.3\textwidth,angle=-90]{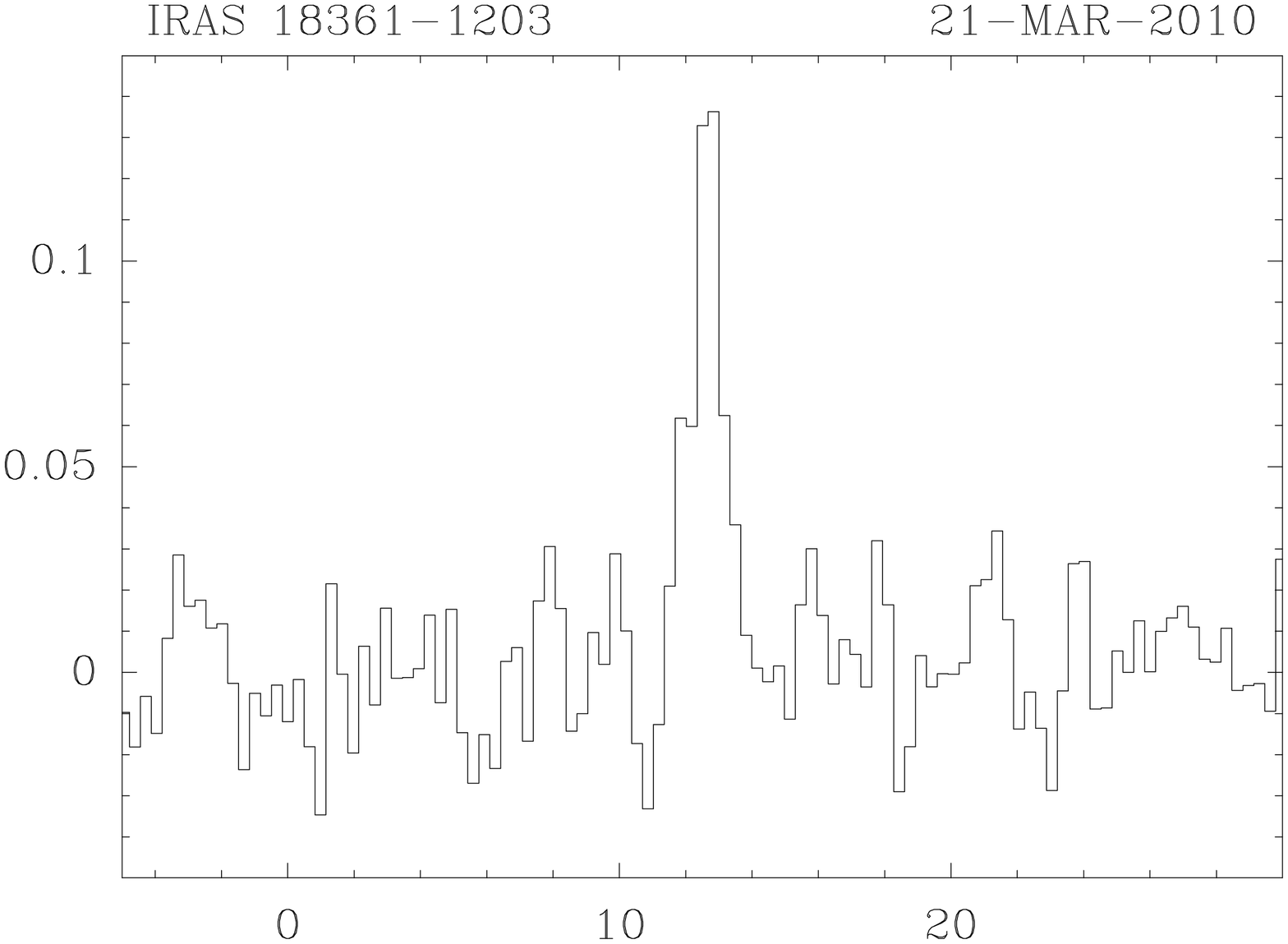}
\includegraphics[width=0.3\textwidth,angle=-90]{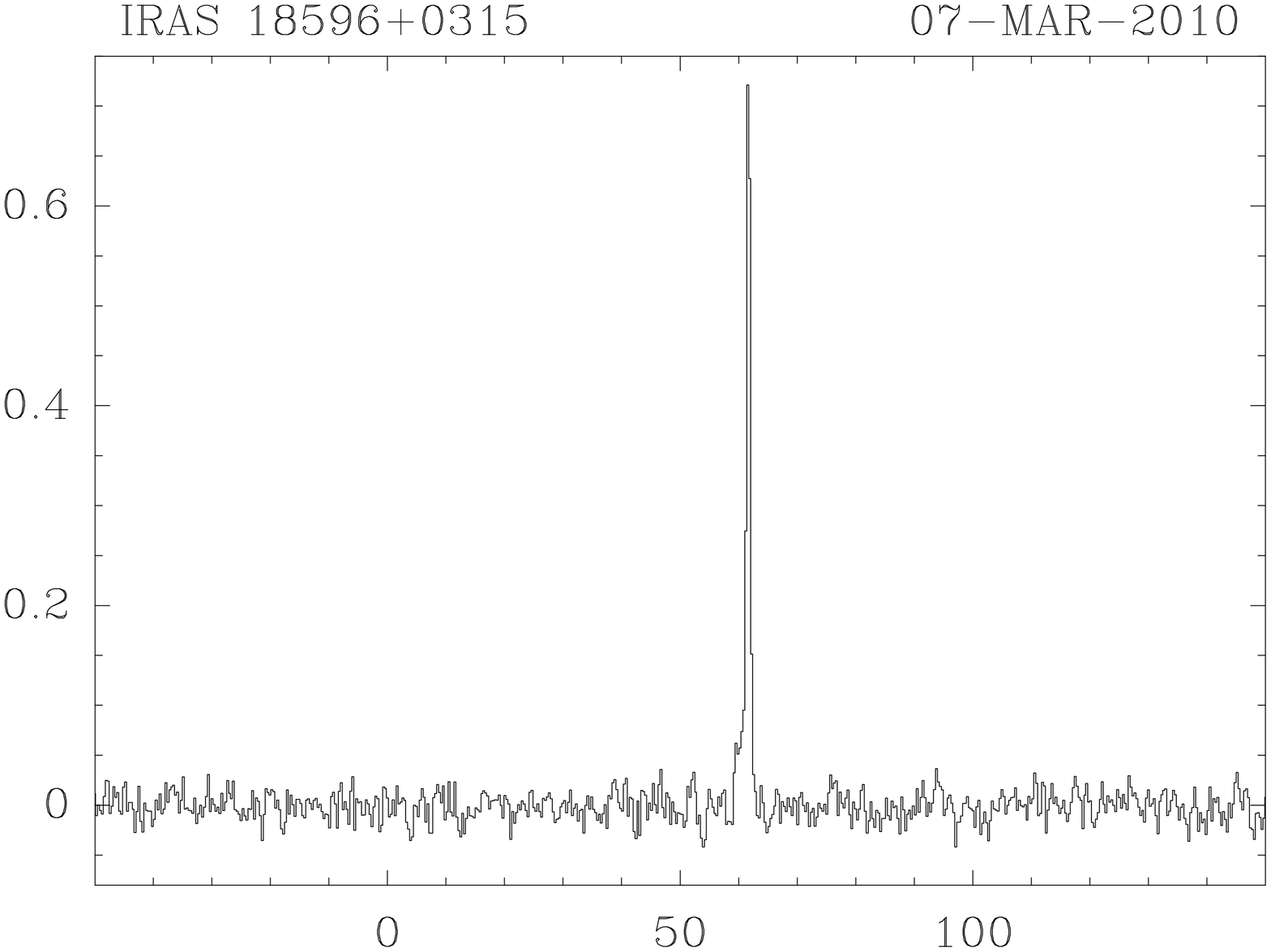}
\includegraphics[width=0.3\textwidth,angle=-90]{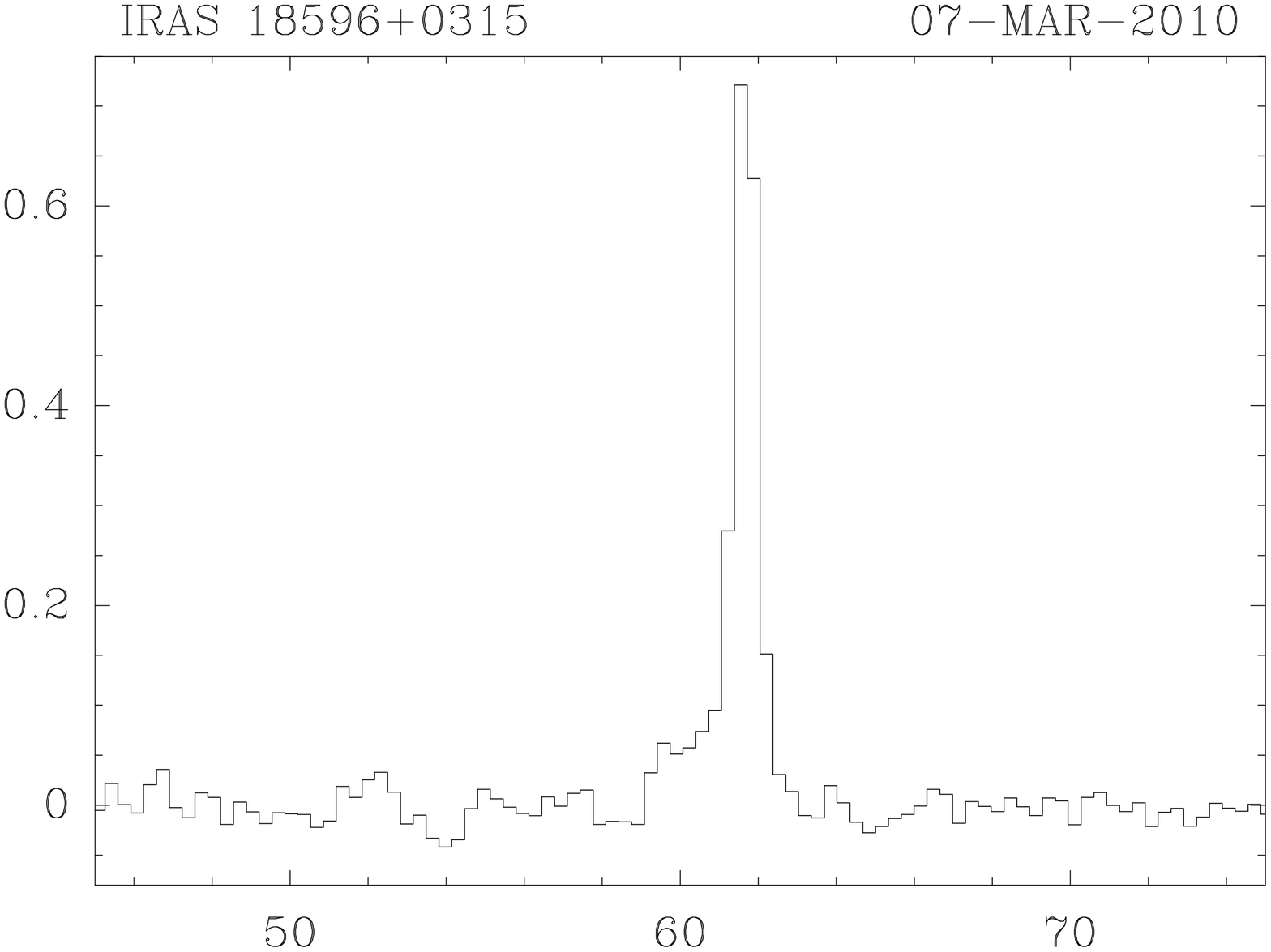}
\includegraphics[width=0.3\textwidth,angle=-90]{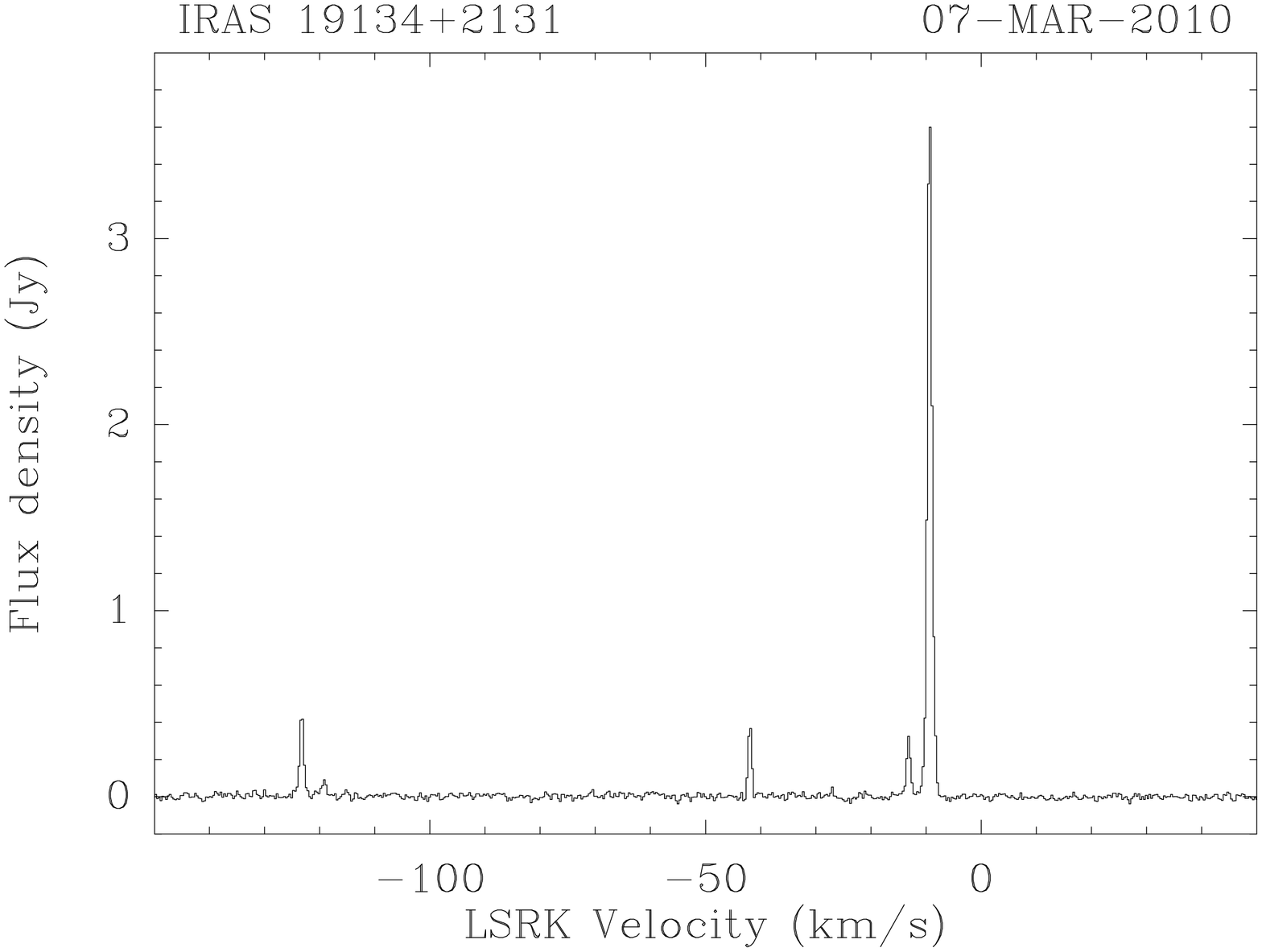}
\includegraphics[width=0.3\textwidth,angle=-90]{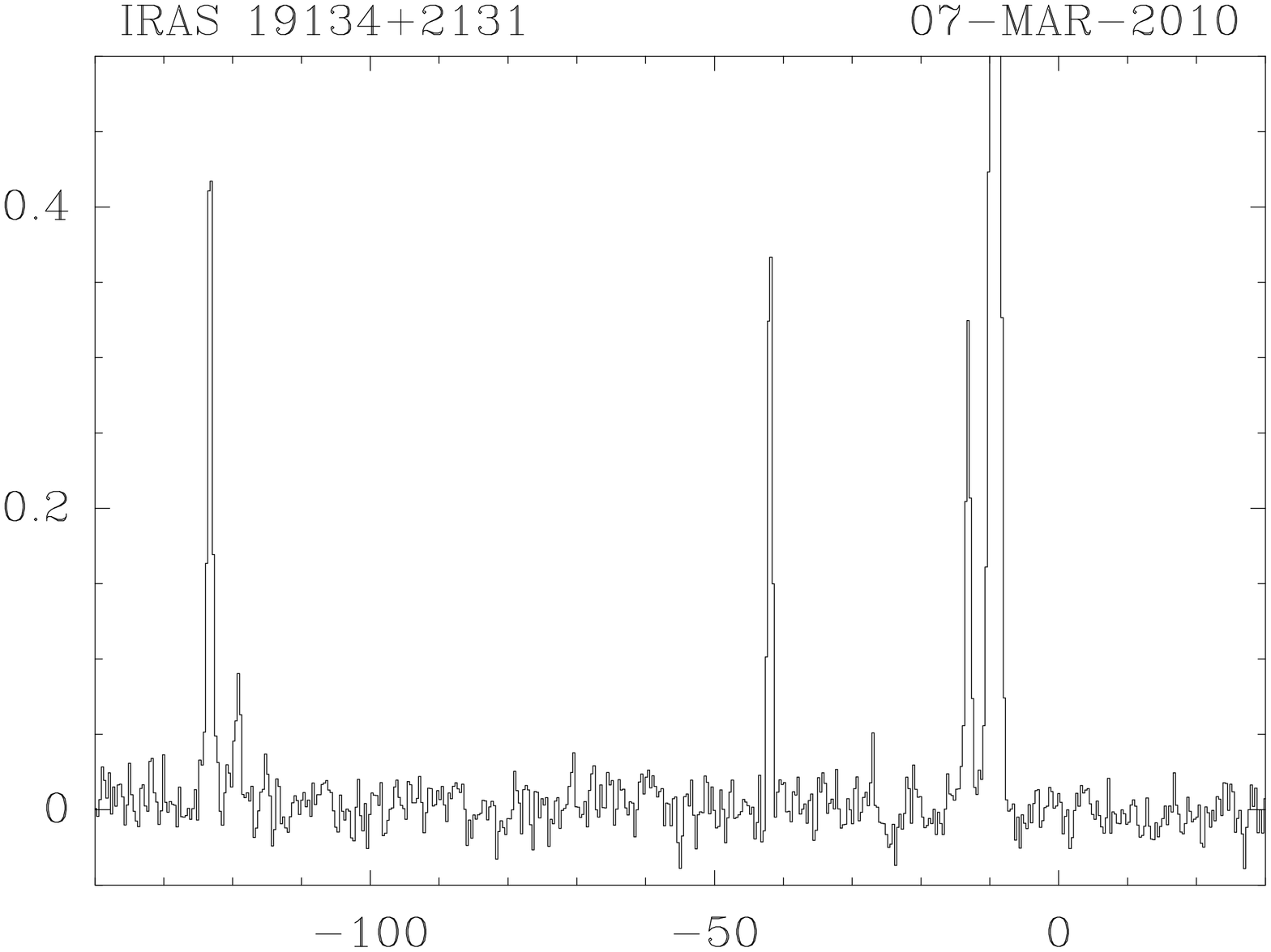}
\caption{
continued.}
\end{figure*}

\begin{figure*}[!t]
\centering
\ContinuedFloat
\includegraphics[width=0.3\textwidth,angle=-90]{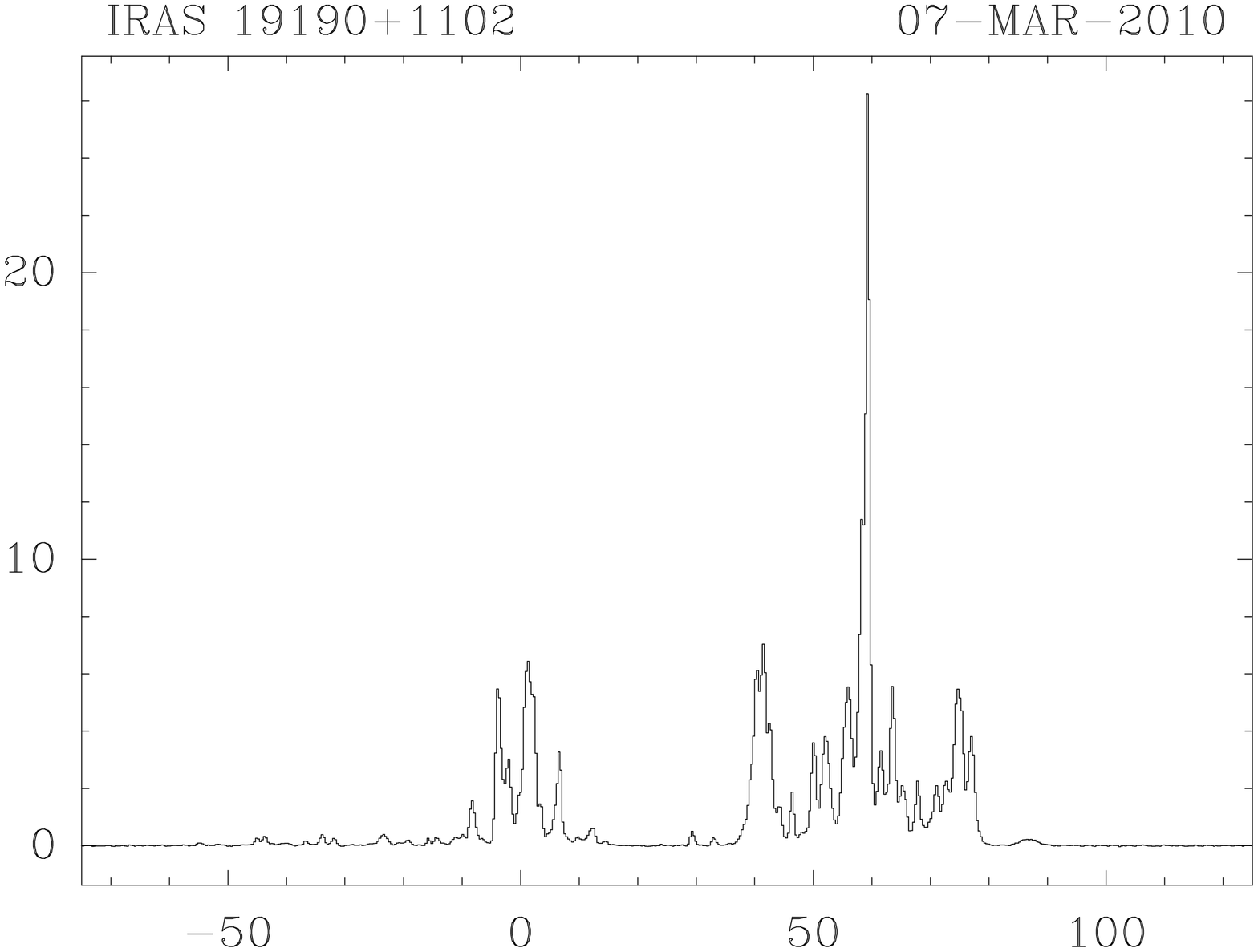}
\includegraphics[width=0.3\textwidth,angle=-90]{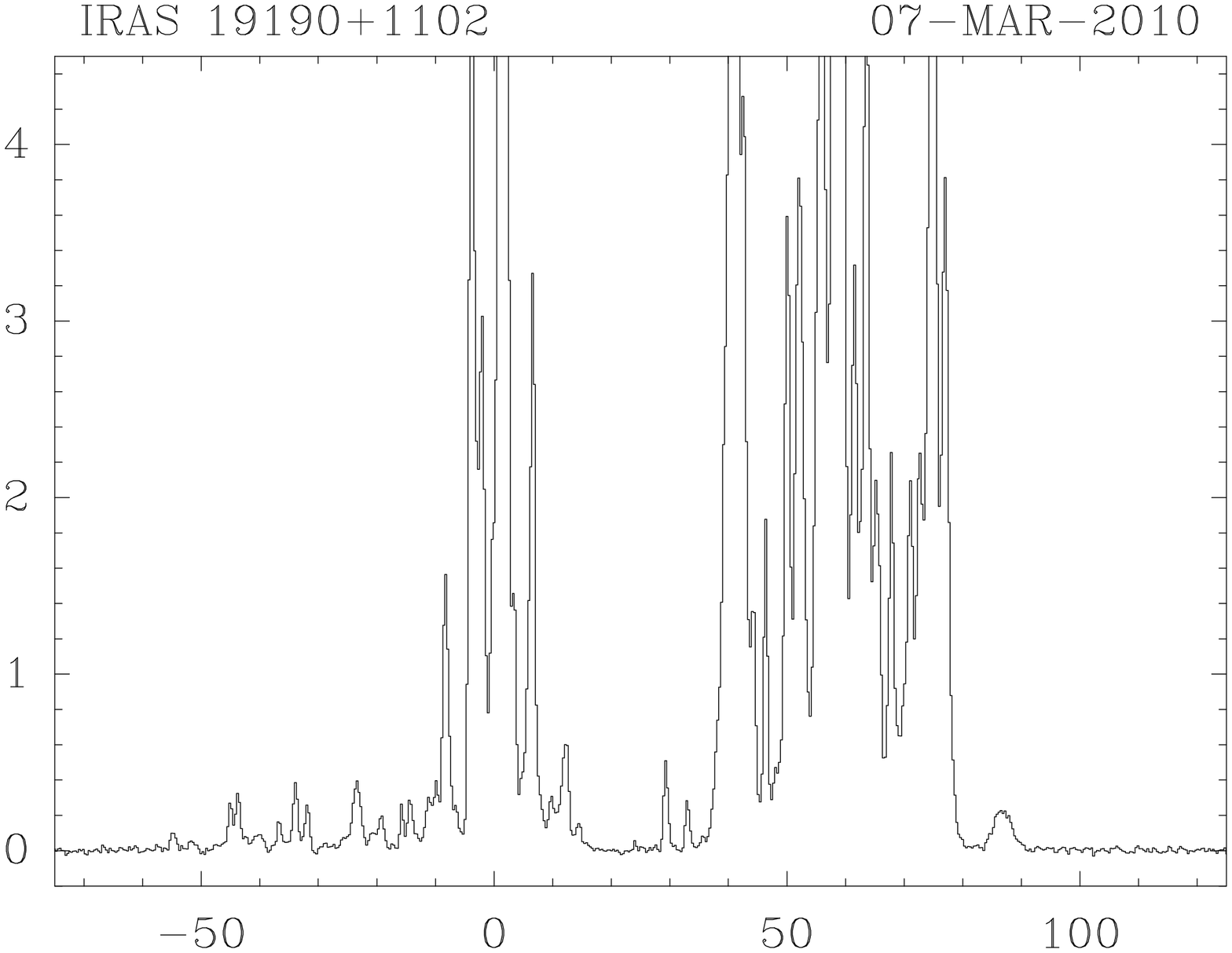}
\includegraphics[width=0.3\textwidth,angle=-90]{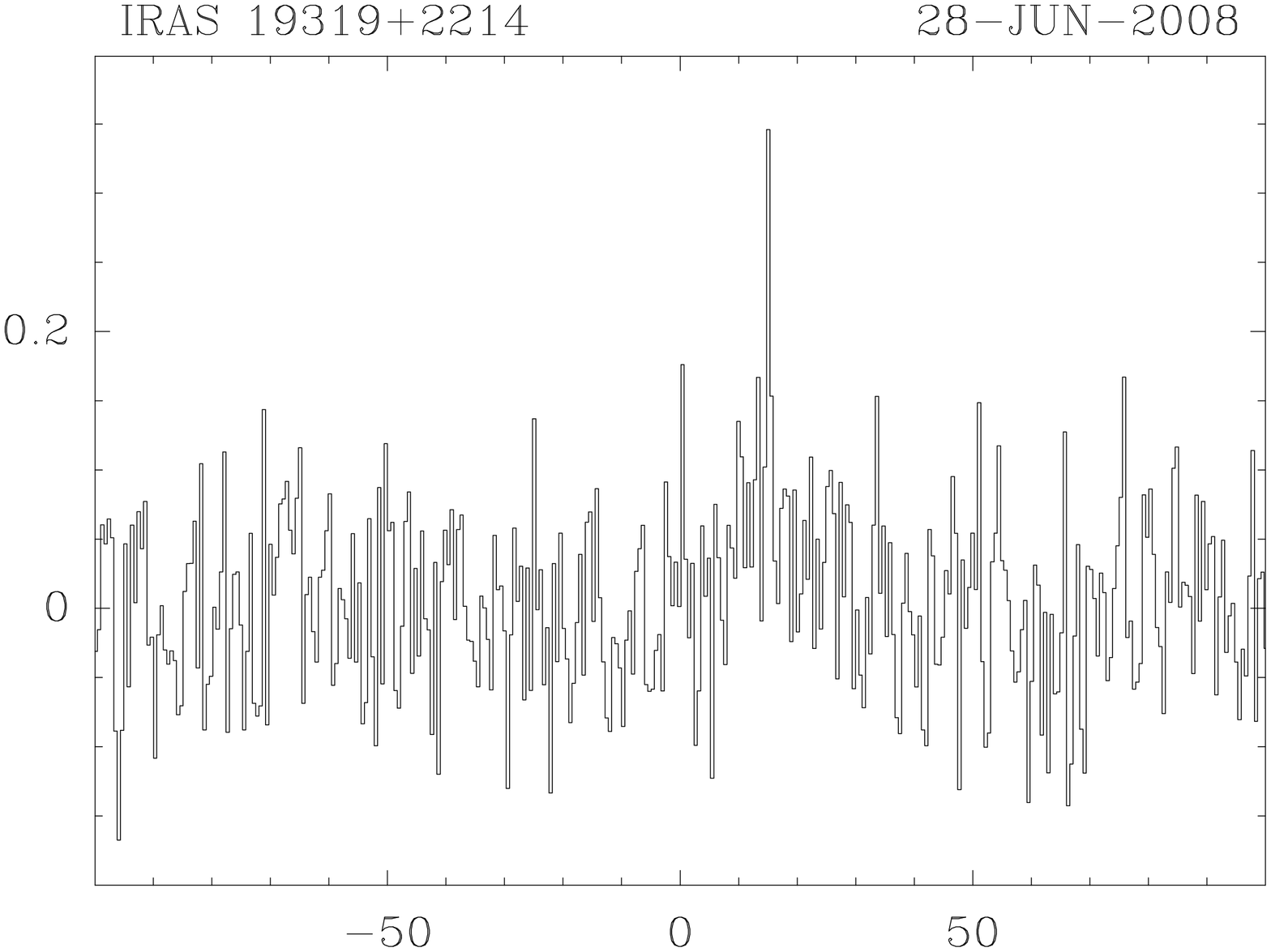}
\includegraphics[width=0.3\textwidth,angle=-90]{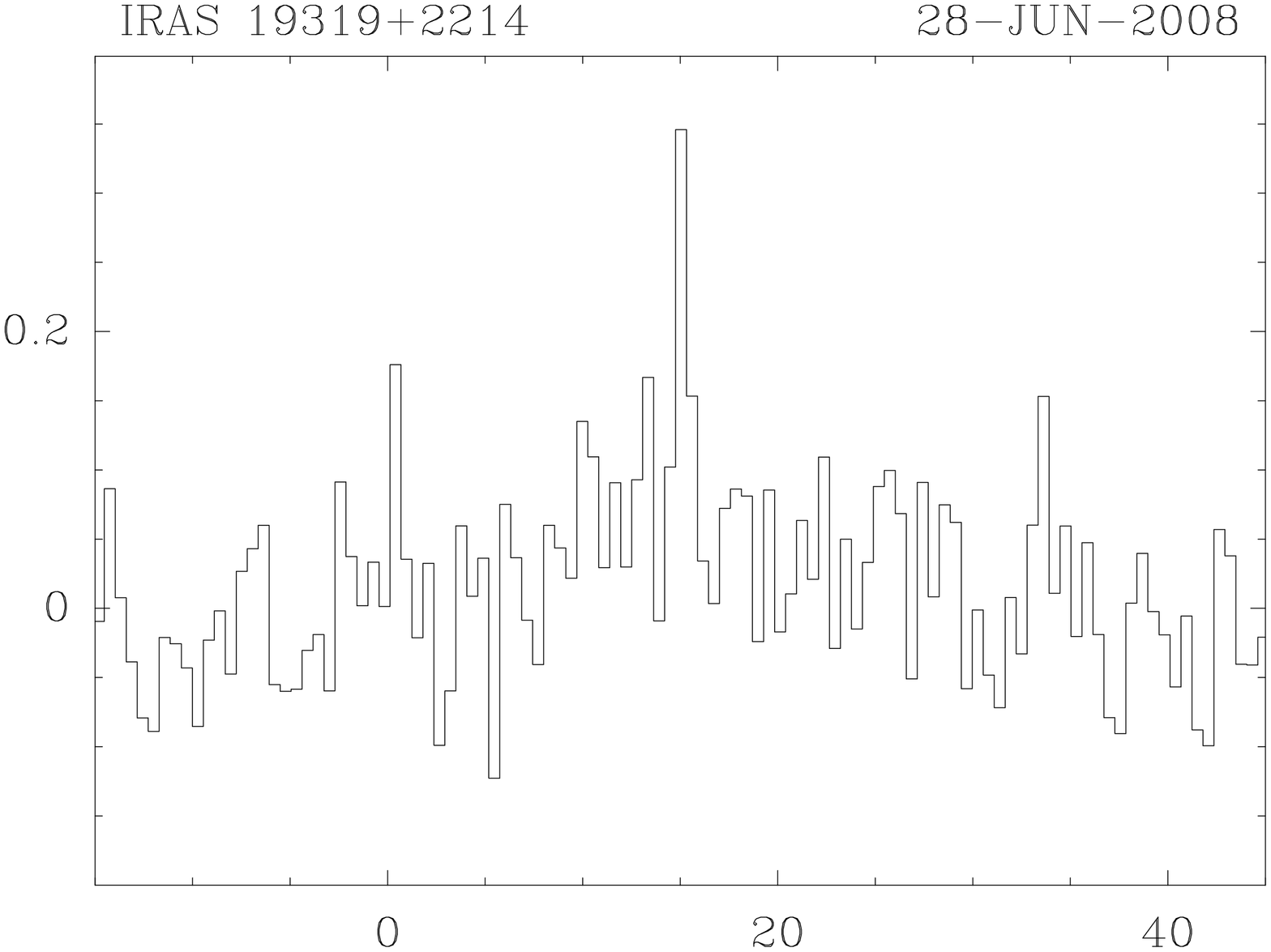}
\includegraphics[width=0.3\textwidth,angle=-90]{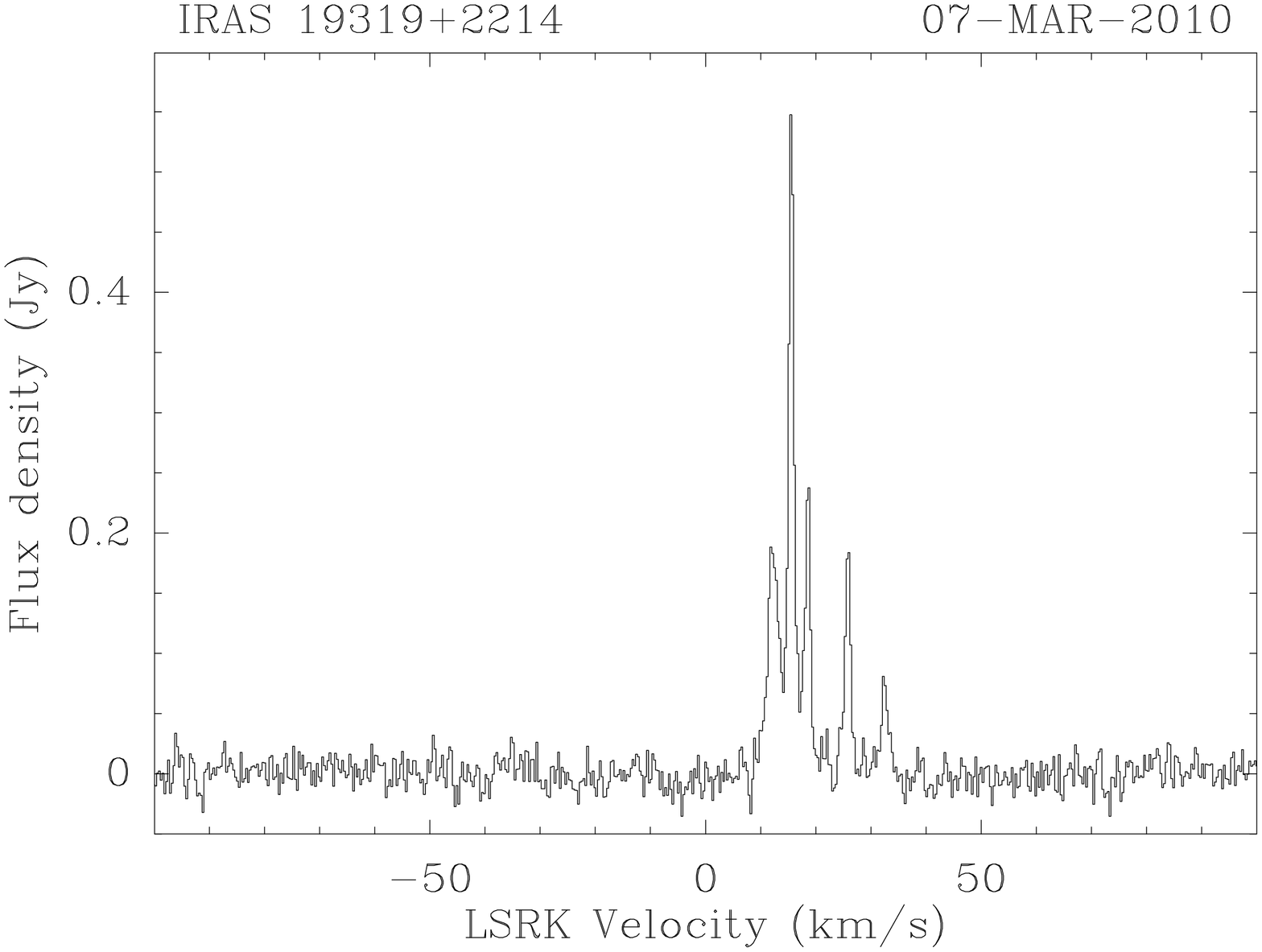}
\includegraphics[width=0.3\textwidth,angle=-90]{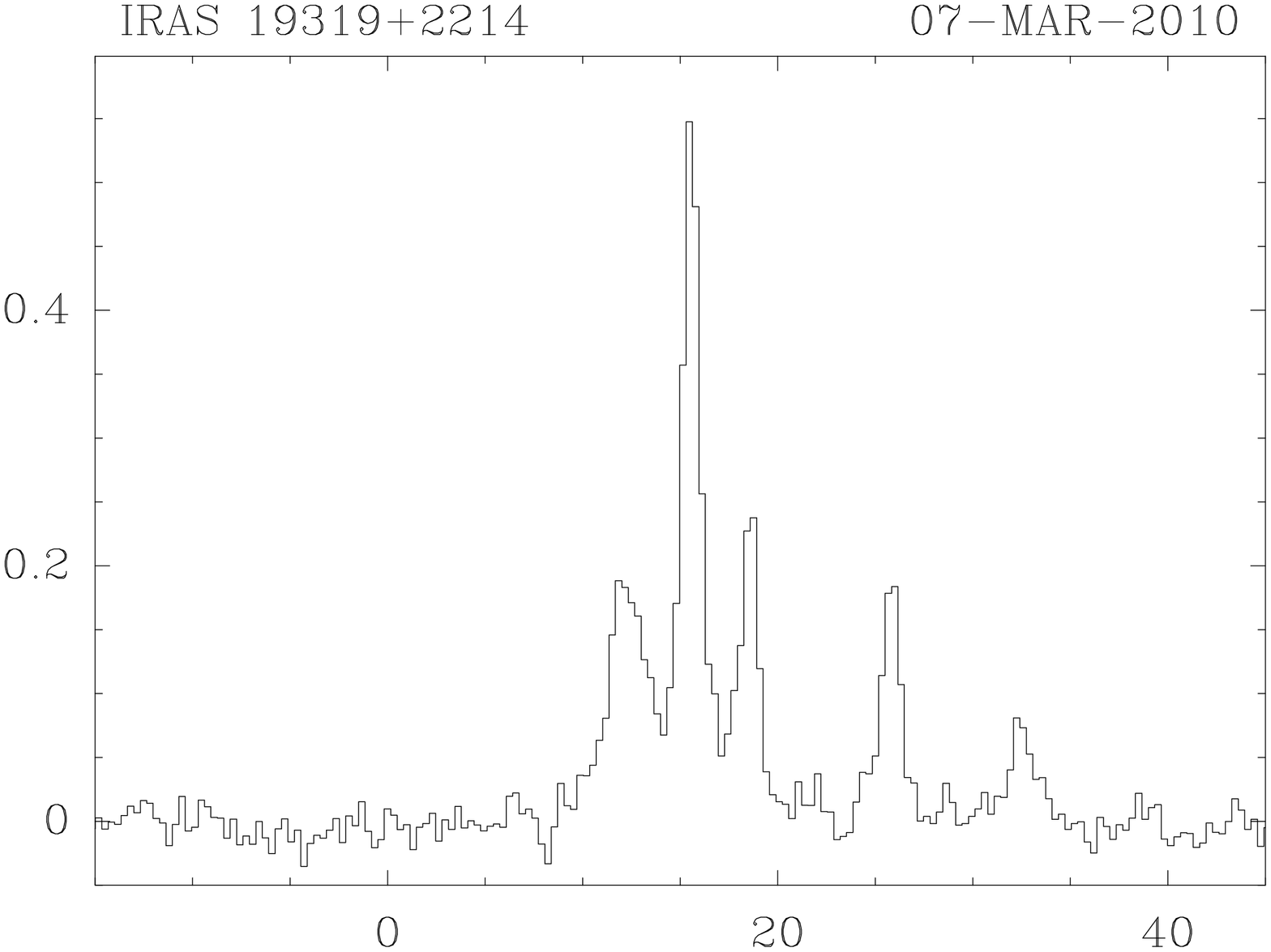}
\caption{
continued.}
\end{figure*}

}

%\subsection{Individual Sources}

\subsection{IRAS 13483$-$5905}

The nature of this source is still unclear. The similar fluxes in the IR between the 2MASS data \citep{rl1} and the observations in \cite{rl2}, should suggest a post-AGB nature. \cite{rl1} report an optical counterpart for this source. However, it is possible that the optical/near-IR source is not the same as  the mid/far-IR one.
The position given by  \cite{rl1} corresponds to the near-IR source 2MASS J13514373-5920155. On the other hand, the WISE data catalog report a mid-IR source, WISE J135144.47-592016.7, which is $\simeq 5.9''$ away from the near-IR position. On close inspection of the WISE images, we could see that the position reported WISE J135144.47-592016.7 corresponds to a strong source seen in the bands at 4.6, 12, and 22 $\mu$m, but there is a source at 3.4 $\mu$m whose position matches that of 2MASS J13514373-5920155.
Therefore, it seems that there are two different nearby sources, and the source seen at $\lambda\geq 4.6$ $\mu$m is an optically obscured source that also corresponds to IRAS 13483$-$5905. This is the most likely source associated to the water maser emission.
% 2mass: 13 51 43.73 -59 20 15.5\\
%IRSA: MSX 13h51m44.30s 	-59d20m17.88s\\
%AKARI FIS: 13h51m44.51s 	-59d20m18.60s\\
%AKARI IRC: 13h51m44.34s 	-59d20m17.74s\\
%WISE catalog: 13h51m44.48s 	-59d20m16.73s. The WISE position correspond to a strong source at 4.6, 12 and 22 micron (bands 2, 3 \& 4). There is a source seen in band 1 (3.4 micron) whose position coincides with that of optical and near-IR, but it is not in the wise catalog. So the wise data may correspond to different sources\\
%I would predict this is the source with the masers.\\
%2MASS J13514373-5920155 is NOT IRAS 13483-5905.
We note that \cite{tel91} reported non-detection of OH masers toward IRAS 13483$-$5905.

The H$_2$O spectrum shows at least ten different components, which seem to cluster into two groups. The blueshifted group comprises the components at $V_{\rm LSR}=-27.8$, $-21.9$, $-18.4$, $-15.2$, $-13.9$, $-6.3$ km s$^{-1}$, while the redshifted one is composed of components at $V_{\rm LSR}=15.2$, 18.5, 24.0, and 28.7 km s$^{-1}$. If we interpret each group as arising from opposite sides of the expanding envelope, the mean velocity of each group would be $\simeq -20.85$ and 21.95 km s$^{-1}$, so the stellar velocity is $\simeq 0.55$ km s$^{-1}$. However, the total velocity spread of the H$_2$O maser components ($\simeq 56$ km s$^{-1}$) would imply an expansion velocity of $\simeq 28$ km s$^{-1}$, which is relatively high for an expanding envelope around an AGB or post-AGB star (see sec. \ref{intro}). Interferometric observations would be needed to ascertain the nature of the structure traced by this H$_2$O  maser emission.

\subsection{IRAS 14249$-$5310}

\cite{rl1} report the absence of an optical counterpart for this source. The observations by \cite{rl2} show a point source in the near-IR. The comparison of infrared fluxes in different epochs \citep{fou92,rl2} shows a difference, with the source being brighter (by 1.65, 1.01, and 0.17 mag in J, H, and K bands, respectively) in the more recent observations. The 2MASS image shown in \cite{rl1} has a lower angular resolution, and it is not able to resolve the object from another nearby one, so it cannot be used to compare the source flux with the other two epochs. The flux variation with time suggests that IRAS 14249$-$5310 may be an AGB star.
This source does not show any detectable OH maser emission \citep{tel91,sil93}.

%But \cite{rl2} (observations July 2008) give $JHK= (11.79\pm 0.05, 9.32\pm 0.04, 7.93\pm 0.03)$, while \cite{fou92} (observations July 1989) give 
%$JHK= (13.44, 10.33, 8.10)$. So, they are brighter in the more recent observations, which suggests variability (AGB star).

Our water maser spectrum shows at least two components at $-19.8$ and $-10.1$  km s$^{-1}$. There are hints of other possible components between 30 and 50 km s$^{-1}$, but it is not possible to confirm them with these data.

\subsection{IRAS 15408$-$5413}

This source has no optical counterpart \citep{rl1}. It shows variability in its near-IR emission of at least 0.8 mag in K and 0.2 in L$'$ \citep[see the different photometric measurements in][]{epc87,leb88,leb90,leb93,pgl97,rl1}. 
However, its light curve does not show the periodic variability typical of OH/IR (AGB) stars \citep{leb93}. 
It has been suggested that the source might be a supergiant \citep{leb90,leb93} or a ``born again'' AGB star \citep{leb88}, although these classifications are uncertain. In any case, its IR variability seems to rule out a post-AGB or PN nature.

%\cite{tel91}
Te Lintel Hekkert et al. (1991) detected OH emission, with several components in emission and absorption. However, the spectrum  does not show the typical double-peaked profile of OH masers in AGB stars. The OH emission may arise from more than one source within their beam or, alternatively, trace independent episodes of mass loss.
There is also SiO maser emission \citep{leb90}, whose spectrum shows  a wide (12.2 km s$^{-1}$ width) and irregular profile, apparently due to the blending of several velocity components, with centroid LSR velocity of $-83.6$ km s$^{-1}$. 

\cite{deg89} failed to detect  H$_2$O maser emission with Parkes,
reaching an rms $=0.5$ Jy. Our spectrum (with an rms noise of 0.09 Jy) shows an asymmetric double-peaked profile, with components at $-82.6$ and $-76.3$ km s$^{-1}$. The velocity of the stronger component is close the the SiO centroid velocity, while that of the secondary peak is outside the bulk of SiO emission. Since SiO emission tends to be close to the stellar velocity, the stronger H$_2$O component seems to trace gas at this stellar velocity, while the secondary component might trace the receding part of the envelope. Both SiO and H$_2$O components are within the velocity range of OH maser emission. The flux density of 53 Jy we obtained should have been detected by \cite{deg89}, which indicates a strong variability of the water maser emission in this source.

\subsection{IRAS 15452$-$5459}

This object is a post-AGB star with an hour-glass shape in infrared images \citep{sah07,rl2}.
Maser emission of SiO, OH, and H$_2$O has been previously detected toward it \citep{tel91,cas98,dea04,dea07,cer13}.

The spectral shape of the H$_2$O maser has changed. While  \cite{dea07} show an irregular profile, our spectrum  and the profile in \cite{cer13}, both taken at later epochs,  tend to cluster in two groups of features, resembling a more regular double-peaked profile. The interferometric SiO, H$_2$O, and OH maser observations by \cite{cer13} show different spatio-kinematical patterns in these species. While OH masers seem to trace outflows along the lobes of the nebula, the SiO and H$_2$O maser components display a velocity gradient perpendicular to it, which could be tracing rotational motions in a circumstellar torus. 

\subsection{IRAS 17021$-$3109}

This is a heavily obscured object \citep{rl2} with no optical counterpart \citep{rl1}. The near-IR fluxes of the most likely counterpart for the IRAS sources are similar in different epochs \citep{rl1,rl2}, suggesting a post-AGB nature. We note that \citet{pgl97} report brighter magnitudes, but they might correspond to a different object in the field.

The spectrum of OH maser emission at 1612 MHz \citep{dav93a} shows at least four peaks between $\sim -30.5$ and -5.5 km s$^{-1}$. OH emission at 1667 MHz has also been detected \citep{dav93b}, with a single peak at -7.5 km s$^{-1}$.

Our H$_2$O maser spectrum shows a single peak at $+13.5$ km s$^{-1}$. It lies well outside the velocity range of the OH emission. Therefore, it is possible that OH and H$_2$O masers arise from different nearby sources  but with a velocity difference between them of $\sim 30$ km s$^{-1}$. Alternatively, if the emission from both maser species arises from the same source, it would indicate the presence of non-spherical mass loss.

\subsection{IRAS 17291$-$2147}

This source has a weak optical counterpart \citep{rl1}. 
Its near-IR fluxes are similar in different epochs \citep{rl1,rl2}, which suggests a post-AGB nature. 
%$(J,H,K)=(13.507\pm 0.021, 12.188\pm 0.022,  11.008\pm 0.019)$ in 2MASS. In \cite{rl2}, $(J,H,K)=13.49\pm 0.04, 12.25\pm  0.04, 11.081\pm 0.026)$. Similar flux, so might be a post-AGB star
No OH maser emission has been detected toward this source \citep{tel91}. 

\cite{yoo14} report the presence of H$_2$O maser emission, with only one velocity component at $V_{\rm LSR}=-16.3$ km s$^{-1}$ having been clearly detected. Our Robledo spectrum  resolves two spectral components around that velocity.
However, the GBT spectrum, with a much higher signal-to-noise ratio, shows at least four components, at $-27.0$, $-16.1$, $-14.5$, and $+69.1$  km s$^{-1}$. Other possible weaker components might be present, but we cannot confirm them at this stage. The total velocity spread of the H$_2$O maser components is large, $\simeq 96$ km s$^{-1}$. This means that IRAS 17291$-$2147 is a candidate water fountain source. Its confirmation as such would require interferometric observations to ascertain that all velocity components observed with single-dish arise from the same source.

\subsection{IRAS 17348$-$2906}

This object has an optical counterpart \citep{rl1}. The near-IR fluxes in the observations of \citet{pgl97} are similar to those in 2MASS data \citep{rl1}, suggesting it is a post-AGB star.

OH maser emission has been reported in this source, between $\sim 8$ and 13 km s$^{-1}$ \citep{tel91,dav93a}. Our H$_2$O maser spectrum shows at least three components: two narrow ones at $\simeq 1.0, 2.6$  km s$^{-1}$ and a wider one between 6 and 9 km s$^{-1}$, which could be the blending of several unresolved ones. Most of the H$_2$O maser emission lies outside the velocity range of OH maser emission, so it is unclear whether this indicates non-spherical mass loss or if the emission from both molecules arises from different sources.

\subsection{IRAS 17393$-$2727 (OH 0.9+1.3)}

This source shows a bipolar morphology in the optical \citep{man11}, and its IR colors suggest that the central star is strongly obscured \citep{rl2}. The presence of bright [Ne II] emission indicates that it is a planetary nebula \citep{gar07}.

IRAS 17393$-$2727 presents both OH maser \citep{zij89,sev97} and radio continuum emission \citep{pot87}, whose spatial association have been confirmed with interferometric observations, making it one of the six confirmed planetary nebulae with OH emission \citep{usc12}. The OH spectrum at 1612 MHz shows two peaks   located at $V_{\rm LSR} =  -122.8$  and $-93$ km s$^{-1}$.

Previous observations of water maser emission toward this source resulted in non-detections 
\citep{gom90,dea07,yoo14}. 
%deacon list rms between 0.04 and 0.13 Jy rms for their observations, but does not give value for each one
% yolanda gives 0.5-2 Jy
However, we detected water maser emission with a single component of $\simeq 1$ Jy at $V_{\rm LSR} = -107.6$ km s$^{-1}$, above the detection threshold of \cite{dea07} and \cite{yoo14}. Although with our single-dish observations we cannot determine whether the water maser emission is associated with IRAS 17393$-$2727, the velocity of the observed component is very close to the mean velocity of the OH peaks ($\sim -107.9$ km s$^{-1}$), which is the expected central velocity of the star. This would be a very unlikely coincidence if the water maser emission arises from a different source. Therefore, we think this is a most probable candidate to also be a water-maser-emitting PN. The presence of both OH and H$_2$O emission and its strong obscuration make it a PN very similar to K 3-35 \citep{mir01}, IRAS 17347$-$3139 \citep{ideg04}, and IRAS  16333$-$4807 \citep{usc14} and, like them, an extremely young PN.

\subsection{IRAS 18039$-$1903}

This source does not have an optical counterpart, and it is very weak in the near-IR \citep{rl2}. Its infrared colors confirm that it is a heavily obscured object \citep{rl2}.

OH maser emission has been detected in this source \citep{tel91,dav93a,dav93b}, showing a double-peaked profile  characteristic of OH/IR stars, with LSR velocities $\simeq 143.3$ and $170.1$ km s$^{-1}$. 
H$_2$O maser emission has been detected by \cite{yoo14}, with two distinct velocity components. Our spectrum is similar, but it shows four components at 153.1, 162.0, 162.9, and 165.2  km s$^{-1}$. The peak emission is slightly blueshifted with respect to the stellar velocity inferred from the OH spectrum ($\simeq 157.6$ km s$^{-1}$), while the other three are redshifted. All H$_2$O maser components are within the velocity range of the OH emission.

\subsection{IRAS 18113$-$2503}

The observations presented here yielded the first detection of water maser emission toward this source, although interferometric follow-up observations have already been reported in  \cite{gom11}. That paper describes the source in detail. As mentioned there, it seems to be a post-AGB star. The water maser spectrum shows a wealth of components, spanning an extremely wide velocity range of $\simeq 500$ km s$^{-1}$. This is the water fountain with the fastest jet known to date.

\subsection{IRAS 18361$-$1203}

This source has a weak optical counterpart in red DSS plates, and its infrared fluxes do not seem to vary significantly \citep{rl1,rl2}, which suggests this is a post-AGB star.
IRAS 18361$-$1203 shows a double-peaked profile in OH \citep{tel91,dav93a} with velocities $\sim 11.77$ and $19.32$ km s$^{-1}$.

\cite{yoo14} report a non-detection of water maser emission, with a $3\sigma$ upper limit of $\sim 1.3$ Jy. Our spectrum shows weak emission, well below their detection limit, with a single confirmed component at $\simeq 12.7$ km s$^{-1}$. 

\subsection{IRAS 18596+0315 (OH 37.1-0.8)}

This object is weak in the near-IR, and it has infrared colors  that indicate strong obscuration \citep{rl2}. 
OH maser emission is present in this source and shows a double-peaked profile with $\sim 28$ km s$^{-1}$ separation between them \citep{win75,bow78,bau85,her85h,her85,che93,gom94,sev01,szy04,en07}. The OH central velocity is close to that of the CO emission \citep{riz13}. OH maser emission at 1665 and 1667 MHz \citep{dic91,lew97,szy04} extends beyond the velocity range of the emission  at 1612 MHz. 

Interferometric observations of OH \citep{bau85,her85,gom94,gom00,sev01,ami11} show that the two peaks are actually the blend of different features. Blue- and redshifted groups are spatially separated by $\sim 150$ mas along the E-W direction. It is unclear whether OH emission arises from a biconical or an equatorial flow.
\cite{jew91} reported a tentative detection of SiO, but it has not been confirmed in subsequent observations \citep{deg04}.
%nym93 and nym98 did not have enough sensitivity.

Water maser emission has been previously detected in this source \citep{eng86,bra94,gom94,eng02,dea07}, showing a variable spectral pattern of multiple components, extending over a range $\sim 57$  km s$^{-1}$, beyond the velocity range of OH. This strongly suggests it is a water fountain, although there is no published map showing that the water maser emission traces a jet. \cite{gom94} proved that OH and water maser emission arise from the same source, confirming that the latter lies outside the velocity range of the former, but they could not determine the spatial distribution of the water masers or even confirm that all the maser emission detected with single dish actually arise from this source, since their interferometric observation only showed a single peak of water maser emission at $V_{\rm LSR}\simeq 65.7$ km s$^{-1}$. Our single-dish data also shows a single component, at $V_{\rm LSR}\simeq 61.6$ km s$^{-1}$.

\subsection{IRAS 19134+2131}

This is a confirmed water fountain, and the only one in which OH emission has not been detected, despite several searches \citep{lew87,hu94,lik89}.

Water maser emission was first detected in this source by  \cite{eng84}, who identified a wide velocity spread ($\sim 110$ km s$^{-1}$) in their components. The water maser emission traces a collimated jet \citep{ima04,ima07b}.
%\cite{lew88,ces88,eng88,lik89,lik92}
Our spectrum shows five distinct components spanning a range of $\simeq 115$ km s$^{-1}$. 

\subsection{IRAS 19190+1102}

This object is a known water fountain, which was first detected by \cite{lik89}, who obtained a water maser spectrum with a wide velocity range of $\geq 70$ km s$^{-1}$, well outside the velocity range covered by the OH maser emission. The distribution and proper motions of the H$_2$O maser emission \citep{day10} confirms that it arises from a collimated jet. The departure of the OH spectrum from the typical double-peaked profile in OH/IR stars \citep{lik89} suggests that this object is already in the post-AGB phase.

Our water maser spectrum shows components spanning $\sim 143$ km s$^{-1}$. We note that this velocity range is slightly more than found by \cite{day10}. In particular, our components with the most extreme velocities ($-55.0$ and $+87.9$ km s$^{-1}$) were not present in their spectra, although they were above their sensitivity limit. While the appearance of these new components could merely be the result of the typical variability of water masers, it might also indicate an acceleration process in the jet.

\subsection{IRAS 19319+2214}

This is an optically obscured source, with a relatively constant near-IR flux \citep{rl1,rl2}, which suggests it is a post-AGB star.

It shows a double-peaked OH spectrum \citep{lew90,dav93a}, with components at $\sim 12$ and $29$ km s$^{-1}$. Water maser emission, spanning a velocity range of $\simeq 12$ km s$^{-1}$, was first detected by \cite{eng96}. Our GBT water maser spectrum shows five components over a wider velocity spread ($\sim 20$ km s$^{-1}$), whose most extreme velocities are similar to those seen in OH.

\section{Discussion and conclusions}

%\subsection{Detection rates}

The detection rate of water masers seems to be higher in obscured
objects (7\% and 14\% in objects with and without an optical
counterpart, respectively). This is consistent with our results in
previous surveys \citep{sua07,sua09}. Another significant trend in our
data is the high detection rate of water masers in bipolar objects. Out of the seven sources in our sample that are clearly bipolar in the near-IR \citep{rl2}, we detected water maser emission in three. This high detection rate (43\%) clearly suggests that water maser emission tends to be associated with collimated mass-loss processes, either directly (tracing jets, as in water fountains) or indirectly (tracing toroidal structures perpendicular to the bipolar nebulae). However, all these detection rates should be taken with care, since the sensitivity of our observations is not homogeneous throughout the sample.

%\subsection{On the nature of new water fountain candidates}

We have identified a water fountain candidate (IRAS 17291$-$2147) with a total velocity spread of $\sim 96$ km s$^{-1}$ in its water maser components. Another two sources (IRAS 17021$-$3109 and IRAS 17348$-$2906) show water maser emission whose velocity lies outside the velocity range covered by OH masers, a criterion that is also used to identify water fountain candidates.  IRAS 17291$-$2147, however, has a very scarce number of maser components, as in the case of IRAS 19134+2131 or IRAS 18596+0315. The spectrum of these water fountain candidates is strikingly different from those of IRAS 18113$-$2503 or IRAS 19190+1102, which are very rich in maser components. This suggests an intrinsically different nature of these sources, although they are collectively included in the ``water fountain'' category based on the wide velocity spread of their masers. Also, IRAS 17291$-$2147 qualifies as a water fountain candidate, because of the detection of a very weak component at $\sim 69$ km s$^{-1}$, which was below the sensitivity of previous observations. Moreover, given the high variability of water maser emission, components at extreme velocities may be above or below the sensitivity of the telescope depending on the epoch of observation. 

If the identification of a source as a water fountain depends so much on the sensitivity or the epoch of the observations, there is thus an observational bias in the definition of this type of source. In fact, post-AGB stars have high-velocity winds of several hundred km s$^{-1}$, as seen in optical and infrared data \citep[e.g.,][]{rie95,wit09,san10}, so the presence of these velocities in their maser spectra should not be surprising. It is possible that most water-maser-emitting post-AGBs will show high-velocity maser components when observed with high enough sensitivity or if monitored for a long enough time. 
Also, a particular orientation (close to the plane of the sky) of the jets traced by water masers  could make components with intrinsically high velocities not appear as such in a spectrum. Therefore, high-velocity jets traced by water masers could be much more widespread in post-AGB stars than previously thought.  If this were the case, the segregation of a group of water-maser-emitting post-AGBs as "water fountains", based only on the velocity spread of the masers, would lose its meaning to a certain extent. We think it would be interesting to further differentiate  those "water fountains" with  high maser luminosities as a special group. This spectral characteristic could be due to a higher mass in their progenitor or to their being in a different evolutionary stage. Monitoring of the water maser spectra of ``water fountains'' or a better characterization of their stellar properties would shed some light on the origin of these spectral differences.

%\subsection{On the nature of the new water-maser-emitting PN candidate}

We also identified IRAS 17393$-$2727 as a possible new H$_2$O-PN. This is the sixth source that could pertain to this category. This PN also  shows OH emission. The PNe K3-35 and IRAS 17347$-$3139 also show maser emission from both molecules. PN emitting OH and/or H$_2$O masers are supposed to be very young PNe. In fact, most of them are optically obscured and spatially compact, with the exception of the NGC 6302 (an OHPN) and IRAS 18061$-$2505 \citep{sua06}, which show extended optical emission. These two sources only show maser emission from one of the species (OH in NGC 6302 and H$_2$O in IRAS 18061$-$2505), while observations of the other yielded negative results \citep{dea07,gom08}. We suggest that the presence of both maser species in a PN is a better tracer of its youth, rather than the presence of just one of them.

Interferometric observations will be needed to confirm the nature of IRAS 17291$-$2147, IRAS 17021$-$3109, and IRAS 17348$-$2906 as water fountains, and of IRAS 17393$-$2727 as an H$_2$O-PN. These observations will indicate whether all maser components arise from these evolved objects.

\begin{acknowledgements}
%J.F.G. and L.F.M. acknowledge partial support from
%Ministerio de Educaci\'on y Ciencia (Spain) grants AYA2005-08523-C03-03
%and AYA2005-01495, respectively. 
%L.F.M, O.S., J.F.G., J.M.T.,  are partially supported
 This work is partially
based on observations carried out at MDSCC (Robledo),
under the Host Country Radio Astronomy program.
The Parkes radio telescope is part of the Australia Telescope National Facility, which is funded by the Commonwealth of Australia for operation as a National Facility managed by CSIRO. The National Radio Astronomy Observatory
is a facility of the National Science Foundation
operated under cooperative agreement by Associated
Universities, Inc. 
JFG, OS, LFM, and JMT are supported by MICINN (Spain) grants AYA2011-30228-C03 and AYA2014-57369-C3, while
MAG is supported by grant AYA2011-29754-C03-02 (all these grants include FEDER funds).  J.R.R. acknowledges support
from MICINN grants AYA2009-07304 and
AYA2012-32032.  AP acknowledges the financial support from UNAM and CONACyT, M\'exico.
G.R.L. acknowledges support
from CONACyT and PROMEP (Mexico).
\end{acknowledgements}

%TABLES


\begin{thebibliography}{}

\bibitem[Amiri et 
al.(2011)]{ami11} Amiri, N., Vlemmings, W., \& van Langevelde, H.~J.\ 2011, \aap, 532, A149 
\bibitem[Baud et al.(1985)]{bau85} Baud, B., Sargent, A.~I., 
Werner, M.~W., \& Bentley, A.~F.\ 1985, \apj, 292, 628 
\bibitem[Bowers(1978)]{bow78} Bowers, P.~F.\ 1978, \aap, 64, 307 
\bibitem[Bl\"ocker(1995a)]{blo95a} Bl\"ocker, T.\ 1995a, \aap, 297, 727 
\bibitem[Bl\"ocker(1995b)]{blo95b} Bl\"ocker, T.\ 1995b, \aap, 299, 755 
\bibitem[Brand et 
al.(1994)]{bra94} Brand, J., Cesaroni, R., Caselli, P., et al.\ 1994, \aaps, 103, 541 
\bibitem[Caswell(1998)]{cas98} Caswell, J.~L.\ 1998, \mnras, 
297, 215 
\bibitem[Cerrigone et al.(2013)]{cer13} Cerrigone, L., 
Menten, K.~M., \& Wiesemeyer, H.\ 2013, \mnras, 434, 542 
\bibitem[Cesaroni et 
al.(1988)]{ces88} Cesaroni, R., Palagi, F., Felli, M., et al.\ 1988, \aaps, 76, 445 
\bibitem[Chengalur et al.(1993)]{che93} Chengalur, J.~N., 
Lewis, B.~M., Eder, J., \& Terzian, Y.\ 1993, \apjs, 89, 189 
\bibitem[Comoretto et 
al.(1990)]{com90} Comoretto, G., Palagi, F., Cesaroni, R., et al.\ 1990, \aaps, 84, 179 
\bibitem[David et 
al.(1993a)]{dav93a} David, P., Le Squeren, A.~M., \& Sivagnanam, P.\ 1993, \aap, 277, 453 
\bibitem[David et 
al.(1993b)]{dav93b} David, P., Le Squeren, A.~M., Sivagnanam, P., \& Braz, M.~A.\ 1993, \aaps, 98, 245 
\bibitem[Day et al.(2010)]{day10} Day, F.~M., Pihlstr{\"o}m, 
Y.~M., Claussen, M.~J., \& Sahai, R.\ 2010, \apj, 713, 986 
\bibitem[Deacon et al.(2004)]{dea04} Deacon, R.~M., Chapman, 
J.~M., \& Green, A.~J.\ 2004, \apjs, 155, 595 
\bibitem[{{Deacon} {et~al.}(2007){Deacon}, {Chapman}, {Green}, \&
  {Sevenster}}]{dea07} {Deacon}, R.~M., {Chapman}, J.~M., {Green}, A.~J., \& {Sevenster}, M.~N. 2007, \apj, 658, 1096
\bibitem[de Gregorio-Monsalvo et al.(2004)]{ideg04} de 
Gregorio-Monsalvo, I., G{\'o}mez, Y., Anglada, G., et al.\ 2004, \apj, 601, 
921 
\bibitem[Deguchi et al.(2004)]{deg04} Deguchi, S., Fujii, T., 
Glass, I.~S., et al.\ 2004, \pasj, 56, 765 
  \bibitem[Deguchi et al.(1989)]{deg89} Deguchi, S., Nakada, 
Y., \& Forster, J.~R.\ 1989, \mnras, 239, 825 
\bibitem[Dickinson 
\& Turner(1991)]{dic91} Dickinson, D.~F., \& Turner, B.~E.\ 1991, \apjs, 75, 1323 
\bibitem[Engels(2002)]{eng02} Engels, D.\ 2002, \aap, 388, 252 
\bibitem[Engels et 
al.(1984)]{eng84} Engels, D., Habing, H.~J., Olnon, F.~M., Schmid-Burgk, J., \& Walmsley, C.~M.\ 1984, \aap, 140, L9 
\bibitem[Engels 
\& Jim{\'e}nez-Esteban(2007)]{en07} Engels, D., \& Jim{\'e}nez-Esteban, F.\ 2007, \aap, 475, 941 
\bibitem[Engels 
\& Lewis(1996)]{eng96} Engels, D., \& Lewis, B.~M.\ 1996, \aaps, 116, 117 
\bibitem[Engels et 
al.(1986)]{eng86} Engels, D., Schmid-Burgk, J., \& Walmsley, C.~M.\ 1986, \aap, 167, 129 
 \bibitem[Engels et 
al.(1988)]{eng88} Engels, D., Schmid-Burgk, J., \& Walmsley, C.~M.\ 1988, \aap, 191, 283 
 \bibitem[Epchtein et 
al.(1987)]{epc87} Epchtein, N., Le Bertre, T., Lepine, J.~R.~D., et al.\ 1987, \aaps, 71, 39 
  \bibitem[Fouque et 
al.(1992)]{fou92} Fouque, P., Le Bertre, T., Epchtein, N., Guglielmo, F., \& Kerschbaum, F.\ 1992, \aaps, 93, 151 
\bibitem[Garc{\'{\i}}a-Hern{\'a}ndez et al.(2007)]{gar07} 
Garc{\'{\i}}a-Hern{\'a}ndez, D.~A., Perea-Calder{\'o}n, J.~V., Bobrowsky, 
M., \& Garc{\'{\i}}a-Lario, P.\ 2007, \apjl, 666, L33 
\bibitem[Garcia-Lario et 
al.(1997)]{pgl97} Garcia-Lario, P., Manchado, A., Pych, W., \& Pottasch, S.~R.\ 1997, \aaps, 126, 479 
\bibitem[G{\'o}mez et al.(2011)]{gom11} G{\'o}mez, J.~F., 
Rizzo, J.~R., Su{\'a}rez, O., et al.\ 2011, \apjl, 739, L14 
\bibitem[G{\'o}mez et al.(2008)]{gom08} G{\'o}mez, J.~F., 
Su{\'a}rez, O., G{\'o}mez, Y., et al.\ 2008, \aj, 135, 2074 
\bibitem[Gomez et al.(2015a)]{gom15a} G{\'o}mez, J.~F., 
Su{\'a}rez, O., Bendjoya, P., et al.\ 2015, \apj, 799, 186 
\bibitem[G{\'o}mez et al.(2015b)]{gom15b} G{\'o}mez, J.~F., 
Uscanga, L., Su{\'a}rez, O.\ 2015b, in 20.DA.10
Research and Teaching in Astrophysics in Guanajuato, ed. M.~T. Trinidad,
H. Andernach, \& M. Avalos (Guanajuato: Univ. Guanajuato), in press
\bibitem[G{\'o}mez et al.(2014)]{gom14} G{\'o}mez, J.~F., 
Uscanga, L., Su{\'a}rez, O., Rizzo, J.~R., 
\& de Gregorio-Monsalvo, I.\ 2014, \rmxaa, 50, 137 
\bibitem[G{\'o}mez et al.(1990)]{gom90} G{\'o}mez, Y., Moran, 
J.~M., \& Rodr{\'{\i}}guez, L.~F.\ 1990, \rmxaa, 20, 55 
\bibitem[G{\'o}mez 
\& Rodr{\'{\i}}guez(2000)]{gom00} G{\'o}mez, Y., \& Rodr{\'{\i}}guez, L.~F.\ 2000, in Asymmetrical Planetary Nebulae II: From Origins to Microstructures, ed.  J.~H. Kastner, N. Soker, \& S.~A. Rappaport, (San Francisco: Astronomical Society of the Pacific), Astronomical Society of the Pacific Conference Series, 199, 75 
\bibitem[G{\'o}mez et al.(1994)]{gom94} G{\'o}mez, Y., 
Rodr{\'{\i}}guez, L.~F., Contreras, M.~E., 
\& Moran, J.~M.\ 1994, \rmxaa, 28, 97 
\bibitem[Han et 
al.(1998)]{han98} Han, F., Mao, R.~Q., Lu, J., et al.\ 1998, \aaps, 127, 181 
\bibitem[Herman et 
al.(1985)]{her85} Herman, J., Baud, B., Habing, H.~J., \& Winnberg, A.\ 1985, \aap, 143, 122 
\bibitem[Herman 
\& Habing(1985)]{her85h} Herman, J., \& Habing, H.~J.\ 1985, \aaps, 59, 523 
\bibitem[Hu et 
al.(1994)]{hu94} Hu, J.~Y., te Lintel Hekkert, P., Slijkhuis, F., et al.\ 1994, \aaps, 103, 301 
\bibitem[Imai et 
al.(2004)]{ima04} Imai, H., Morris, M., Sahai, R., Hachisuka, K., \& Azzollini F., J.~R.\ 2004, \aap, 420, 265 
\bibitem[Imai(2007a)]{ima07a} Imai, H.\ 2007, in IAU Symp. 242, Astrophysical Masers and Their Environments,
ed. J. M. Chapman \& W. A. Baan (Cambridge: Cambridge Univ.
Press), 279
\bibitem[Imai et al.(2007b)]{ima07b} Imai, H., Sahai, R., 
\& Morris, M.\ 2007b, \apj, 669, 424 
\bibitem[Jewell et 
al.(1991)]{jew91} Jewell, P.~R., Snyder, L.~E., Walmsley, C.~M., Wilson, T.~L., \& Gensheimer, P.~D.\ 1991, \aap, 242, 211 
\bibitem[Le Bertre et 
al.(1988)]{leb88} Le Bertre, T., Heydari-Malayeri, M., \& Epchtein, N.\ 1988, \aap, 197, 143 
\bibitem[Le Bertre 
\& Nyman(1990)]{leb90} Le Bertre, T., \& Nyman, L.-A.\ 1990, \aap, 233, 477 
\bibitem[Le 
Bertre(1993)]{leb93} Le Bertre, T.\ 1993, \aaps, 97, 729 
\bibitem[Lewis(1989)]{lew89} Lewis, B.~M.\ 1989, \apj, 338, 
234 
\bibitem[Lewis(1997)]{lew97} Lewis, B.~M.\ 1997, \apjs, 109, 
489 
\bibitem[Lewis et al.(1987)]{lew87} Lewis, B.~M., Eder, J., 
\& Terzian, Y.\ 1987, \aj, 94, 1025 
\bibitem[Lewis et al.(1990)]{lew90} Lewis, B.~M., Eder, J., 
\& Terzian, Y.\ 1990, \apj, 362, 634 
\bibitem[Lewis 
\& Engels(1988)]{lew88} Lewis, B.~M., \& Engels, D.\ 1988, \nat, 332, 49 
\bibitem[Likkel(1989)]{lik89} Likkel, L.\ 1989, \apj, 344, 
350
\bibitem[Likkel et 
al.(1992)]{lik92} Likkel, L., Morris, M., \& Maddalena, R.~J.\ 1992, \aap, 256, 581 
\bibitem[Manteiga et al.(2011)]{man11} Manteiga, M., 
Garc{\'{\i}}a-Hern{\'a}ndez, D.~A., Ulla, A., Manchado, A., 
\& Garc{\'{\i}}a-Lario, P.\ 2011, \aj, 141, 80 
\bibitem[Migenes et al.(1999)]{mig99} Migenes, V., Horiuchi, 
S., Slysh, V.~I., et al.\ 1999, \apjs, 123, 487 
\bibitem[Miranda et al.(2001)]{mir01} Miranda, L.~F., 
G{\'o}mez, Y., Anglada, G., \& Torrelles, J.~M.\ 2001, \nat, 414, 284 
%\bibitem[Nyman et 
%al.(1993)]{nym93} Nyman, L.-A., Hall, P.~J., \& Le Bertre, T.\ 1993, \aap, 280, 551 
%\bibitem[Nyman et 
%al.(1998)]{nym98} Nyman, L.-A., Hall, P.~J., \& Vlofsson, H.\ 1998, \aaps, 127, 185 
\bibitem[Pottasch et 
al.(1987)]{pot87} Pottasch, S.~R., Bignell, C., \& Zijlstra, A.\ 1987, \aap, 177, L49 
  \bibitem[Ramos-Larios et 
al.(2009)]{rl1} Ramos-Larios, G., Guerrero, M.~A., Su{\'a}rez, O., Miranda, L.~F., \& G{\'o}mez, J.~F.\ 2009, \aap, 501, 1207 
  \bibitem[Ramos-Larios et 
al.(2012)]{rl2} Ramos-Larios, G., Guerrero, M.~A., Su{\'a}rez, O., Miranda, L.~F., \& G{\'o}mez, J.~F.\ 2012, \aap, 545, A20 
\bibitem[Reid 
\& Moran(1981)]{rei81} Reid, M.~J., \& Moran, J.~M.\ 1981, \araa, 19, 231 
\bibitem[Reid et al.(1977)]{rei77} Reid, M.~J., Muhleman, 
D.~O., Moran, J.~M., Johnston, K.~J., 
\& Schwartz, P.~R.\ 1977, \apj, 214, 60 
\bibitem[Riera et 
al.(1995)]{rie95} Riera, A., Garc\'{\i}a-Lario, P., Manchado, A., Pottasch, S.~R., \& Raga, A.~C.\ 1995, \aap, 302, 137 
\bibitem[Rizzo et 
al.(2013)]{riz13} Rizzo, J.~R., G{\'o}mez, J.~F., Miranda, L.~F., et al.\ 2013, \aap, 560, A82 
\bibitem[Sahai et al.(2007)]{sah07} Sahai, R., Morris, M., 
S{\'a}nchez Contreras, C., \& Claussen, M.\ 2007, \aj, 134, 2200 
\bibitem[S{\'a}nchez Contreras et al.(2010)]{san10} 
S{\'a}nchez Contreras, C., Cortijo-Ferrero, C., Miranda, L.~F., 
Castro-Carrizo, A., \& Bujarrabal, V.\ 2010, \apj, 715, 143 
\bibitem[Sevenster et 
al.(1997)]{sev97} Sevenster, M.~N., Chapman, J.~M., Habing, H.~J., Killeen, N.~E.~B., \& Lindqvist, M.\ 1997, \aaps, 122, 79 
\bibitem[Sevenster et 
al.(2001)]{sev01} Sevenster, M.~N., van Langevelde, H.~J., Moody, R.~A., et al.\ 2001, \aap, 366, 481 
\bibitem[Silva et 
al.(1993)]{sil93} Silva, A.~M., Azcarate, I.~N., Poppel, W.~G.~L., \& Likkel, L.\ 1993, \aap, 275, 510 
\bibitem[Su{\'a}rez et 
al.(2006)]{sua06} Su{\'a}rez, O., Garc{\'{\i}}a-Lario, P., Manchado, A., et al.\ 2006, \aap, 458, 173 
\bibitem[Su{\'a}rez et 
al.(2009)]{sua09} Su{\'a}rez, O., G{\'o}mez, J.~F., Miranda, L.~F., et al.\ 2009, \aap, 505, 217 
\bibitem[Su{\'a}rez et 
al.(2007)]{sua07} Su{\'a}rez, O., G{\'o}mez, J.~F., \& Morata, O.\ 2007, \aap, 467, 1085 
\bibitem[Szymczak 
\& G{\'e}rard(2004)]{szy04} Szymczak, M., \& G{\'e}rard, E.\ 2004, \aap, 423, 209 
\bibitem[te Lintel Hekkert et 
al.(1991)]{tel91} te Lintel Hekkert, P., Caswell, J.~L., Habing, H.~J., et al.\ 1991, \aaps, 90, 327 
\bibitem[Uscanga et 
al.(2012)]{usc12} Uscanga, L., G{\'o}mez, J.~F., Su{\'a}rez, O., \& Miranda, L.~F.\ 2012, \aap, 547, A40 
\bibitem[Uscanga et al.(2014)]{usc14} Uscanga, L., G{\'o}mez, 
J.~F., Miranda, L.~F., et al.\ 2014, \mnras, 444, 217 
\bibitem[Vassiliadis 
\& Wood(1993)]{vas93} Vassiliadis, E., \& Wood, P.~R.\ 1993, \apj, 413, 641 
\bibitem[Vassiliadis 
\& Wood(1994)]{vas94} Vassiliadis, E., \& Wood, P.~R.\ 1994, \apjs, 92, 125 
\bibitem[Winnberg et 
al.(1975)]{win75} Winnberg, A., Nguyen-Quang-Rieu, Johansson, L.~E.~B., \& Goss, W.~M.\ 1975, \aap, 38, 145 
\bibitem[Witt et al.(2009)]{wit09} Witt, A.~N., Vijh, U.~P., 
Hobbs, L.~M., et al.\ 2009, \apj, 693, 1946 
\bibitem[Yoon et al.(2014)]{yoo14} Yoon, D.-H., Cho, S.-H., 
Kim, J., Yun, Y.~j., \& Park, Y.-S.\ 2014, \apjs, 211, 15 
\bibitem[Yung et al.(2013)]{yun13} Yung, B.~H.~K., Nakashima, 
J.-i., Imai, H., et al.\ 2013, \apj, 769, 20 
\bibitem[Zijlstra et 
al.(1989)]{zij89} Zijlstra, A.~A., te Lintel Hekkert, P., Pottasch, S.~R., et al.\ 1989, \aap, 217, 157
\bibitem[Zuckerman 
\& Lo(1987)]{zuc87} Zuckerman, B., \& Lo, K.~Y.\ 1987, \aap, 173, 263  
\end{thebibliography}
\end{document}